\documentclass[twoside,12pt]{book}
\usepackage{graphicx,amssymb,amsfonts,amsmath,amsthm,eucal}
\usepackage{epsfig,psfrag,multicol}
\usepackage{color}
\input{epsf}
\makeindex

\begin{document}

\thispagestyle{empty} {\large
\renewcommand{\baselinestretch}{1.5}

$$$$

\vspace{-3cm}

\small
\centerline{Institute f\"ur Theoretische Physik}

\centerline{Fakult\"at Mathematik und Naturwissenschaften}

\centerline{Technische Universit\"at Dresden}

\vspace{1.6cm}

\renewcommand{\baselinestretch}{1}
\huge
{\bf
\centerline{Wavefunction-based method for}

\centerline{excited-state electron correlations in}

\centerline{periodic systems}
}

\renewcommand{\baselinestretch}{1.5}
{\large \bf
\centerline{--- application to polymers ---}
}

\vspace{1.3cm}
\normalsize

\centerline{Dissertation}

\centerline{zur Erlangung des}

\centerline{Doktorgrades der Naturwissenschaften}

\centerline{(Doctor rerum naturalium)}

\vspace{1.6cm}

\centerline{vorgelegt von}

\centerline{\large \bf Viktor Bezugly}

\centerline{geboren am 7. Juli 1976 in Charkow}

\vspace{5.8cm}

\centerline{\large Dresden~~~~2004}

\vspace{5mm}

\centerline{ \makebox{\psfig{figure=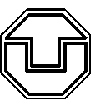,width=0.35in}}}

\newpage

\thispagestyle{empty}

$$$$
\vspace{15.8cm}
\normalsize

Eingereicht am 17. Dezember 2003

\vspace{0.3cm}

1. Gutachter: Prof. Dr. Peter Fulde

2. Gutachter: Dr. habil. Uwe Birkenheuer

3. Gutachter: Prof. Dr. Hermann Stoll

\vspace{5mm}

Verteidigt am 25. Februar 2004

} \setlength{\textheight}{8.7in}
\newpage

\thispagestyle{empty}

\centerline{\Large Abstract}

$$$$

In this work a systematic method for determining
correlated wavefunctions
of extended systems in the ground state as well as in excited states is presented.
It allows to fully exploit
the power of quantum-chemical
programs designed for correlation calculations of finite
molecules.

Using localized Hartree--Fock (HF) orbitals (both occupied and virtual ones),
an effective Hamiltonian which can easily be
transferred from finite to infinite systems is built up.
Correlation corrections to the matrix elements of the effective Hamiltonian
are derived from clusters using an incremental scheme.
To treat the correlation effects, multireference configuration
interaction (MRCI) calculations with singly and doubly excited
configurations (SD) are performed.
This way one is able to generate both valence and conduction bands where all
correlation effects in the excited states as well as in the ground state of the system
are taken into account. An appropriate size-extensivity correction to the 
MRCI(SD) correlation energies is developed
which takes into account the open-shell character of the excited states.

This approach is applicable to a wide range of polymers and crystals.
In the present work {\it trans}-polyacetylene is chosen as a test system.
The corresponding band structure is obtained with the correlation of 
all electrons in the system being included on a very high level of
sophistication. The account of correlation
effects leads to substantial shifts of the "center-of-mass"
positions of the bands (valence bands are shifted upwards and
conduction bands downwards) and a flattening of all bands
compared to the corresponding HF band structure.
The method reaches the quantum-chemical level of accuracy.

Further an extention of the above approach to excitons
(optical excitations) in crystals is developed
which allows to use standard quantum-chemical methods to
describe the electron-hole pairs and to finally obtain
{\it excitonic} bands.

\newpage
\thispagestyle{empty}
$$$$

\newpage

\pagestyle{plain}
\pagenumbering{roman}
\setcounter{page}{1}
\tableofcontents

\newpage
\clearpage
\pagestyle{plain}
\pagenumbering{roman}
\setcounter{page}{3}

{\huge \bf List of abbreviations}

\vspace{15mm}

\begin{tabular}{lll}

{\bf 1D}        & & one-dimensional \\
{\bf CC}        & & Coupled-cluster \\
{\bf cc-p}      & & correlation consistent (basis set) including polarization functions \\
{\bf CCSD}      & & Coupled-cluster with singly and doubly excited configurations \\
{\bf CI}        & & Configuration interaction \\
{\bf DFT}       & & Density functional theory \\
{\bf EA}        & & Electron affinity \\
{\bf GTO}       & & Gaussian-type orbital \\
{\bf HF}        & & Hartree--Fock \\
{\bf HOMO}      & & Highest occupied molecular orbital \\
{\bf IP}        & & Ionization potential \\
{\bf IS}        & & Incremental scheme \\
{\bf LCAO}      & & Liner combination of atomic orbitals \\
{\bf LDA}       & & Local density approximation \\
{\bf LMO}       & & Localized molecular orbital \\
{\bf LUMO}      & & Lowest unoccupied molecular orbital \\
{\bf LME}       & & Local matrix element \\
{\bf MP2}       & & M\o{}ller--Plesset perturbation theory in second order \\
{\bf MR}        & & Multireference \\
{\bf MRCI}      & & Multireference configuration interaction\\
{\bf SD}        & & Single and double excitations\\
{\bf SCF}       & & Self-consisted field \\
{\bf tPA}       & & {\it trans}-polyacetylene \\
{\bf VDZ}       & & Valence double zeta \\
{\bf VTZ}       & & Valence triple zeta \\
{\bf WO}        & & Wannier orbital \\

\end{tabular}

\newpage
\thispagestyle{empty}
$$$$

\clearpage

\pagenumbering{arabic} \setcounter{page}{1}

\addtocontents{toc}{\addvspace{20pt}}

\renewcommand{\baselinestretch}{1}\normalsize
\pagestyle{headings}

\setcounter{chapter}{0}

\chapter{Introduction}

The study of electronic properties of solids was constantly
a hot topic in physics over many decades. Various different approximations,
e.g. the free-electrons model, the tight-binding method and others
(see, e.g. \cite{Ziman}),
were tried to get qualitative description of the electron states
in periodic systems. More sophisticated methods aiming already at a
quantitative description have been
developed since computers have become a common tool of
researchers. Among these methods one could distinguish a number of
approaches based on the local density approximation (LDA) to
density functional theory (DFT) \cite{Hohenberg64}, \cite{Kohn65}
which are widely used presently in solid-state physics. They allow,
in particular, to account for electron correlation 
effects by locally treating electrons as a
homogeneous gas. These methods have become very popular as they
provide quantitatively correct results for the ground state of
many systems. However, these methods fail to describe correctly
the band structure of solids, e.g. they usually give too
narrow band gaps for non-conducting systems. To overcome this
problem, further improvements of the formalism have been developed.
For instance in the last decade the most successive methods
became those based on the so-called $GW$ approximation \cite{Hedin65}.

One of main advantages of methods based on DFT is
their relatively small computational cost that opens the possibility 
to routinely apply them to periodic systems with large unit
cells. However, all DFT-based methods are focused on computing
physical ground-state properties of the systems which are related to
electron density and avoid calculation
of the many-body wavefunction which would provide us with
explicite information on the electron correlation in particular
materials. Another serious disadvantage of DFT methods is the
impossibility to control the accuracy of the obtained results
and to improve them in a systematic manner. Thus, in the
long perspective, an alternative way to study electron states
in periodic systems is required.

In parallel to the development of DFT-based approaches,
methods which solve the Hartree--Fock (HF) equation for solids 
were established. They provide single-particle wavefunctions and
energies obtained self-consistently with
the many-body wavefunction being a superposition
of Bloch waves. For this purpose powerful program packages
like CRYSTAL \cite{Pisani88}, \cite{CRYSTAL} have been designed.
The use of the variational method to solve the HF equation allows
one to reach the desired accuracy within a given set of parameters
(such as basis sets, crystal geometry, thresholds for integral
values etc.) and a systematic improvement of the 
basis sets and a reducing of the thresholds leads to further
improvement of the results. However, in this approach electrons
are considered to experience only a mean interaction field and 
thus the electron correlation is missing totally. As
a result of this, the fundamental band gap of non-conducting 
materials is substantially overestimated by the HF method.

Let us now turn to quantum chemistry. In this field, an
intensive development of methods which start with the HF
solution for finite systems and accurately account for electron
correlation
took place. These methods yield {\it correlated} wavefunctions
and {\it correlated} energies of finite molecules in good
agreement with experimental data. A detailed description of
such approaches can be found in, e.g. \cite{Szabo} and
\cite{Lindgren}. Among these methods one can distinguish
the configuration interaction (CI) methods, the coupled
cluster (CC) methods and methods based on the many-body
perturbation theory (MBPT) such as the second-order
M\o{}ller--Plesset (MP2) method which have become standard
in quantum chemistry and are implemented in quantum-chemical
program packages like MOLPRO \cite{MOLPRO} and GAUSSIAN
\cite{GAUSSIAN}.

At first glance, there is no possibility to apply these
quantum-chemical methods (which operate with single-electron
wavefunctions defined in real space) to extended
systems where the states are usually defined in momentum space.
However, if one reformulates the HF solution in terms of
localized Wannier functions, which are defined in the real
space, and considers finite clusters of the crystal, then
one can account for correlation effect within this cluster.
An electron placed in a localized orbital interacts strongly
with electrons in neighbor localized orbitals trying to
avoid coming too close to each other. Electrons in farther
orbitals are already screened and "felt" via the mean field
established in the HF description.
Thus, electron correlation is predominantly a local effect, and
therefore the electron correlation effects in an entire crystal
can be derived from
finite clusters. Here, very accurate quantum-chemical
methods can be employed. One only needs a scheme to transfer 
the results from clusters to periodic systems which in
fact resembles a Wannier back transformation. This way one
gets the {\it correlated} wavefunction of extended system
with which any physical quantity of a system can be
calculated. The scheme described above preserves the spirit
of {\it wavefunction-based} methods which consist of a sequence
of {\it well-controlled} approximations and thus allowing to
{\it estimate} and {\it improve} the accuracy of the final results.

The implementation of this idea has started in the late eighties
of the last century and from that time has reached a valuable
success. Ground-state and cohesive energies of many covalent
and ionic systems have been calculated until now (see e.g.
\cite{Stoll92a}, \cite{Stoll92b}, \cite{Paulus95}, 
\cite{Paulus96}, \cite{Doll97}, \cite{Kaldova97}, 
\cite{Abd00a} and \cite{Abd00b}). The approach was also used
to study structural properties of polymers (\cite{Yu97} and
\cite{Abd99}). In parallel to the study of ground-state properties
of extended systems, an approach for the hole states
in periodic infinite systems has been developed. To the present
moment it was addressed to the valence bands of
simple semiconductors, such as diamond, silicon and germanium
(\cite{Graef93}, \cite{Graef97} and \cite{Albrecht00}).
Further information on the current state of the art of
wavefunction-based approaches for extended systems which employ
quantum-chemical methods can be found in the recent review
\cite{Fulde02}.

Another promising wavefunction-based method appeared recently
and has successively been applied to the band structure of LiF
(see \cite{Albrecht01}, \cite{Albrecht02} and \cite{Albrecht02b}).
In the post-HF step it uses Green's function formalism
which is usually not considered as a standard quantum-chemical 
method. Therefore, we only mention it here briefly.

As one can notice, the wavefunction-based methods for solids 
described in \cite{Graef93}, \cite{Graef97} and \cite{Albrecht00}
study correlation of electrons in occupied orbitals.
However, to also obtain correlated {\it conduction} bands one needs
a proper description of correlation effects for states with an 
additional electron in the virtual space as well. Without this one 
can not evaluate one of the most fundamental electronic property of
insulators and semiconductors, namely the band
gap. The main reason of the previous approaches to fail here is the
impossibility to construct localized virtual orbitals in clusters
(in contrast to occupied orbitals) by standard localization schemes 
used in quantum chemistry. Therefore, a different approach to
localization of virtual orbitals was needed.

Conceptual works \cite{Marzari97} and \cite{Marzari01} on 
obtaining maximally localized orbitals in periodic systems and
implementations of similar ideas in CRYSTAL code
(\cite{Baranek01} and \cite{Zicovich01}) convinced us to
use localized virtual orbitals obtained in a periodic system
as a starting point for the construction of localized virtual 
molecular orbitals for sufficiently large cluster models of
periodic systems. This way we could properly
describe states with an attached electron in crystals by means
of finite clusters. Thus, we can calculate not only
valence but also conduction bands and, therefore, the band gap
provided electron correlation effects are accurately and systematically
taken into account. This we regard as the main motivation
for the present work.

In fact, having obtained the {\it correlated} band structure we 
extract from it not only the band gap but also band widths, the
ionization potential and the electron affinity of the system with
electron correlation being taken into account. The correlated 
wavefunction is used implicitely to calculate these quantities.
Our approach also allows us to estimate the error bar for
each calculated quantity and it provides a way to improve the 
accuracy further.

\begin{figure}
\centerline{ \psfig{figure=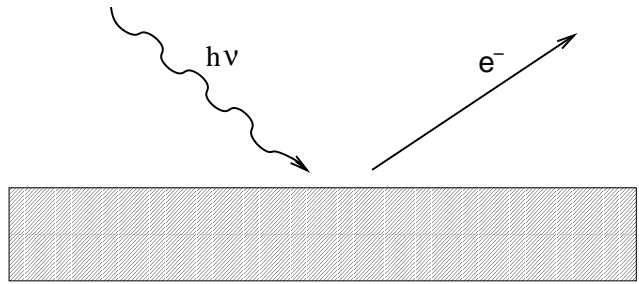,width=3.3in} }
\caption{A schematic view of photoionization process.}
\label{ip}
\end{figure}

\begin{figure}
\centerline{ \psfig{figure=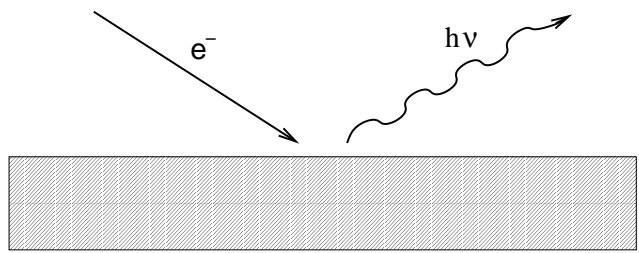,width=3.3in} }
\caption{A schematic view of inverse photoionization process.}
\label{ea}
\end{figure}

All the quantities calculated here can also be measured directly
in experiments. To study the band structure of crystals photoionization
(Fig.~\ref{ip}) and inverse photoionization experiments
(Fig.~\ref{ea}) can be performed (see e.g. \cite{Pople}).

Ionization potentials IP$_n$ (the valence-band energies) are
the difference between the energy of ($N-1$)-electron system with
a hole in band $n$ and the energy of the neutral $N$-electron
system. This quantity can be measured by irradiation of light
with known frequency on the surface of a crystal and measuring
the kinetic energy of the photoemitted electrons:
$E_{\rm kin}= h\nu + (E_0 \: - \; ^{N-1} \! E_n) = h\nu - {\rm IP}_n$.

The inverse process is used to measure the electron affinities EA$_m$
which correspond to the energy which a neutral system gains
when an extra electron is accepted in the conduction band $m$
and an ($N+1$)-electron state is formed.
Electrons with known kinetic energy are shined onto the surface
and the frequency of the emitted light is measured:
$h\nu_m = E_{\rm kin} + (E_0 \: - \; ^{N+1} \! E_m) = E_{\rm kin} + {\rm EA}_m$.
This two experiments give band energies which we aim to calculate
theoretically.

Another motivation of this study is to extend the borders of the
wavefunction-based approaches to correlated band structures. By this
we mean that we would like i) to investigate a system with more
complex unit cell than was done before, ii) to use larger basis
sets (valence triple zeta) which are necessary in quantum chemistry
to obtain indeed accurate results and which are appropriate for correlation
calculation in ($N+1$)-electron systems and iii) to use one of the most
accurate and powerful correlation method (the multireference
configuration interaction method restricted to singly and doubly
excited configurations) available in standard quantum-chemical
program packages. The implementation of all the ideas mentioned above
put our approach far ahead with respect to other methods for band 
structure calculations in non-conducting systems.

Yet another goal of our study is to develop an approach which
allows to even
describe well-bound excitons in crystals. These electron-hole
pairs made up by neutral excitations of the electronic structure are
rather compact objects and therefore can be treated in real
space. Occupied and virtual Wannier orbitals are used for that purpose
such that the electron-hole pair can be described in finite clusters
of the crystal by standard quantum-chemical methods. Provided, the localized
orbitals in such a cluster represent the Wannier orbitals of
the crystal well, an appropriate transformation allows to transfer
the relevant data on the localized excitonic states from the clusters to the periodic
system and excitonic bands can be calculated.

As a test system we have chosen {\it trans}-polyacetylene (tPA). This
is a semiconducting polymer which consists of weakly interacting
single chains. Therefore, its highly anisotropic electronic properties
can be studied in a one-dimensional system, a tPA single chain.
This is a very useful feature for the implementation of our
approach to conduction bands in particular because the corresponding 
localized virtual one-particle orbitals are in fact rather diffuse objects. 
To study the decay of the correlation effects with the distance of the 
involved localized orbitals we need a linearly scaling system.

Also polymers themselves are intriguing systems providing researchers
with surprising results. The discovery and development of electrically
conductive polymers were appreciated by the Nobel Prize in Chemistry for
2000. One of the experimentally well-studied systems is doped
{\it trans}-polyacetylene. Thus, the sound understanding of the electronic
structure of tPA is highly desired. In our work we also aim to answer
the question whether electron correlation in polymers can be described 
properly in a single chain to give results which agree with experimental
data for the bulk system. For this purpose, in section 3.1.1, we describe
in details the investigated material and further, in section 4.4, we compare
results obtained for tPA single chains with existing experimental data on
bulk tPA crystals.

The thesis is organized as follows. In Chapter 2 the basic theoretical aspects
of our approach are discussed. At the beginning of the chapter we briefly
sketch the approximations used in {\it ab initio} methods.
In first sections we consider the solution of the HF equation for infinite
systems in terms of Bloch waves and Wannier orbitals and define the local
quantities which also show up in infinite systems and clusters and which can
be transferred from one to another.
In the last two sections we discuss why electron correlations are necessary
to be taken into account, the physical meaning of these effects in both neutral
and charged systems, and how these correlation effects can be evaluated by 
the MRCI(SD) method.

Chapter 3 is dedicated to the development of our wavefunction-based
approach. First we introduce the structural and electronic properties of
the test system. Next we discuss in details how to get localized HF
occupied and virtual orbitals in clusters and define the corresponding
local matrix elements (LMEs) which are used to calculate
the SCF band structure of tPA. In section~3.3 we discuss how
to evaluate the correlation corrections to the LMEs by MRCI(SD) method, 
and the use of an approximate but well controlled scheme for this purpose 
(namely incremental scheme) is
presented in section~3.4. The possibility to reduce the
size-extensivity error of CI(SD) is explored in section~3.5.
In the last section of Chapter 3 the way to handle the problem
of the emergence of satellite states in cluster models is presented.

In Chapter 4 we summarize and discuss numerical results for
{\it trans}--polyacetylene. There we emphasize the effect of
electron correlation on the key quantities of our method, namely the local
matrix elements. We use correlation corrected LMEs to plot the correlated
band structure from which we get the band gap, IPs, EAs, and band widths
of linear tPA. The accuracy of the whole method is explored and
the error bars for the calculated quantities are estimated.
In section~4.4 we compare our results obtained for the one-dimensional system
to experimental data on bulk {\it trans}--polyacetylene.

In Chapter 5 we introduce a new approach to excitonic states in crystals
in terms of singly excited configurations. The method is established
and some preliminary results are presented.

In Chapter 6 we conclude and discuss perspectives of our approach.

\chapter{Local approach to wavefunctions}

In this chapter we would like to discuss the basic theoretical approaches
to our study. The ultimate goal of the physicist studying theoretically
electronic properties of an atom, a molecule or even a bulk material 
would be to get a many-body wavefunction which maximally approaches the 
exact one and contains all the information about the electronic structure, 
the forces on the nuclei, and other parameters of the system in all its 
possible states. Unfortunately, obtaining the exact many-body wavefunction 
is unrealistic. In order to proceed one has either to make some 
approximations for the wavefunctions and then to calculate the desired 
quantities with this approximated wavefunctions (and to estimate the 
emerging errors) or one tries to obtain those quantities directly (avoiding 
the calculation of wavefunctions) using some tricks or semiempirical 
methods. In the latter case the question about the achievable accuracy 
often remains open. When one is interested not only in a qualitative 
description of a system and the effects in it, but also in accurate 
quantitative results then the first way seems to be more attractive, even 
though more complicated. Having chosen this way, one choses the most 
accurate among all feasible approximations in order to get a realistic 
description of the system and to reduce the errors to a minimum.

In our study we are aiming at a {\it correlated} wavefunction of a periodic 
system (e.g. {\it trans}-polyacetylene). This means that we introduce some 
controlled approximations at the very beginning, then get the wavefunction 
which is assumed to be very close to the real one, and finally calculate 
the ground state energy and energies of one-particle states (the band 
structure). By the term "controlled approximation" we mean that we can both 
estimate the error bar for the calculated quantities and improve our results 
to (in principle) any desired accuracy by increasing the computational 
efforts. Here we mention these approximations.

The first one is the Born--Oppenheimer approximation. The system under 
consideration consists of nuclei vibrating around their equilibrium 
positions and electrons filling the space between the nuclei. The 
Hamiltonian of the system and, correspondingly, its wavefunction depends
on coordinates of both all nuclei and all electrons. As electrons are much 
lighter than nuclei they move much faster and, therefore, can be considered 
as moving in the field of fixed nuclei (for details we reffer to \cite{Ziman}, 
chapter 6.11). Then solving only the "electronic" part of the Hamiltonian one
obtains electron eigenfunctions and eigenvalues which depend parametrically on
the nuclear coordinates. Having the solution for the electron problem one 
considers the nuclei as slowly moving in the averaged field of the fast moving 
electrons and solves the problem of vibration, rotation and translation of the
system consistently with the help of electronic solutions. In our work we are
concentrating on the electronic problem and consider the nuclei as fixed at
their equilibrium positions which we can define e.g. by using experimental data
on the structure of a given material. By this we exclude the vibrational and 
rotational degrees of freedom of finite systems baring in mind, however, the
possibility to obtain them by looking at the electronic structure for different
positions of the nuclei. This approximation is valid as long as electron-phonon 
interaction in the system is negligible.

The second point to discuss is the use of basis sets for the electron 
wavefunctions which are necessary for numerical calculations of electronic 
structures. We use atomic basis sets since quantum chemistry has demonstrated
for many decades that for atoms and molecules this representation gives very
accurate results as compared to experimental data. We also have to notice that 
this basis sets are finite. To set up an analytic expression for the 
wavefunction of an electron we use Gaussian-type atomic orbitals (GTO) which 
are very favourable from the computational point of view. With these GTO basis 
sets very accurate results can be obtained and the accuracy depends on the 
number of atomic orbitals in the basis set for each atom and on the number of
Gaussians per atomic orbital. Enlarging these numbers leads to tremendous
computational efforts when one wants to treat big molecules which in fact
limits the size of basis sets. However, at this point we want to emphasize that
finite but sufficiently large GTO basis sets can provide reasonable accuracy and 
are still convenient to use.

In our work we consider systems consisting of light atoms whose atomic numbers
are not larger than six. Therefore, we can neglect relativistic effects and
consider the spin coordinates of electrons as independent from their spatial 
ones. Spin-orbital coupling only becomes not negligible for more heavy atoms
starting from the atomic number around 30 (see, e.g. \S72 in \cite{Landau} and 
chapter~3.11 in \cite{Ziman}). In zero-oder approximation we consider the 
wavefunction of a system as being the one of an independent particles system 
where electrons occupy spin orbitals (a function of electron position vector 
${\bf{r}}_i$ and spin coordinate $\sigma_i$) or alternatively we can say that
spatial orbitals (which depend only on ${\bf{r}}_i$) are occupied by pairs of 
electrons one having spin up and the other having spin down. Such spin orbitals
are solutions of a single-particle Hamiltonian in the self-consisted field (SCF)
approximation and are represented as linear combinations of atomic orbitals (LCAO). 
Then, the many-body wavefunction of the whole system is given as the 
antisymmetrized product of all occupied spin orbitals, the so-called Slater 
determinant (see e.g. \cite{Szabo}). The ansatz to write molecular orbitals as
LCAO is also an approximation and is based on the fact that when an electron is
placed close to one nucleus its interaction energy with this nucleus is much larger 
than interaction with all other nuclei. In quantum chemistry the LCAO approximation
is very common and considered as being proved empirically to yield very good 
agreement between experimental data and theoretical results for molecules,
once electron correlations are included.

Despite the essentially unavoidable approximations mentioned so far we do not
make any approximation for the nuclear potentials and the Coulomb interaction 
between electrons. The systems are treated as realistic as possible keeping by 
this the sense of {\it{ab initio}} calculations.

As mentioned above we start from the wavefunction of the closed-shell system 
obtained in the SCF (or HF) approximation. In this approximation the one-particle 
Hamiltonian is constructed, treating an electron as moving in the field
of the fixed nuclei and the averaged field of all the remaining electrons. 
The set of electron orbitals is obtained as the solution of this problem. 
Electrons are placed in this orbitals starting from the one with the lowest 
energy and filling the $N/2$ energetically lowest spatial orbitals where $N$
denotes the number of electrons in the system. All the other orbitals remain 
unoccupied. The solution is obtained iteratively with the desired precision 
for the chosen basis sets. We do not want to discuss here the quality of
results obtained with such a wavefunction as it is well known from both solid 
state physics and quantum chemistry that the Hartree--Fock approximation 
provides a reasonable but not accurate wavefunction and that for some
properties it only gives qualitative results. However, it is also known that 
this wavefunction may serve as a good starting point for further
improvement towards the exact wavefunction as long as weakly correlated 
systems are considered. So, our aim is start from the HF approximation and to
go beyond it in order to obtain some good approximation to the correlated 
wavefunctions of a system and to calculate the ground state energy and 
energies of the excited states.

Moreover, the system which we refer to is periodic and infinite. Our approach 
will allow to transfer the data obtained in clusters with standard
quantum-chemical program packedges to the infinite system exploiting
the local character of electron correlations. For this purpose we have to 
define local quantities (electron orbitals and matrix elements) which have 
identical meaning in the clusters and in the infinite systems and to esteblish
a scheme to transfer them.

In the first Section we consider the HF solution for infinite systems, namely, 
Bloch waves, the ground state energy and the band structure. In Section~2.2 we 
redefine these quantities in terms of Wannier orbitals. In Section~2.3 we show 
how those local quantities may be obtained for clusters and how they are 
transferred to infinite systems to give the band structure. In Section~2.4 we
discuss why electron correlation should be taken into account, the physical 
meaning of this effect and how it is incorporated in our method. There we also 
define the correlated wavefunction. In Section~2.5 the quantum-chemical 
correlation method is described: why we have chosen the particular method, its 
advantages and disadvantages as compeared to other methods, the results which 
we get by using it and their quality.

\section{The Hartree--Fock approximation}

The electronic {\it ab initio} Hamiltonian of a solid or molecule in the 
Born--Oppenheimer approximation is written in the form

\begin{equation}
\label{H_el}
H_{\rm el} = - \sum_{i} \frac{1}{2}\nabla_i^2 - \sum_{i,A} \frac{Z_A}{\vert {\bf r}_A - {\bf r}_i \vert}
+\sum_{i<j} \frac{1}{\vert {\bf r}_i - {\bf r}_j \vert}
+\sum_{A<B} \frac{Z_A Z_B}{\vert {\bf r}_A - {\bf r}_B \vert}
\end{equation}
where ${\bf r}_i$ and ${\bf r}_j$ are the coordinates of electrons, ${\bf r}_A$ and $Z_A$ are
the fixed coordinates and the atomic number of nucleus $A$. Atomic (Hartree) units are used 
here where electron charge $e$ is set to 1.
The last term in Eq.~(\ref{H_el}) is the interaction energy between the fixed nuclei and does not affect the electronic
eigenfunctions. We omit this term in further consideration bearing in mind that this constant
for given geometry must be added to the system total energy at the end.

The many-body wave function for $N$ independent electrons may be written
as an antisymmetrized product (or the Slater determinant) of the spin orbitals 
$\psi_i({\bf r}\sigma)$:

\begin{eqnarray}
\label{Slaterdet}
\Phi({\bf r}_1\sigma_1; \ldots ;{\bf r}_N\sigma_N) &
= & \frac{1}{\sqrt{N!}}
\begin{vmatrix}
\psi_1({\bf r}_1\sigma_1) & \psi_2({\bf r}_1\sigma_1) & \cdots & \psi_N({\bf r}_1\sigma_1) \\
\psi_1({\bf r}_2\sigma_2) & \psi_1({\bf r}_2\sigma_2) & \cdots & \psi_N({\bf r}_2\sigma_2) \\
\vdots                    & \vdots                    & \ddots & \vdots                    \\
\psi_1({\bf r}_N\sigma_N) & \psi_1({\bf r}_N\sigma_N) & \cdots & \psi_N({\bf r}_N\sigma_N) \\
\end{vmatrix}
\nonumber \\
 & & \nonumber \\
 &\equiv & \frac{1}{\sqrt{N!}}
\det [\psi_1({\bf r}_1\sigma_1) \ldots \psi_N({\bf r}_N\sigma_N)].
\end{eqnarray}
Let us denote $\Phi({\bf r}_1\sigma_1; \ldots ;{\bf r}_N\sigma_N)$ as $\Phi$ for the convenience.
The expectation value of the ground-state energy of the system is defined as
$E_0 = \langle \Phi | H_{\rm el} | \Phi \rangle$. By optimizing the spin orbitals $\psi_i$ one obtains the
lowest value for $E_0$ under the constraint that spin orbitals $\psi_i$ remain orthogonal to each other
and normalized: $\langle \psi_i | \psi_j \rangle = \delta_{ij}$. As the solution of this
variational problem the HF single-determinant wavefunction of the many-electron system is
obtained. It consists of spin orbitals being eigenfunctions of the self-consistent single-particle
Fock operator $F$:

\begin{equation}
\label{Fock_equat}
F | \psi_i \rangle = \varepsilon_i | \psi_i \rangle
\end{equation}
where $F$ takes a diagonal form in the representation
of the canonical orbitals $\psi_i$ and has the following matrix elements:

\begin{equation}
\label{Fock_matr}
f_{ij} = \langle \psi_i | h_1 | \psi_j \rangle +
\sum_{l=1}^N \bigl( \langle \psi_i \psi_l | v_{12} | \psi_j \psi_l \rangle -
\langle \psi_i \psi_l | v_{12} | \psi_l \psi_j \rangle \bigr).
\end{equation}
and electron orbitals $\psi_i$ are obtained self-consistently. In Eq.~(\ref{Fock_matr}) index $l$
counts the occupied orbitals and operators $h_1$ and $v_{12}$ are defined as

\begin{equation}
\label{h}
h_1 = -\frac{1}{2} \nabla_1^2 - \sum_A \frac{Z_A}{\vert {\bf r}_A - {\bf r}_1 \vert}
\end{equation}
and

\begin{equation}
\label{v}
v_{12} = \frac{1}{\vert {\bf r}_1 - {\bf r}_2 \vert}.
\end{equation}
The derivation of the equation (\ref{Fock_equat}) can be found in many textbooks
(e.g., in \cite{Szabo}).

Getting an analytic solution of the Fock equation (\ref{Fock_equat}) is not
possible in general case and numerical methods have to be used for this purpose. 
In quantum-chemical program
packages the Fock eigenfunctions are obtained iteratively in the form of linear
combinations $\psi_i = \sum_j \chi_j \alpha_{ji}$
within a chosen atomic basis sets $\{ \chi_j \}$ by varying the coefficients $\alpha_{ji}$
and minimize the ground-state energy of the system.
In our study solving the Fock equation is a routine procedure and we do not
want to pay too much attention to the details of Hartree--Fock theory. We only want to point out the physical
meaning of the eigenvalues in Eq.~(\ref{Fock_equat}). The quantity $\varepsilon_i$ represents
an energy of an electron occupying the spin orbital $\psi_i$. It consists of the electron
kinetic energy, its potential energy in the field of nuclei and in the mean field
of the other electrons. The latter in its turn consists of the Coulomb repulsion
energy and the exchange energy which are determined correspondingly by the two last terms
in the right-hand side of Eq.~(\ref{Fock_matr}). The $N$ spin orbitals with the lowest
orbitals energies are occupied in the ground state. Further we use indices
$a$, $b$, $c,\ldots$ for these occupied orbitals. The remaining eigenfunctions from
(\ref{Fock_equat}) stay unoccupied in the ground state
and are labeled by $r$, $s$, $t,\ldots$. If a statement or a formula is valid for both,
occupied and virtual orbitals, we use the indices $i$, $j$, $k,\ldots$. Also for the sake of
convenience we use $| i \rangle$ instead of $| \psi_i \rangle$. Multiplying Eq.~(\ref{Fock_equat})
by $\langle i |$ from the left and taking into account Eqs.~(\ref{Fock_matr})--(\ref{v}) we get an 
expression for orbitals energies

\begin{equation}
\label{orb_en}
\varepsilon_i = \langle i | h_1 | i \rangle +
\sum_{b=1}^N \bigl( \langle i b | v_{12} | i b \rangle -
\langle i b | v_{12} | b i \rangle \bigr) \equiv
\langle i | h | i \rangle +
\sum_{b=1}^N \bigl( \langle i b | i b \rangle -
\langle i b | b i \rangle \bigr)
\end{equation}
where the potential $v_{12}$ is omitted in the matrix element for convenience. 
In particular we can write the energies for the occupied orbitals as

\begin{equation}
\label{e_a}
\varepsilon_a = \langle a | h | a \rangle +
\sum_{b\neq a} \bigl( \langle a b | a b \rangle -
\langle a b | b a \rangle \bigr)
\end{equation}
and that for the unoccupied orbitals as

\begin{equation}
\label{e_r}
\varepsilon_r = \langle r | h | r \rangle +
\sum_{b=1}^N \bigl( \langle r b | r b \rangle -
\langle r b | b r \rangle \bigr).
\end{equation}

These two quantities are of special interest in our study. Let us look more
closely at expressions (\ref{e_a}) and (\ref{e_r}). The energy of an electron
in an occupied orbital $\psi_a$ (\ref{e_a}) consists of its kinetic energy,
the energy of attraction to the nuclei plus the Coulomb $(\langle r b | r b \rangle)$
and exchange $(-\langle r b | b r \rangle)$ interaction with each of the $N-1$ remaining
electrons of the system in the ground state. Therefore we can refer to the value
$-\varepsilon_a$ as to an energy which is necessary to pay in order
to remove an electron from the occupied orbital $\psi_a$ without changing the 
rest of the system, that is the ionization
potential of the system in the frozen orbital approximation.
An unoccupied orbital energy is the sum of kinetic energy of
an electron on this orbital, the energy of its attraction to the nuclei and energies of
Coulomb and exchange interaction with all $N$ electrons of the system in the ground
state as if an additional ($N+1$)-th electron would have been adiabatically added to the
system (i.e. in the frozen orbital approximation) on the $r$-th orbital. 
Thus, we can regard the value $-\varepsilon_r$ as
the energy one gains when adding one electron to the neutral system
and placing it into the orbital $\psi_r$ that is by definition the electron
affinity. These two statements are known as Koopmans' theorem \cite{Koopmans}.

To demostrate how these simple relations (\ref{e_a}) and (\ref{e_r}) arise
from the frozen orbital approximation we provide the proof for this theorem. 
The ionization potential is the difference
between the energy of the system with $N-1$ electron (excited state) and the
energy of $N$-electron system (its ground state)

\begin{equation}
\label{IP}
{\rm IP} = \; ^{N-1} \! E - E_0
\end{equation}
and the electron affinity is the difference between the ground state energy and
the energy of ($N+1$)-electron system

\begin{equation}
\label{EA}
{\rm EA} = E_0 - \; ^{N+1} \! E.
\end{equation}
It is convenient to define state eigenfunctions in the second quantization formalism
(especially in the case of infinite systems). Then the ground state wavefunction
is written as

\begin{equation}
\label{PhiN}
|\Phi \rangle = \prod_{a} c^{\dagger}_a | 0 \rangle
\end{equation}
where $c^{\dagger}_a$ is the creation operator for an electron in an occupied spin orbital 
$\psi_a$ and $|0\rangle$ is the vacuum state. The ($N-1$)-electron state
in the frozen orbital approximation is given by

\begin{equation}
\label{PhiN-1}
| \; ^{N-1} \! \Phi_c \rangle =  c_c | \Phi \rangle
\end{equation}
and the ($N+1$)-electron state by

\begin{equation}
\label{PhiN+1}
| \; ^{N+1} \! \Phi_r \rangle =  c^{\dagger}_r | \Phi \rangle
\end{equation}
where the operator $c_c$ destroys an electron in occupied spin orbital $\psi_c$ and the
operator $c^{\dagger}_r$ creates an electron in virtual spin orbital $\psi_r$. Let us note
that $^{N-1} \! \Phi_c$ and $^{N+1} \! \Phi_r$ are singe determinants constructed from
$N-1$ and $N+1$ spin orbitals with $1/\sqrt{(N-1)!}$ and $1/\sqrt{(N+1)!}$ as normalization
factors, respectively. The energies $E_0$, $^{N-1} \! E_c$ and $^{N+1} \! E_r$ are obtained
as the expectation values of these single determinants:

\begin{equation}
\label{E_0}
E_0 = \langle \Phi \: | \: H | \: \Phi \rangle =
\sum_{a=1}^N \langle a | h| a \rangle +
\frac{1}{2} \sum_{a=1}^N \sum_{b=1}^N \bigl( \langle a b | a b \rangle -
\langle a b | b a \rangle \bigr),
\end{equation}

\begin{equation}
\label{E_c}
^{N-1} \! E_c = \langle ^{N-1} \! \Phi_c \: | \: H \: | \: ^{N-1} \! \Phi_c \rangle =
\sum_{a \neq c} \langle a | h| a \rangle +
\frac{1}{2} \sum_{a \neq c} \sum_{b \neq c} \bigl( \langle a b | a b \rangle -
\langle a b | b a \rangle \bigr)
\end{equation}
and

\begin{eqnarray}
\label{E_r}
^{N+1} \! E_r & = & \langle ^{N+1} \! \Phi_r \: | \: H \: | \: ^{N+1} \! \Phi_r \rangle =
\sum_{a=1}^N \langle a | h| a \rangle + \langle r | h| r \rangle +        \nonumber \\
& & \frac{1}{2} \sum_{a=1}^N \sum_{b=1}^N \bigl( \langle a b | a b \rangle -
\langle a b | b a \rangle \bigr) +
\sum_{b=1}^N \bigl( \langle r b | r b \rangle - \langle r b | b r \rangle \bigr).
\end{eqnarray}

The energy for taking an electron out from the spin orbital $\psi_c$ of the neutral
system is given by difference between (\ref{E_c}) and (\ref{E_0})

\begin{eqnarray}
\label{IP_c}
{\rm IP}_c & = & \sum_{a \neq c} \langle a | h| a \rangle +
\frac{1}{2} \sum_{a \neq c} \sum_{b \neq c} \bigl( \langle a b | a b \rangle -
\langle a b | b a \rangle \bigr)    \nonumber \\
& & - \sum_{a=1}^N \langle a | h| a \rangle -
\frac{1}{2} \sum_{a=1}^N \sum_{b=1}^N \bigl( \langle a b | a b \rangle -
\langle a b | b a \rangle \bigr)        \nonumber \\
& = & - \langle c | h | c \rangle -
\sum_{b\neq c} \bigl( \langle c b | c b \rangle -
\langle c b | b c \rangle \bigr)  =
- \varepsilon_c
\end{eqnarray}
and the energy obtained from adding an electron to the neutral system in the spin
orbital $\psi_r$ is given by difference between (\ref{E_0}) and (\ref{E_r})

\begin{equation}
\label{EA_r}
{\rm EA}_r = -\langle r | h | r \rangle -
\sum_{b=1}^N \bigl( \langle r b | r b \rangle -
\langle r b | b r \rangle \bigr) =
- \varepsilon_r.
\end{equation}

We have described the ($N+1$)- and ($N-1$)-electron states by single-determinant wavefunctions
constructed with frozen (non-relaxed) spin orbitals which are obtained as a solutions of the equation (\ref{Fock_equat})
for the ground state of $N$ electron system. For the case of finite molecules one
should not expect spin orbitals of a neutral system to be the same as in cases of the
charged one. For the more apropriate description of the ($N+1$)- and ($N-1$)-electron
states in molecules, the so-called $\Delta$SCF approach, one has to solve the 
corresponding HF equations for the charged open-shell systems and get energies 
$^{N-1} \! E_c$ and $^{N+1} \! E_r$ from these solutions. Thus, the expressions derived above
for $\rm{IP}_c$ and $\rm{EA}_r$ in the frozen orbital approximation only give approximate
results for finite molecules. However, formulas (\ref{IP_c}) and (\ref{EA_r}) are correct for
extended systems where the electrons are delocalized over an infinite volume and hence no orbital
relaxation takes place. Here we mean "correct" in the sense of the $\Delta$SCF approach which is 
purely based on the HF energies.

In a periodic infinite system the single-electron wave functions must obey the translation symmetry
of the crystal. This leads to Bloch's theorem which states that each electron eigenfunction translated
by a distance $R = n a_l$ along a lattice vector should be equal to the original one within an 
exponential prefactor

\begin{equation}
\label{Bloch}
\psi_k ({\bf r} + R \: {\bf l}_x) = {\rm e}^{ikR} \psi_k ({\bf r})
\end{equation}
where, for simplicity, we only consider a one-dimensional crystal with lattice constant $a_l$
(aligned along the $x$-axis), $n$ is
an integer and the electron states are labeled by a crystal-momentum quantum numbers
$k \in [-\pi / a_l ; + \pi / a_l]$.
The $k$-dependent eigenvalues of the states form energy bands and the quantity $\varepsilon_{k\nu}$
refers to the energy of an electron in a Bloch state with crystal-momentum $k$ and from the band $\nu$.
In the ground state of an insulator or a semiconductor the energetically lowest bands are completely filled
(valence bands) and are separated by a gap from the others which are empty (conduction bands). 
For metals partially filled bands exist in the ground state. We explicitly exclude
metals for our approach to correlated wavefunctions in solids and consider
systems with band gaps only.

As Koopmans' theorem holds for extended systems the electron energy $\varepsilon_{k\nu}$ of 
a Bloch wave $\psi_{k\nu}$ is (up to a sign) the ionization potential for taking
out an electron from the system (${\rm IP}_{k\nu}$) in case of valence electrons and
the energy gain from putting an extra electron into a conduction band which is the electron affinity
(${\rm EA}_{k\nu}$). As defined in (\ref{IP}) and (\ref{EA}) they are the differences of the energies of
($N-1$)- or ($N+1$)-particle system respectively and the neutral $N$-particle system. These energies are
defined as the expectaction values of Hamiltonians for the corresponding single-determinant wavefunctions.
Since $N$, in this case, approaches infinity such a representation of the wavefunction becomes cumbersome
and the use of the second quantization formalism seems to be more appropriate here. Then we
define the ground-state wavefunction of the $N$-particle system as

\begin{equation}
\label{PhiNk}
|\Phi \rangle = \prod_{k \nu}^{\rm occ} c^{\dagger}_{k\nu \uparrow} c^{\dagger}_{k\nu \downarrow} | 0 \rangle
\end{equation}
where operator $c^{\dagger}_{k\nu \uparrow}$ ($c^{\dagger}_{k\nu \downarrow}$) creates a spin-up
(spin-down) electron in a Bloch wave, $k$ runs
over the first Brillouin zone and $\nu$ counts only the valence bands. The energy of this state is $E_0$
and is proportional to the volume of the extended system.
The ($N-1$)-particle-state wavefunction is obtained by applying an annihilation operator for an electron
with crystal-momentum $k$ in the valence band $\nu$ and with spin $\sigma$ to the wavefunction (\ref{PhiNk})

\begin{equation}
\label{PhiN-1k}
| k \nu \sigma \rangle = c_{k \nu \sigma} \: | \: \Phi \rangle
\end{equation}
and the ($N+1$)-particle-state wavefunction by applying a creation operator
$c^{\dagger}_{k\mu \sigma}$ to (\ref{PhiNk})

\begin{equation}
\label{PhiN+1k}
| k \mu \sigma \rangle = c^{\dagger}_{k \mu \sigma} \: | \: \Phi \rangle
\end{equation}
where the index $\mu$ is used to label conduction bands. In this new notation we 
can define bands as
energy differences which depend on $k$. The valence bands are given by

\begin{equation}
\label{val_band}
-{\rm IP}_{k\nu} = - \langle k \nu \sigma \: | \: H \: | \:k \nu \sigma \rangle + E_0
\end{equation}
and the conduction bands by

\begin{equation}
\label{cond_band}
-{\rm EA}_{k\mu} = \langle k \mu \sigma \: | \: H \: | \:k \mu \sigma \rangle - E_0.
\end{equation}

At first glance, such a definition of the energy bands looks more complicated as compared 
to the energies of single-particle states in a crystal $\varepsilon_{k \nu}$ and 
$\varepsilon_{k \mu}$. However, it will give us the possibility to express ${\rm IP}_{k\nu}$
and ${\rm EA}_{k\mu}$ as a kind of Fourier transformation of some local matrix elements and
to include electron correlation by affecting these matrix elements which would be not 
possible if electrons are treated as Bloch waves.

Some alternative approaches to correlated band structure (e.g. based on DFT) are focused on
finding an equation analogous to Eq.~(\ref{Fock_equat}) but already containing ground-state 
correlations. In such schemes the band energies $\varepsilon_{k \nu}$ are usually simply 
defined as the eigenvalues of a single-particle equation similar to the HF equations. 
However, these values are not strictly energies of correlated holes or attached electrons
and in the DFT framework this approximation usually leads to an underestimation of the band 
gap of insulators and semiconductors. In contrast, in our method, we calculate separately 
and explicitly correlated ($N-1$)-, ($N+1$)- and $N$-particle states and this allows us to 
see how the system reacts in detail on the presence of an extra charge.

\section{Wannier orbitals}

In the HF approximation an electron is considered as moving in the averaged
self-consistent field of all the other electrons. But in reality each particular
electron tries to avoid coming too close to its nearest neighbors due to
Coulomb repulsion. This effect can not be
described on the HF level and is called electron correlation.

In our study we describe the process of adding of an electron to an infinite
system or removing it which is equivalent to producing a positive charge.
These extra charges are totally delocolized in the extended system
since extra electrons (or holes) in crystals are Bloch waves and the response of
a system with infinite number of electrons to these extra particles is a pure
correlation effect. Alternatively, we can use some mathematically constructed
localized orbitals for the electrons and holes and restrict the response of an
area around the localized extra charge (localized electron or hole). As a
consequence, the correlation problem is reduced to the consideration of a limited
number of electrons in a finite part of the crystal. The obtained correlation
effect can subsequently be transferred back to Bloch states. This way we are
able to get correlated wavefunctions and correlated band structures.
In the present Section we will introduce the localized crystalline orbitals
which can be used for that purpose.

The natural choice would be the atomic basis functions however they are not mutually
orthogonal in the crystal and are not solutions of the HF problem which we want
to start from. The HF eigenfunction of the system, obtained in the form of the
single determinant, in turn, has the nice property to be invariant (up to a phase
factor) under any unitary transformation of the set of single-particle orbitals
which compose it. Hence, a set of spatially localized orthonormal orbitals can be
obtained as linear combination of canonical (delocalized) ones

\begin{equation}
\label{locorb}
\varphi_i = \sum_j \psi_j u_{ji}
\end{equation}
and several localization schemes (finding the unitary matrix
$u_{ji}$ which provides the less extended orbitals) have been
proposed in quantum chemistry for finite molecules, e.g.
\cite{Boys}, \cite{Ruedenberg}, \cite{Pipek}. Then, the
wavefunction of the whole system can be written as a single
determinant over $\varphi_i$ instead of $\psi_i$. The total energy
of the system obtained as the expectation value of the Hamiltonian
with the transformed wavefunction remains unchanged.

In periodic systems the Wannier functions \cite{Wannier37} are 
known to yield localized electron orbitals if a suitable unitary 
transformation of the Bloch waves is used:

\begin{equation}
\label{addformula}
\varphi_{Rn} = \int_{-\pi/a_l}^{\pi/a_l} {\rm e}^{-ikR} 
\sum_{\nu = \nu_1, \ldots , \nu_m} U_{\nu n}^{(k)} \psi_{k\nu} \: {\rm d}k
\end{equation}
and details of this step can be found in \cite{Eschrig}. The $m$
Wannier functions $\varphi_{0m}$ which represent $m$ energy bands
$\nu_1, \ldots \nu_m$ are referred to some unit cell (for instance
by their centers) and repeated over the whole crystal by primitive
translation vectors $R$. Pairs of spin-up and spin-down electrons
from the $m$ bands are said to occupy these repeated Wannier
orbitals (WOs). Now the ($N-1$)- and ($N+1$)-particle-state 
wavefunctions are
obtained from the canonical ones $| k \nu \sigma \rangle$
introduced in Section~2.1 by a transformation totally analogous to
the Wannier transformation of the Bloch orbitals. The wavefunction
of the state with a localized hole is then given by

\begin{equation}
\label{Wannier_hole}
| R n \sigma \rangle = c_{Rn\sigma} | \Phi \rangle
\end{equation}
where an electron with spin $\sigma$ is removed from the $n$-th
Wannier orbital centered at the unit cell with lattice vector $R$
(we refer to a one-dimensional crystal here) and the ground-state
wavefunction $\Phi$ is constructed from Wannier orbitals. Note
that due to translational invariance the vector $R$ is not
uniquely defined. However by fixing once the reference (or zero)
unit cell we eliminate this ambiguity. Analogously, the
($N+1$)-electron states are defined as

\begin{equation}
\label{Wannier_el}
| R m \sigma \rangle = c^{\dagger}_{Rm\sigma} | \Phi \rangle
\end{equation}
where $m$ is used to indicate the use of WOs from the conduction bands.

Now we show how energy bands can be obtained from quantities defined in
terms of $| R n \sigma \rangle$ and $| R m \sigma \rangle$. Analogously to
our discourse in Section~2.1 we define "local" matrices of ionization potentials

\begin{equation}
\label{IPnn}
{\rm IP}_{R^\prime-R, nn^\prime} = \langle R n \sigma | H |
R^\prime n^\prime \sigma \rangle -
\delta_{RR^\prime} \delta{nn^\prime} E_0
\end{equation}
and electron affinities

\begin{equation}
\label{EAmm}
{\rm EA}_{R^\prime-R, mm^\prime} = - \langle R m \sigma | H |
R^\prime m^\prime \sigma \rangle +
\delta_{RR^\prime} \delta{mm^\prime} E_0.
\end{equation}
Following the results (\ref{IP_c}) and (\ref{EA_r}) their diagonal
elements can be regarded as the energy of an electron occupying
some Wannier orbital $n$ (or $m$) in any cell and the off-diagonal
elements are considered as hopping matrix elements between
orbitals $n$ in the cell $R$ and $n^\prime$ in the cell
$R^\prime$. However, one should always keep in mind, that in
reality an electron can not be removed from (or added to) some
Wannier orbital. Nevertheless, these matrix elements can be used
to obtain the band structure.

\begin{figure}
\centerline{
\psfig{figure=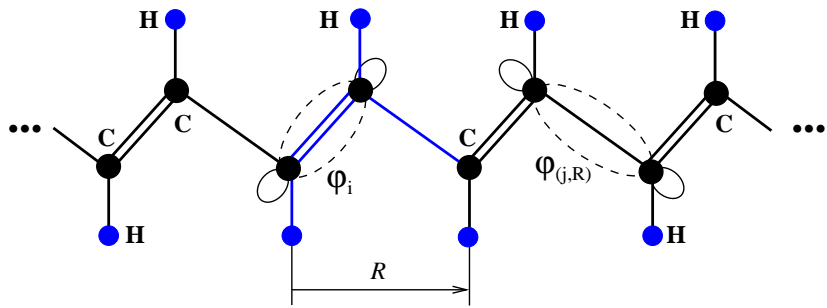,height=4cm}
}
\caption{Schematic picture of two Wannier orbitals in {\it trans}-polyacetylene.}
\label{WOfig}
\end{figure}

Since the quantities (\ref{IPnn}) and (\ref{EAmm}) depend only on
the difference $R^\prime-R$ we fix $R$ at the zero cell
($R \equiv 0$). In Fig.~\ref{WOfig} two occupied Wannier
orbitals in {\it trans}-polyacetylene are shown schematically. The
WO $\varphi_i$ is centered at the zero unit cell (marked by blue
bonds). The energy of an electron in this orbital is $-{\rm
IP}_{0,ii}$. The hopping matrix element between WO $\varphi_i$ and
WO $\varphi_j$ from the cell at distance $R$ from the reference
cell is $-{\rm IP}_{R,ij}$. From now on we will refer to ${\rm
IP}_{R,ij}$ and ${\rm EA}_{R,ij}$ as the "local matrix elements"
(LMEs).

The energy bands (one-particle energies as a function of the crystal-momentum $k$)
are obtained as eigenvalues of the "back-transformed" matrices

\begin{equation}
\label{IPn}
{\rm IP}_{nn^\prime} (k) = \sum_R {\rm e}^{ikR}
{\rm IP}_{R,nn^\prime}
\end{equation}
and

\begin{equation}
\label{EAm}
{\rm EA}_{mm^\prime} (k) = \sum_R {\rm e}^{ikR}
{\rm EA}_{R,mm^\prime}.
\end{equation}

The matrix elements ${\rm IP}_{R,nn^\prime}$ and ${\rm
EA}_{R,mm^\prime}$ are the key quantities in our method for
correlated band structures. Being calculated on the HF level
they reproduce the HF band structure. In fact, only a relatively
small amount of matrix elements is needed for doing this as the
hopping matrix elements decay rapidly with increasing distance $R$
and therefore the infinite summation in Eqs.~(\ref{IPn}) and
(\ref{EAm}) can be terminated at some $R_{\rm cut}$ such that the rest of
the sum is negligible with respect to the terms summed up so far.

The effect of electron correlation leads to a change of the matrix
elements ${\rm IP}_{R,nn^\prime}$ and ${\rm EA}_{R,mm^\prime}$ as
will be shown in the following chapters. This means that the
energy of a correlated hole or attached electron in some localized
orbital and also the hopping matrix elements between different
sites are not the same as in the HF approximation but contain all
the information about the response of the charged ($N-1$)- or
($N+1$)-particle system and the correlation in the ground state.
Thus, substituting "correlated" matrix elements ${\rm IP}^{\rm
corr}_{R,nn^\prime}$ and ${\rm EA}^{\rm corr}_{R,mm^\prime}$ in
Eqs.~(\ref{IPn}) and (\ref{EAm}) gives the band structure with
electron correlations included.

The correlation effect is determined by a variational method in
which all unoccupied orbitals are used to construct the
variational space. In a crystal there is an infinite amount of
virtual orbitals which would lead to an infinitely large
variational space and make the calculation not feasible. To arrive
at a finite variational space all the unoccupied Bloch waves can
be transformed to Wannier orbitals and ordered by their distance
from the localized charge and then only WOs up to a certain
distance $R_{\rm cut}$ are taken into account. However, firstly,
the localization (Wannier transformation) of {\it all} virtual
Bloch orbitals becomes very expensive for rich basis sets and,
secondly, WOs corresponding to energetically high Bloch states are
spatially rather extended and therefore the cut-off radius $R_{\rm
cut}$ has to be chosen large enough that may again lead to
infeasible correlation calculations. As an alternative to the set
of unoccupied WOs, one can use atomic orbitals with one projected
out of the occupied space. This idea is being efficiently
implemented in quantum-chemical codes (see e.g. \cite{Schuetz99}
and \cite{Werner03}). These orbitals are perfectly localized and
the cut-off radius is reduced to the minimum value but they are
(i) not mutually orthogonal and (ii) can not be used to reproduce
the HF bands by a suitable unitary transformation.

In our approach we avoid the problem of an infinite variational
space by performing the correlation calculations in finite
molecules representing a small piece of the periodic system. This
implies a finite number of virtual HF orbitals in these molecules.

\section{One-particle configurations}

Here we have come to the crucial point of our method. The matrix
elements ${\rm IP}_{R,ij}$ and ${\rm EA}_{R,ij}$ defined in
(\ref{IPnn}) and (\ref{EAmm}) can be obtained (approximately) from
calculations in finite molecules. Moreover, electron correlation
can be easily incorporated by performing standard quantum-chemical
finite-cluster correlation calculations. In the present Section we
focus our attention on the first statement.

For understanding this statement the molecular-orbital theory may
provide a hint. A bond orbital $\varphi_a$ between two atoms $A$
and $B$ in a molecule (a localized one-particle wavefunction) is
obtained as a linear combination of atomic basis functions with
the main contribution centered on these two atoms:

\begin{equation}
\label{mol_bond} \varphi_a = \sum_i \alpha_{A,i} \chi_{A,i} +
\sum_i \alpha_{B,i} \chi_{B,i} + \sum_{i,C\neq A,B} \alpha_{C,i}
\chi_{C,i}
\end{equation}
with $|\alpha_{A,i}|$, $|\alpha_{B,i}| \gg \: ^{\;\; \large \rm max}_{C\neq A,B}
\: |\alpha_{C,i}|$ where $\chi_{A,i}$ is the $i$-th atomic
basis function of atom $A$ and $|\alpha_{C,i}|$ decays rapidly with
the distance between the $C$-th atom and the bond. The same bond
one expects to find in a covalent extended system when the same
two atoms at the same distance and with similar atomic structure
in the neighborhood are considered. Therefore one expects to
obtain occupied Wannier orbitals in a crystal which are very
similar to the localized occupied spin orbitals in a molecule when
this molecule represents some finite part of the crystal provided
the same atomic basis sets and the same localization criteria are
used for both the molecule and the crystal. The larger the
molecule the closer the localized molecular orbital (LMO) in the
central part of the molecule are to the relevant WO of the
crystal. The LMOs at the edges of the molecule are less similar to
the corresponding WOs as the surrounding of the former differs
strongly from the surrounding in the crystal.

The matrix elements IP$_{R, nn^\prime}$ (\ref{IPnn}) are defined
in terms of localized crystalline orbitals which are derived from
HF solutions for the closed-shell system in the ground state. The
diagonal elements $-{\rm IP}_{0,nn}$ are the energies
$\varepsilon_{0,nn}$ of an electron in the $n$-th orbital of the
reference unit cell which constitutes a generalization of
Koopmans' theorem to the case of localized orbitals. The quantity
$\varepsilon_{0,nn}$ depends on the shape of the orbital itself
and on the atomic and electronic structure of the close surrounding. If they are
similar in the molecule and in the crystal the corresponding
orbital energies should coincide.The same statement is valid for
the hopping matrix elements in (\ref{IPnn}) when the two localized
orbitals and the structure between them do not differ. Thus all
the matrix elements which are needed in the restricted summation
(\ref{IPn}) can be obtained from calculations in large enough but
finite molecules.

Similar argumentation holds for the unoccupied WOs corresponding
to the energetically lowest conduction bands with the distinction
that localized unoccupied orbitals in molecules can not be
obtained by standard quantum-chemical localization methods, in
general. However, in this case the WOs from the periodic system
can directly be projected onto a molecule when the structure and
the basis sets are identical and can be used as unoccupied LMOs.
This technics is described in details in chapter 3.2.2. Here we
only emphasize that local matrix elements EA$_{R,mm^\prime}$
(\ref{EAmm}) can be obtained from calculations in finite molecules
as well.

To get the matrix elements (\ref{IPnn}) and (\ref{EAmm}) from
molecular calculations an appropriate molecules have to be chosen.
The structure of such molecules must contain the appropriate
clusters of a crystal under consideration, e.g. those unit cells
which are needed to represent the corresponding WOs without
noticeable loss of accuracy. Such clusters cut from covalent
crystals have dangling bonds which can be saturated by hydrogen
atoms. To be more specific let us consider the occupied WOs
schematically shown in the Fig.~\ref{WOfig}. There, four unit
cells of {\it trans}-polyacetylene are drawn. The two WOs define
the three matrix elements ${\rm IP}_{0,ii}$, ${\rm IP}_{R,ij}$ and
${\rm IP}_{R,jj} \equiv {\rm IP}_{0,jj}$. To obtain these matrix
elements from a molecule calculation the shown cluster C$_8$H$_8$
(with four unit cells) is appropriate and therefore for
calculations one can use a C$_8$H$_{10}$ molecule with the same
geometry as the C$_8$H$_8$ cluster but with the two cut C--C bonds
being substituted by C--H bonds with the same orientation as the
C--C bond and a typical C--H bond length. In our investigation of
{\it trans}-polyacetylene we use C$_6$H$_8$, C$_8$H$_{10}$,
C$_{10}$H$_{12}$ and C$_{12}$H$_{14}$ molecules which represent
different-size clusters of the infinite chain all being
constructed to contain full unit cells such that they can be
terminated by single C--H bonds. The size of the chosen cluster is
defined by the distance between the relevant WOs and should be as
small as possible for cheaper calculations but still big enough to
preserve the shape of the WOs and by this the accuracy of the
results. These criteria hold for the selection of proper molecules for
all kinds of covalent materials, e.g. diamond, silicon, gallium
arsenide, polyethylene etc.

Thus, having chosen appropriate molecules we can formally redefine
the matrix elements (\ref{IPnn}) and (\ref{EAmm}) in terms of
localized molecular orbitals

\begin{equation}
\label{IPmol} {\rm IP}^{\rm mol}_{IJ} = \langle I | H^{\rm mol} |
J \rangle - \delta_{IJ} E^{\rm mol}_0
\end{equation}
with

\begin{equation}
\label{I}
| I \rangle = c_I |\Phi^{\rm mol} \rangle
\end{equation}
and

\begin{equation}
\label{EAmol}
{\rm EA}^{\rm mol}_{I^\ast J^\ast} = - \langle I^\ast | H^{\rm mol} | J^\ast
\rangle + \delta_{I^\ast J^\ast} E^{\rm mol}_0
\end{equation}
with

\begin{equation}
\label{I*}
| I^\ast \rangle = c^\dagger_{I^\ast} |\Phi^{\rm mol} \rangle
\end{equation}
where $c_I$ destroys an electron in an occupied localized
molecular orbital named $I$ (a bonding orbital) and
$c^\dagger_{I^\ast}$ creates an electron in an unoccupied
localized molecular orbital $I^\ast$ (an anti-bonding orbital).

Starting from this point we shall refer to local matrix elements
as obtained from calculations in molecules. Also we adopt the
notations for many-body wavefunctions used in quantum chemistry.
In the HF approximation the electronic wavefunction of a molecule
in the ground state is written as a Slater determinant
(\ref{Slaterdet}). Since the Slater determinant is essentially (up
to a phase factor) invariant under any unitary transformation of
the orbitals we rewrite it in terms of the localized spin orbitals
(\ref{locorb})

\begin{equation}
\label{Phi_gs}
\Phi = \frac{1}{\sqrt{N!}} \det [\varphi_1, \ldots , \varphi_N]
\end{equation}
where $N$ is the number of electrons in the molecule. In the HF
ground state configuration all occupied spin orbitals $\varphi_a$
contain an electron and all virtual spin orbitals $\varphi_r$ are
empty. By taking out an electron from an orbital $\varphi_a$ we
produce an $(N-1)$-particle state with the wavefunction

\begin{equation}
\label{Phi_N-1} \Phi_a = \frac{(-1)^{N-a}}{\sqrt{(N-1)!}} \det
[\varphi_1, \ldots , \varphi_{a-1}, \varphi_{a+1}, \ldots ,
\varphi_N].
\end{equation}
When an electron is added to the system and placed into the
virtual spin orbital $\varphi_r$ an $(N+1)$-particle state is
produced and its wavefunction is

\begin{equation}
\label{Phi_N+1}
\Phi_r = \frac{1}{\sqrt{(N+1)!}} \det [\varphi_1, \ldots , \varphi_N, \varphi_r].
\end{equation}
We shall call the wavefunctions (\ref{Phi_N-1}) and (\ref{Phi_N+1})
"one-particle configurations" as they correspond to states with
one particle (a hole or an electron) being added to the system in
the ground state.

Now we would like to show explicitly how local matrix elements are
obtained from molecular quantum-chemical calculations on the HF
level. As follows from Eqs.~(\ref{IPmol})--(\ref{Phi_N+1}) the
LMEs are given in terms of the one-particle configurations as

\begin{equation}
\label{IPab} {\rm IP}_{ab} = \langle \Phi_a | H^{\rm mol} | \Phi_b
\rangle - \delta_{ab} E_0^{\rm mol},
\end{equation}

\begin{equation}
\label{EArs} {\rm EA}_{rs} = - \langle \Phi_r | H^{\rm mol} |
\Phi_s \rangle + \delta_{rs} E_0^{\rm mol}
\end{equation}
where

\begin{equation}
\label{E0mol} E_0^{\rm mol} = \langle \Phi | H^{\rm mol} | \Phi
\rangle.
\end{equation}

In the following we omit the superscript "mol" since our
calculations are always performed in molecules except of the
localization of the unoccupied orbitals and there we explicitly
distinguish infinite and finite systems. From the HF calculations
with the program package MOLPRO \cite{MOLPRO} the total
ground-state energy $E_0$, the canonical HF spin orbitals $\psi_i$
and the Fock matrix (\ref{Fock_matr}) are obtained. In the
representation of the canonical orbitals the Fock matrix has
diagonal form

\begin{equation}
\label{f_ij} f_{ij} = \langle \psi_i | F | \psi_j \rangle =
\delta_{ij} \varepsilon_i
\end{equation}
where $\varepsilon_i$ are the canonical orbital energy
(\ref{orb_en}). By a localization procedure applied to the
canonical orbitals one obtains the LMOs (\ref{locorb}) and the
corresponding unitary matrix $u_{ji}$. Then using
Eqs.~(\ref{locorb}), (\ref{E_c}), (\ref{IP}) and  (\ref{IP_c}) one
can express matrix elements IP$_{ab}$ in terms of calculated
quantities:

\begin{eqnarray}
\label{IPab2} {\rm IP}_{ab} & = & \langle \Phi_a | H | \Phi_b
\rangle - \delta_{ab} E_0 \; =  \; \langle \sum_i^{\rm occ} u_{ia} \; ^{N-1} \! \Phi_i |
H | \sum_j^{\rm occ} u_{jb} \;  ^{N-1} \! \Phi_j \rangle - \delta_{ab} E_0 \nonumber \\
& = & \sum_i^{\rm occ} \sum_j^{\rm occ} u_{ia} u_{jb} \langle ^{N-1} \! \Phi_i | H | ^{N-1} \! \Phi_j
\rangle - \delta_{ab} E_0 \;  = \; \sum_i^{\rm occ} \sum_j^{\rm occ} u_{ia} u_{jb}
\; ^{N-1}E_i \delta_{ij} - \delta_{ab} E_0 \nonumber \\
& = & \sum_i^{\rm occ} u_{ia} u_{ib} (E_0 - \varepsilon_i) - \delta_{ab} E_0
\;  = \; \sum_i^{\rm occ} u_{ia} \varepsilon_i u_{ib}.
\end{eqnarray}
(Here we omitted complex conjugation as molecular HF orbitals
are usually real.)
Thus, the local matrix elements IP$_{ab}$ obtained on the HF level
are in fact the elements of the Fock matrix (with the opposite
sign) which is defined in terms of the localized occupied
orbitals:

\begin{equation}
\label{IP_as_Fock_matr} {\rm IP}_{ab}  = - f_{ab} = - \langle
\varphi_a | F | \varphi_b \rangle.
\end{equation}
Similarly one gets

\begin{equation}
\label{EArs2} {\rm EA}_{rs}  = - \sum_i^{\rm vir} u_{ir} u_{is} (E_0 +
\varepsilon_i) + \delta_{rs} E_0 = - \sum_i^{\rm vir} u_{ir} \varepsilon_i
u_{is}.
\end{equation}
which simply is

\begin{equation}
\label{EA_as_Fock_matr} {\rm EA}_{rs}  = - f_{rs} = - \langle
\varphi_r | F | \varphi_s \rangle.
\end{equation}
In (\ref{IPab2}) index $i$ runs over all occupied canonical
orbitals and in (\ref{EArs2}) it counts only relevant virtual ones
which means all virtual canonical orbitals which were used for the
generating of the virtual LMOs $\varphi_r$.

The crystalline matrix elements (\ref{IPnn}) and (\ref{EAmm})
which enter the restricted summations (\ref{IPn}) and (\ref{EAm})
can be replaced by the matrix elements (\ref{IPab2}) and
(\ref{EArs2}) obtained in molecules, and by this the HF band
structure of the corresponding crystal can be calculated. Such a
band structure, obtained by LMEs from molecules, can be compared
to that obtained from canonical HF calculations for the infinite
system, e.g. done by the program package CRYSTAL \cite{CRYSTAL}
using identical basis sets and geometries of the crystal and the
clusters. Also the LMEs themselves can be compared directly if
elements of the Fock matrix in the representation of the WOs can
be obtained from the calculations of the infinite system. Such
comparisons show how accurate the cluster approach is. In fact,
there should be small deviations between the band structure by
LMEs on the HF level and that calculated by CRYSTAL because of the
truncation in the summations (\ref{IPn}) and (\ref{EAm}). When the
correlated band structure is obtained by LMEs including
correlation corrections one can account for these small deviations
as is explained in chapter 4.2.

\section{Electron correlation effects}

Above we have showed how the many-body wavefunction in the HF
approximation can be obtained. This approximation serves as a good
starting point for the study of electronic properties both in
finite molecules and in extended systems providing single-electron
orbitals with corresponding orbital energies, the ground state
energy of the system, the energies of the ($N+1$)- and
($N-1$)-particle system in different excited states, ionization
potentials and electron affinities. However, all these results can
only be considered as approximate since the effect of electron
correlation is missing on the HF level.

Let us consider an extended periodic system (a crystal or a
polymer chain) in the ground state for which the HF solution is
known, in particular the ground-state energy $E_0$ and the
localized electron orbitals (Wannier orbitals). $E_0$ accounts for
the Coulomb repulsion between the electrons only in an averaged
way (mean-field approximation). In reality electrons move in
accordance to the movement of their neighbor electrons trying to
avoid coming too close to each other and to reduce by this the
Coulomb repulsion. Consequently, the total energy of the system,
$E_0$, is reduced noticeably. Alternatively, we can describe this
effect as charge fluctuations. In a neutral system a dipole may
emerge due to local fluctuation of the charge. The system reacts
to this by inducing another dipole and gains the energy of the
dipole-dipole interaction.

If an additional electron is added to a localized virtual orbital
(or removed from some occupied WO) the system reacts on the
presence of the extra charge by producing a polarization cloud
around it. The main effect of electron rearrangement happens in
the close surrounding and decays with increasing distance. Of
course, the effect, described above for the neutral system, also
takes place in the charged one. Both these effects lead to a
substantial reduction of the energy $^{N+1}E$ of the
($N+1$)-particle system and the energy $^{N-1}E$ of the
($N-1$)-particle system.

As we have seen, electron correlations in the ground state of a
system are of van der Waals type and the contribution to the
system energy falls rapidly with the distance between dipoles.
Therefore, only rather small clusters are needed to properly
estimate the correlation contribution to the ground-state energy.
Significantly more efforts requires the proper description of
electron correlations in the charged system. An accurate treatment
of the polarization cloud around the localized extra charge is
necessary as this is the main contribution to the correlation
corrections to the LMEs. Since electron-dipole interaction decays
more slowly than dipole-dipole interaction, larger clusters have
to be chosen. Also, one can account separately for the (relatively
small) contribution coming from the polarization of the farther
surrounding which is not present in the cluster models. For that
purpose the crystal is considered as a polarizable continuum in
the presence of a point charge and the approximate formula for the
corresponding contribution to the correlation energy is presented
in chapter~4.3. Thus, electron correlations have mainly local
character and the corresponding corrections to the ground-state
energy and the LMEs can be obtained from calculations on finite
clusters.

Now we would like to show explicitly how electron correlation is
included in quantum-chemical post-HF methods. Solving the
Hartree--Fock equations (\ref{Fock_equat})--(\ref{v}) not only
provides an approximate ground-state wavefunction in the form of a
single determinant (\ref{Slaterdet}) but also one obtains a set of
$2 \times N_{\rm bas}$ spin orbitals (or  $N_{\rm bas}$ spatial
orbitals) where  $N_{\rm bas}$ is the number of atomic basis
functions $\{ \chi_i \}$, $i=1, \dots , N_{\rm bas}$ for a given
molecule. Using these spin orbitals one can select a huge number
of different orbital configurations which is of the order of
$\bigl(^{2N_{\rm bas}}_N \bigr)$ where $\bigl(^a_b \bigr)$ is the
binomial coefficient. Each of these orbital configurations
represents a single determinant referred to simply as
"configuration". The HF wavefunction $\Phi$ is usually defined as
that configuration which gives the lowest (self-consistent) energy
$E_0$ for a given number of electrons in the molecule.

To incorporate correlations in the wavefunction one takes
advantage of the fact that the entire set of configurations
$\Phi_I$ form a basis of the Hilbert space. Hence one can write
the "correlated" wave function of the system as the linear
combination of single-determinant configurations:

\begin{equation}
\label{Phi_corr}
\Phi^{\rm corr} = \Phi + \sum_I \alpha_I \Phi_I
\end{equation}
(here written in the intermediate normalization without a
prefactor for $\Phi$). By varying the coefficients $\alpha_I$ one
can minimize the energy corresponding to the correlated
wavefunction (\ref{Phi_corr})

\begin{equation}
\label{E_corr}
E_0^{\rm corr} = \frac{\langle \Phi^{\rm corr} | H | \Phi^{\rm corr} \rangle}
{\langle \Phi^{\rm corr} | \Phi^{\rm corr} \rangle}.
\end{equation}
Of course, in practice, one has to truncate the set of
configurations somehow. This is the essence of the so-called
"configuration interaction" (CI) methods. The obtained energy
$E_0^{\rm corr}$ is lower than $E_0$ from the HF calculations. The
same argumentation holds for the wavefunction and the energy of
($N-1$)- and ($N+1$)-particle systems when the ($N-1$)- and
($N+1$)-electron configurations are used, respectively, to
construct the multi-determinant wavefunctions.

In fact only for small molecules with a few atoms and small basis
sets a variation of all configurations is possible (full CI). For
larger systems a drastic reduction of the variational space (the
number of involved configurations) must be done. The excited
configurations denoted as $\Phi_I$ in (\ref{Phi_corr}) can be
sorted by the number of substitutions of occupied spin orbitals in
the ground-state configuration by unoccupied ones and the
correlated wavefunction (\ref{Phi_corr}) can be rewritten as

\begin{equation}
\label{Phi_corrCI}
\Phi^{\rm corr} = \Phi + \sum_{ar} \alpha_a^r \Phi_a^r
+\sum_{a<b, \: r<s} \alpha_{ab}^{rs} \Phi_{ab}^{rs}
+\sum_{a<b<c, \: r<s<t} \alpha_{abc}^{rst} \Phi_{abc}^{rst}
+ \ldots
\end{equation}
where $\Phi_a^r$, $\Phi_{ab}^{rs}$ and $\Phi_{abc}^{rst}$ are the
excited configurations, subscripts denote the spin orbitals
excluded from $\Phi$ and superscripts denote the substituting
virtual spin orbitals.

The contributions to the correlated wavefunction, associated with
configurations with three and more substitutions (so-called
excitations), are negligibly small all together compared to those
of first three terms on the right-hand side of
Eq.~(\ref{Phi_corrCI}). Therefore, we take into account only the
ground-state configuration (the dominant contribution to the
correlated wavefunction) and the singly and doubly excited
configurations (single-double CI). Note that the singly excited
configurations do not contribute directly to the correlation energy of the
ground state as they do not couple to the HF ground-state
configuration: $\langle \Phi_a^r | H | \Phi \rangle = 0$.
This is the statement of Brillouin's theorem
(see, e.g., chapter 3.3.2 in \cite{Szabo}) which holds as long as
the HF spin orbitals are determined self-consistently. Thus, the
second sum in Eq.~(\ref{Phi_corrCI}) including the double
excitations gives the main contribution
to correlation energy and includes dipole-dipole interactions.

The correlated wavefunctions for the charged systems can be
constructed similarly. For instance, the wavefunction of a system
with a {\it static} hole in a {\it frozen} orbital $\varphi_a$
would be written as

\begin{equation}
\label{Phi_a^corr_stat} \Phi_a^{\rm corr} = \Phi_a + \sum_{b\neq
a,r} \tilde{\alpha}_{ab}^{\; \; r} \Phi_{ab}^{\; \; r}
+\sum_{b<c\neq a, \: r<s} \tilde{\alpha}_{abc}^{\; \; rs}
\Phi_{abc}^{\; \; rs}
\end{equation}
Here both sums give contribution to the energy of the
($N-1$)-electron system. Matrix elements $\langle \Phi_{ab}^{\; \;
r} | H | \Phi_a \rangle$ do not vanish and describe the
polarization cloud around an extra charge.

Of course, in reality a hole is delocalized over the entire system
and a more general ansatz for the correlated wavefunction of the
($N-1$)-particle system has to be made:

\begin{equation}
\label{Phi_a^corr} ^{N-1}\Phi^{\rm corr} = \sum_a \alpha_a
\Phi_a + \sum_{a, b\neq a,r} \alpha_{ab}^{\; \; r} \Phi_{ab}^{\;
\; r} +\sum_{a,b<c\neq a, \: r<s} \alpha_{abc}^{\; \; rs}
\Phi_{abc}^{\; \; rs}
\end{equation}

The number of configurations (the number of single determinants)
required for this is huge even though we restricted ourselves to
singly and doubly excited configurations only. However, we do not
need to store the explicitly correlated wavefunction consisting of
billions of single determinants since we are rather interested in
physical quantities which are obtained with this wavefunction,
namely, LMEs like the ones introduced on the HF level (\ref{IPab})
and (\ref{EArs}). Below we show how this is done.

First, we turn back to the representation of the HF one-particle
configurations in terms of canonical orbitals in which the Fock
matrix is diagonal. We denote these configurations as $\Psi_i$
here with an electron being removed from (or added to) the
canonical spin orbital $\psi_i$. The configurations $\Psi_i$ can
also be obtained as the linear combinations of the "local"
configurations $\Phi_i$: $\Psi_i = \sum_j \Phi_j \:
(u^{-1})_{ji}$, because the canonical spin orbitals can be
obtained from the localized ones with the help of the unitary
matrix $u_{ji}$ from the localization procedure (\ref{locorb}):
$\psi_i = \sum_j \varphi_j \: (u^{-1})_{ji}$. The correlated
counterpart of the HF wavefunction $\Psi_i$ is that {\it
eigenstate} $\Psi_i^{\rm corr}$ of the charged system which
resembles the HF configuration $\Psi_i$ the most. It will
essentially be composed by local one-particle configurations like
the ones in the first sum on the right-hand side of Eq.~(\ref{Phi_a^corr}) for
$^{N-1}\Phi^{\rm corr}$. The wavefunctions $\Psi_i^{\rm corr}$
of the ($N-1$)-electron or ($N+1$)-electron system are associated
with {\it total} energies $E_i^{\rm corr}$:

\begin{equation}
\label{Schr_eq}
H \Psi_i^{\rm corr} = E_i^{\rm corr} \Psi_i^{\rm corr}.
\end{equation}

Next, we formally introduce a one-to-one correspondence between the HF
many-particle wavefunctions and their correlated counterparts by the
use of a {\it wave operator} $\Omega$ which produces the correlated
wavefunctions $\Psi_i^{\rm corr}$ from the HF ones $\Psi_i$

\begin{equation}
\label{Omega}
\Psi_i^{\rm corr} = \Omega \Psi_i, \qquad \qquad i=1,\ldots,d.
\end{equation}
Then, we divide the full space of local configurations $\Phi_j$,
$\Phi_{ja}^{\; \; r}$ and $\Phi_{jab}^{\; \; rs}$ used to
construct the correlated wavefunctions $\Psi_i^{\rm corr}$ into a
small subspace $\cal{P}$ of dimension $d$ consisting of all SCF
one-particle configurations $\Psi_i$ (or equivalently of all local
configurations $\Phi_j$) with $P$ being the projector onto this
space and its orthogonal complement $\cal{Q}$ (with the projector
$Q$), so $P+Q=1$. The $\cal{P}$-space is often called the {\it
model space} (or the space of {\it model configurations}) and the
$\cal{Q}$-space is called the {\it external space}, and
accordingly the configurations which belong to the $\cal{Q}$-space
are called {\it external configurations}. Thus, the wave operator
$\Omega$ acts on configurations $\Psi_i$ from the $\cal{P}$-space
and yields the exact wavefunctions $\Psi_i^{\rm corr}$ from the
full Hilbert space.

\begin{figure}
\centerline{
\psfig{figure=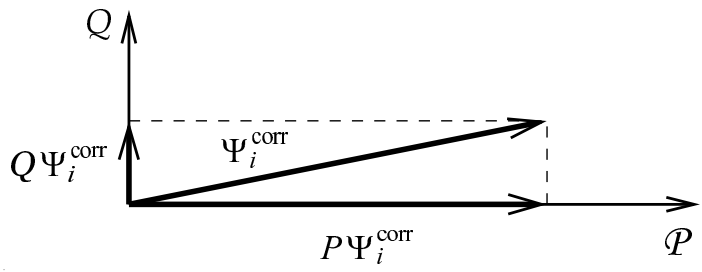,height=3.5cm}
}
\caption{Schematic representation of the correlated wavefunction.}
\label{PQfig}
\end{figure}

The HF wavefunctions $\Psi_i$ can be obtained from the correlated
counterparts by the back transformation

\begin{equation}
\label{project}
\Psi_i = \Pi \: \Psi_i^{\rm corr}.
\end{equation}
$\Pi$ is an operator acting from the $d$-dimensional space spanned
by the correlated counterparts $\{ \Psi_i^{\rm corr} \}_{i=1,\ldots,d}$
to the model space $\cal{P}$. It is similar but not identical to the
projector $P$ which can be written as

\begin{equation}
\label{P}
P = \sum_{i=1}^d |\Psi_i \rangle \langle \Psi_i |
\end{equation}
because the projections $P \Psi_i^{\rm corr}$ are neither
perfectly normalized nor mutually orthogonal. In Fig.~\ref{PQfig}
we schematically show a correlated wavefunction $\Psi_i^{\rm
corr}$. Both, its projection $P \Psi_i^{\rm corr}$ and the HF
counterpart $\Psi_i$ lie in the $\cal{P}$-space but differ
slightly.

Let us substitute (\ref{Omega}) into the left-hand side of Eq.~(\ref{Schr_eq})
and act on the obtained equation by the operator $\Pi$ from the left

\begin{equation}
\label{deriv_Heff} \Pi H \Omega  \Psi_i = \Pi H  \Psi_i^{\rm corr}
= E_i^{\rm corr} \Pi \Psi_i^{\rm corr} =  E_i^{\rm corr} \Psi_i.
\end{equation}
This way we get an eigenvalue equation for $\Psi_i$ and $E_i^{\rm
corr}$. Inspired by this relation we define an {\it effective}
Hamiltonian as

\begin{equation}
\label{Heff}
H^{\rm eff} = \Pi H \Omega.
\end{equation}
It operates fully in the $\cal{P}$-space and its eigenvalues are those of
the {\it exact} Hamiltonian (\ref{Schr_eq}):

\begin{equation}
\label{Schr_eq_Heff}
H^{\rm eff} \Psi_i = E_i^{\rm corr} \Psi_i.
\end{equation}

As the unitary transformation (\ref{locorb}), which gives the localized HF
spin orbitals of a molecule in terms of the canonical ones, is known from
the localization procedure we can readily get the matrix elements of
$H^{\rm eff}$ in terms of the localized one-particle configurations
$\Phi_i$:

\begin{eqnarray}
\label{Heff_ij}
H^{\rm eff}_{ij} & = & \langle \Phi_i | H^{\rm eff} | \Phi_j \rangle =
\langle \sum_{i^\prime} u_{i^\prime i}\Psi_{i^\prime} | H^{\rm eff} |
\sum_{j^\prime} u_{j^\prime j}\Psi_{j^\prime} \rangle \nonumber \\
& = & \sum_{i^\prime} \sum_{j^\prime} u_{i^\prime i} \: u_{j^\prime j}
\: E_{i^\prime}^{\rm corr} \: \delta_{i^\prime j^\prime} =
\sum_{i^\prime} u_{i^\prime i} \: E_{i^\prime}^{\rm corr} \: u_{i^\prime j}
\end{eqnarray}
where the energies $E_{i^\prime}^{\rm corr}$ are those obtained in
our correlation calculations. To be more precise, these energies 
as well as the elements of the matrix $u_{ij}$ were calculated
by the MOLPRO program package \cite{MOLPRO}, \cite{Werner88}, \cite{Knowles88},
\cite{Knowles92}. The matrix elements $H^{\rm eff}_{ij}$ defined in
Eq.(\ref{Heff_ij}) differ from the corresponding ones on the HF level
$H_{ij}$ and contain electron correlation as far as it is included
in the energies $E_{i^\prime}^{\rm corr}$. Together with the
correlated energy of the ground state $E_0^{\rm corr}$
(\ref{E_corr}) the matrix elements of the effective Hamiltonian
define the correlated LMEs:

\begin{equation}
\label{IP_corr}
{\rm IP}_{ab}^{\rm corr} = H_{ab}^{\rm eff} - \delta_{ab} E_0^{\rm corr}
\end{equation}
and

\begin{equation}
\label{EA_corr}
{\rm EA}_{rs}^{\rm corr} = - H_{rs}^{\rm eff} + \delta_{rs} E_0^{\rm corr}
\end{equation}
which are to be used in (\ref{IPn}) and (\ref{EAm}) instead of the
SCF ones to give the correlated band structure. Note the
similarity of the equations (\ref{Heff_ij})--(\ref{EA_corr}) with
the corresponding ones for the matrix elements IP$_{ab}$ and
EA$_{rs}$ on the HF level (\ref{IPab2}) and (\ref{EArs2}).

It should be mentioned here that one can also use $P$ instead of $\Pi$ and
a corresponding wave operator $W$ (which is linked to $\Omega$ by
$W = \Omega (P \Omega)^{-1}$) to define an effective Hamiltonian

\begin{equation}
\label{Heff^prime} \tilde{H}^{\rm eff} = P H W.
\end{equation}
Despite the fact that $\tilde{H}^{\rm eff}$ is non-Hermitian it
has the same properties as $H^{\rm eff}$.

At the end of this section we would like to say a few more words
about the wave operator $\Omega$ and its inverse $\Pi$. They enter
the {\it definition} of the effective Hamiltonian (\ref{Heff})
however they are not used for obtaining its matrix elements
$H^{\rm eff}_{ij}$. The latter ones are calculated using equation
(\ref{Heff_ij}). Hence, $\Omega$ and $\Pi$ (or $P$ and $W$) are
only introduced formally in our method. Nevertheless, $P$ and $W$
can be constructed by the use of perturbation theory up to any
desired accuracy as is comprehensively explained in chapter~9 of
Ref.~\cite{Lindgren}. Then the correlated wavefunctions
(\ref{Phi_a^corr}) can be obtained by applying $W$ to the
projection $P \Psi_i^{\rm corr}$ which is obtained perturbatively
as well.

\section{Multireference configuration interaction method}

Above we have used the CI ansatz to write the correlated
wavefunction of a many-body system (\ref{Phi_corrCI}) and
(\ref{Phi_a^corr}). However, there are also other correlation methods
which use the HF wavefunction as a starting point and excited
configurations to construct its correlated counterpart. Here
we would like to mention two of them, namely the single and double
excitation coupled-cluster (CCSD) method and the second-order
M\o{}ller--Plesset perturbation (MP2) method. The CCSD method is more
successful in ground-state correlation calculations
than the single and double configuration interaction (CI(SD))
and both have the advantage that they are size-consistent in
contrast to CI(SD). Unfortunately, CCSD and MP2 are not suitable
for excited-state calculations.

The CCSD ansatz for the correlated ground-state wavefunction is

\begin{equation}
\label{Phi_corrCC} \Phi^{\rm corr}_{CC} = \exp \biggl( \sum_{ar}
t_a^r c_a c_r^\dagger +\sum_{a<b, \: r<s} t_{ab}^{rs} c_a c_b
c_r^\dagger c_s^\dagger \biggr) \Phi
\end{equation}
where the operators $c_a$ and $c_b$ ($c_r^\dagger$ and
$c_s^\dagger$) destroy (create) an electron in an occupied
(unoccupied) spin orbitals $\varphi_a$ and $\varphi_b$
($\varphi_r$ and $\varphi_s$) of the ground-state configuration,
respectively. The coefficients $t_a^r$ and $t_{ab}^{rs}$ in
Eq.~(\ref{Phi_corrCC}) are obtained from a set of algebraic
equations. The exponential form of $\Phi^{\rm corr}_{CC}$ implies
that the correlated wavefunction of a system consisting of some
well separated subsystems is the product of the wavefunctions of
these subsystems which is the crucial prerequisite for
size-consistency. There is no simple extension of the CC method to
multireference (MR) calculations which are needed to describe
excited states. Though such an extension of CC became a hot topic
in the literature in the last decade, e.g. \cite{Bartlett99},
\cite{Ivanov}, \cite{Pittner03}, the MRCC method seems still to be
far from the stage of implementation in standard quantum-chemical
program packages at the present moment.

The second-order M\o{}ller--Plesset perturbation method can be generalized to
the multireference case. However, in the perturbation expansion for the correlated
wavefunctions and energies of the ($N-1$)- and ($N+1$)-electron
systems there are terms which are proportional to

\begin{equation}
\label{denom}
\frac{1}{E_i - E_{ja}^{\; \; r}}
\end{equation}
Here $E_i$ is the SCF energy of some hole (or attached electron)
state and $E_{ja}^{\; \; r}$ is the energy of an SCF configuration
with one hole (or attached electron) state plus one electron-hole
excitation. It may happen, now, that $E_i$ gets very close to
$E_{ja}^{\; \; r}$ in the case where $E_i$ corresponds to an
energetically deep lying hole and $E_{ja}^{\; \; r}$ corresponds
to an energetically high lying hole and a close electron-hole
excitation. This situation takes place in systems with relatively
small band gap and a strong dispersion of the bands. Such small
denominators blow up the correlated energies and lead to wrong
results. Therefore, the MP2 method can only be applied to systems
with big band gaps and rather flat bands \cite{Albrecht98}. Also
deep valence bands and high conduction bands can not be treated
properly. This problem was discussed in, e.g. \cite{Sun96} and
\cite{Ayala01} where the MP2 method was generalized for
crystal-momentum space and could be applied only to the $\pi$
bands of {\it trans}-polyacetylene single chains.

As a variant, a hybrid method was used to determine
the correlated valence bands of diamond, silicon and germanium in
Ref.~\cite{Graef93}, \cite{Graef97} and \cite{Albrecht00} where
single-excited configurations were incorporated by MRCI(S) and for
the double excitations a separate calculation was performed
employing quasi-degenerate variational perturbation theory
\cite{Cave88}. In that case small denominators do not
show up. However, the size-consistency problem of MRCI(S) is still
present (though reduced as compared to MRCI(SD)) and the
contribution of coupling terms between singly- and doubly-excited
configurations to the correlation energy is missing totally. Also
such a hybrid method is technically more complicated
than pure MRCI(SD).

Hence, for our correlation calculations we have chosen the
multireference version of the single and double configurations
interaction method (MRCI(SD)) to calculate the correlated
ground-state energy (\ref{E_corr}) and the matrix elements of
the effective Hamiltonian (\ref{Heff_ij}) on both HF and
correlation level. The size-consistency problem has been
accounted for by the development of a special size-consistency
correction (for details see Section~3.5).

Let us first consider the CI(SD) ansatz for correlated
ground-state wavefunction $\Phi^{\rm corr}$. The HF ground-state
wavefunction $\Phi$ is written as the Slater determinant composed
from localized occupied spin orbitals. In general all singly- and
doubly-excited configurations are provided to the CI(SD)
expansion. However, one can further restrict the number of excited
configurations by forbidding to take out an electron from some
particular spin orbitals. For instance, we can "freeze" the
artificial C--H bonds which are used to saturate the dangling
bonds of the clusters. By this technique we exclude artificial
contributions to the correlation energy associated with these C--H
bonds. Also, we can freeze the majority of the bond orbitals of a
molecule and allow excitations from a few particular bonds only.
By this, we can replace an expensive correlation calculation for a
big molecule by a sequence of cheaper ones and sum up, in a proper
way, the correlation contribution to the ground-state energy
coming from the several parts of the molecule. This is the idea of
the method of local increments introduced in \cite{Stoll92a} and
\cite{Stoll92b} which we have adopted and which will be discussed
in more details in Section~3.4.

Thus, the correlated ground-state wavefunction of a molecule with $N$
electrons and $n$ bonds (or 2$n$ spin orbitals) open for constructing
excited configurations is written as

\begin{equation}
\label{Phi0_corrCI}
\Phi^{\rm corr} = \Phi + \sum_{a=1}^{2n} \sum_r \alpha_a^r \Phi_a^r
+\sum_{a,b=1}^{2n} \sum_{r,s} \alpha_{ab}^{rs} \Phi_{ab}^{rs}.
\end{equation}
The coefficients $\alpha_{a}^{r}$ and $\alpha_{ab}^{rs}$ are determined
by minimization of the ground-state energy of the system given by
Eq.~(\ref{E_corr}) where they play a role of variational parameters.
The resulting eigenvalue problem can be obtained by multiplying the
Schr\"odinger equation

\begin{equation}
\label{Scr_eq_corrCI}
H |\Phi^{\rm corr} \rangle = E_0^{\rm corr} |\Phi^{\rm corr} \rangle
\end{equation}
successively by $\langle \Phi |$, $\langle \Phi_{c}^{t} |$ and
$\langle \Phi_{cd}^{tu} |$ from the left:

\begin{eqnarray}
\label{CI_eq_set}
\langle \Phi | H |\Phi \rangle
+ \sum_{a,r} \langle \Phi | H |\Phi_{a}^{r} \rangle \alpha_{a}^{r}
+ \sum_{a,b,r,s} \langle \Phi | H |\Phi_{ab}^{rs} \rangle \alpha_{ab}^{rs}
& = & E_0^{\rm corr} \\
\langle \Phi_{c}^{t} | H |\Phi \rangle
+ \sum_{a,r} \langle \Phi_{c}^{t} | H |\Phi_{a}^{r} \rangle \alpha_{a}^{r}
+ \sum_{a,b,r,s} \langle \Phi_{c}^{t} | H |\Phi_{ab}^{rs} \rangle \alpha_{ab}^{rs}
& = & E_0^{\rm corr} \alpha_{c}^{t} \\
\langle \Phi_{cd}^{tu} | H |\Phi \rangle
+ \sum_{a,r} \langle \Phi_{cd}^{tu} | H |\Phi_{a}^{r} \rangle \alpha_{a}^{r}
+ \sum_{a,b,r,s} \langle \Phi_{cd}^{tu} | H |\Phi_{ab}^{rs} \rangle \alpha_{ab}^{rs}
& = & E_0^{\rm corr}  \alpha_{cd}^{tu}.
\label{CI_eq_set2}
\end{eqnarray}
This eigenvalue problem is too large to be solved directly and an
iterative procedure (according to Davidson \cite{Davidson75}) is
employed. The correlated ground-state energy can be obtained with
any desired precision in this procedure. In our calculations the
threshold for energy change form one iteration to the next is set
to $10^{-7}$ eV. However, this does not mean that we get the full
correlation energy with such an accuracy. That would only be the
case for a full CI calculation (with all possible excited
configurations included). Rather it defines an upper bound for the
exact correlation energy since truncated CI is a variational
method.

Now we would like to show how the CI(SD) ansatz can be extended to
get the matrix elements of the effective Hamiltonian
(\ref{Schr_eq}). Suppose we want to get energies of $n$ states
with an electron removed from one of the $n$ bonds of a molecule
and also the hopping matrix elements between these bonds. As bonds
are spatial orbitals containing two electrons with opposite spins
per bond in the ground state, two spin-degenerate states are
possible for a hole in some particular bond. To resolve the
problem we consider only states of a particular spin symmetry,
e.g. only one particle configurations obtained by annihilation of
a {\it spin-up} electron. Then we have to provide $n$
corresponding spin-adapted configurations $\Phi_i$ ($i=$1, 2,
\ldots, $n$) into the model space. They define an $n \times n$
matrix $H_{ij} = \langle \Phi_i | H | \Phi_j \rangle$ on the HF
level. The model wavefunctions $\Phi_j$ are first transformed to
diagonalize this matrix:

\begin{equation}
\label{Psi_i} \Psi_i = \sum_{j=1}^n (u^{-1})_{ji} \: \Phi_j =
\sum_{j=1}^n u_{ij} \: \Phi_j, \qquad \langle \Psi_i | H |\Psi_j
\rangle = E_i \: \delta_{ij}.
\end{equation}
The configurations $\Psi_i$ are called {\it reference
configurations} or {\it reference states} and $E_i$ are the
energies of these states. In our calculations we get the SCF
energies $E_i$ and coefficients $u_{ij}$ from a preliminary step
of the MRCI calculations. Then the SCF matrix elements are
obtained as

\begin{equation}
\label{H_ij_SCF} H_{ij} = \sum_{k=1}^n u_{ki} u_{kj} E_k
\end{equation}
in close analogy to Eq.~(\ref{Heff_ij}) for 
$H^{\rm eff}_{ij}$.

To produce excited configurations for the CI expansion of the
correlated wavefunctions $\Psi_i^{\rm corr}$, pair excitation
operators $c_a c_r^\dagger$ and $c_a c_b c_r^\dagger c_s^\dagger$
are applied to each of the reference states $\Psi_i$, yielding the
following ansatz for the $n$ correlated wavefunctions:

\begin{equation}
\label{Psi_i^corr1} \Psi_i^{\rm corr} = \sum_{j=1}^n \biggl(
t_j(i) + {\sum_{a,r}} ^\prime t_{ja}^{\; \; r}(i) c_a c_r^\dagger
+ {\sum_{a,b,r,s}}^{\! \! \prime} t_{jab}^{\; \; rs}(i) c_a c_b
c_r^\dagger c_s^\dagger \biggr) \Psi_j
\end{equation}
which (in case of an ($N-1$)-particle system) is equivalent to

\begin{equation}
\label{Psi_i^corr} \Psi_i^{\rm corr} = \sum_{j=1}^n  \biggl(
\alpha_j(i) \Phi_j + {\sum_{a,r}} ^\prime \alpha_{ja}^{\; \; r}(i)
\Phi_{ja}^{\; \; r} +{\sum_{a,b,r,s}}^{\! \! \prime}
\alpha_{jab}^{\; \; rs}(i) \Phi_{jab}^{\; \; rs} \biggr)
\end{equation}
since the $n$ reference states $\Psi_j$ and the $n$ spin-adapted
configurations $\Phi_j$ span the same (reference) space. In
Eq.~(\ref{Psi_i^corr}) prime over sums
means that electrons are annihilated from the $n$ active bonds and
only the remaining spin-down electron can be removed from the
localized orbital $j$. These $n$ eigenfunctions (\ref{Psi_i^corr})
correspond to $n$ eigenvalues of the full Hamiltonian:

\begin{equation}
\label{Scr_eq_corrMRCI} H |\Psi_i^{\rm corr} \rangle = E_i^{\rm
corr} |\Psi_i^{\rm corr} \rangle.
\end{equation}
The amplitudes $t_j(i)$, $t_{ja}^{\; \; r}(i)$ and $t_{jab}^{\; \;
rs}(i)$ or equivalently the coefficients $\alpha_j(i)$,
$\alpha_{ja}^{\; \; r}(i)$ and $\alpha_{jab}^{\; \; rs}(i)$ are
determined iteratively from a set of equations obtained by
multiplying both sides of the $n$ equations
(\ref{Scr_eq_corrMRCI}) successively by $\langle \Phi_j |$,
$\langle \Phi_{jc}^{\; \; t} |$ and $\langle \Phi_{jcd}^{\; \; tu}
|$ from the left. For details we refer to \cite{Werner88}.

Once the iterations are converged the $n$ correlated energies of
($N-1$)-electron system $E_i^{\rm corr}$ can be evaluated. Being
used in (\ref{H_ij_SCF}) instead of the HF energies $E_k$ they
determine the matrix elements of the effective Hamiltonian
$H_{ij}^{\rm eff}$ as have been discussed in the previous section
(see Eq.~(\ref{Heff_ij})).

In the case of attached-electron states the wavefunctions $\Psi_j$
in Eqs.~(\ref{Psi_i^corr1}) correspond to the $m$ ($N+1$)-particle
states given by linear combinations (\ref{Psi_i}) of the local
($N+1$)-particle configurations $\Phi^l$ with an extra (spin-down) 
electron being added to one of the $m$ localized
virtual orbitals. The MRCI-ansatz for the $m$ correlated
wavefunctions is written in a form similar to
Eq.~(\ref{Psi_i^corr}):

\begin{equation}
\label{Psi_k^corr} \Psi_k^{\rm corr} = \sum_{l=1}^m  \biggl(
\alpha^l(k) \Phi^l + {\sum_{a,r}} ^\prime \alpha_{\; a}^{lr}(k)
\Phi_{\; a}^{lr} +{\sum_{a,b,r,s}}^{\! \! \prime} \alpha_{\; \;
ab}^{lrs}(k) \Phi_{\; ab}^{lrs} \biggr)
\end{equation}
and here primes mean that only spin-up electron can be put into a
localized virtual orbital $\varphi_l$ if it is already occupied by the
attached spin-down electron. The obtained $m$ correlated energies
of the ($N+1$)-electron system $E_k^{\rm corr}$ can be used in
Eq.~(\ref{H_ij_SCF}) instead of the corresponding $m$ HF energies
to calculate the matrix elements of the effective Hamiltonian
$H_{ij}^{\rm eff}$ for the case of attached-electron states.

Thus, equations (\ref{Psi_i})--(\ref{Psi_k^corr}) define how the
MRCI(SD) method is used to evaluate the matrix elements of the
effective Hamiltonian in the case of both ($N-1$)- and
($N+1$)-particle states.

\chapter{Realization of the method}

While we have outlined the general idea of our approach
to the correlated band structures in the previous chapter
we describe in detail the {\it actual} realization of this
method here. The concepts we have developed to make such
calculations possible are presented in the given chapter.
We will refer to {\it trans}-polyacetylene (tPA) single chains
as a test system. In
section 3.1 the geometry and electronic properties of tPA on the
HF level are described. In section 3.2 we show how to generate
suitable localized occupied and virtual HF orbitals in the
clusters. The corresponding HF local matrix elements are used to
calculate the SCF band structure of tPA. In section 3.3 we discuss
how to evaluate the correlation corrections to the LMEs by the
MRCI(SD) method and the use of an approximate but well controlled
scheme for this is presented in section 3.4. The possibility to
reduce the size-extensivity error of the MRCI(SD) method is
explored in section 3.5 and the analytic formula for the
corresponding correction to the MRCI(SD) correlation energy is
derived for open-shell systems. The occurrence of satellite states
in cluster models and the handling of this problem by an
appropriate construction of the model space in the MRCI
calculation are discussed in section 3.6.

\section{The investigated system}

In Chapter 2 we have described the general idea of our
wavefunction-based method for obtaining band structures of
insulators and semiconductors with electron correlation effects
being properly included. This approach was used in
Ref.~\cite{Graef97} and \cite{Albrecht00} to get the correlated
valence bands of three structurally similar semiconductors
(diamond, silicon and germanium). In the present work we make the
first attempt to treat systematically not only the valence but
also the conduction bands of a periodic system. For this purpose
we introduce {\it localized} virtual HF orbitals which are, in
contrast to localized occupied orbitals, usually extended over
several unit cells of the crystal. Therefore, rather big clusters
have to be used for a proper treatment of the ($N+1$)-electron
states. Moreover, even larger clusters are needed to investigate
the decay of the correlation effects with the distance from the
localized extra charge. On the other hand, since rich basis sets
are required to describe the anionic ($N+1$)-electron states
properly and a powerful but rather expensive correlation method is
employed, we reach the limits of our computational abilities when
the size of the employed clusters reaches 30 atoms. In a bulk
crystal a virtual WO extend in three dimensions and the number of
atoms covered by the localized virtual orbital scales as $r^3$
with the size of the WO whereas in one-dimensional (1D) chains it
only scales as $r^1$. Thus, for our test calculations, a linear
system has been chosen.

\subsection{Polyacetylene}

\begin{figure}
\centerline{
\psfig{figure=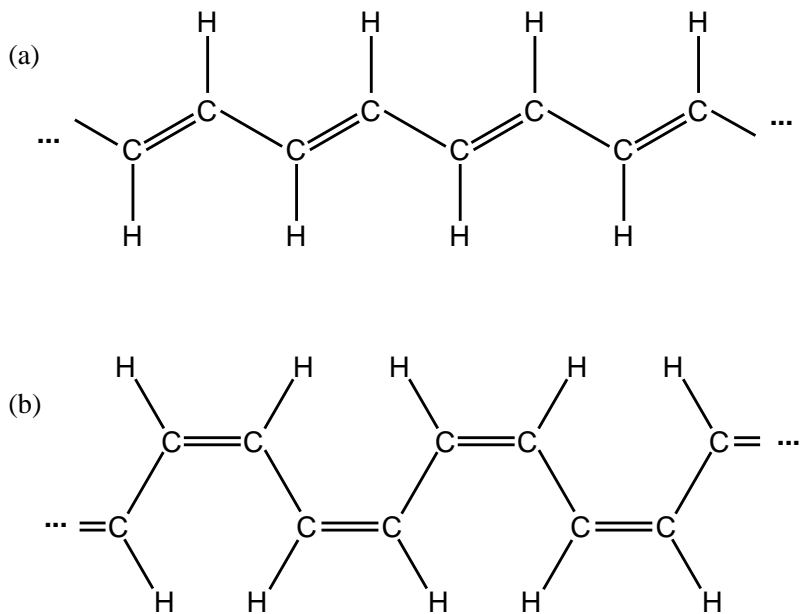,height=8cm}
}
\caption{Two conformations of polyacetylene: (a) {\it trans}-PA and
(b) {\it cis}-PA.}
\label{conform}
\end{figure}

A natural examples of 1D periodic systems are infinitely long
linear polymers. One of the simplest among them is {\bf
polyacetylene} (PA). It is a plane periodic molecule which
consists of repeating C$_2$H$_2$ units and exists in two basic
conformations (Fig.~\ref{conform}). The first one exhibits a
zigzag carbon chain with alternating single (or long) and double
(or short) C--C bonds; the C--H bonds are pointing outwards this
chain nearly perpendicular to the molecular axis. This is {\it
trans}-polyacetylene which is schematically shown on
Fig.~\ref{conform}a. The second one is the {\it cis} conformation
and one of its realization is drawn in Fig.~\ref{conform}b
(another realization can be obtained by the interchange of single
and double C--C bonds).

When a polyacetylene sample is synthesized it consists of finite
molecules (oligomers) of both conformations (at room temperature).
The {\it trans} conformation is thermodynamically more stable and
the isomerisation of the sample is achieved by heating it to
temperatures above 150~$^\circ$C for a few minutes \cite{SI79/80}.
The {\it cis}--{\it trans} conversion of polyacetylene was
rigorously studied in Ref.~\cite{Robin83} where the sample was
annealed at 100~$^\circ$C for more than 1 day. After 3~h it
contained more than 50{\%} of {\it trans} species and more than
80{\%} after 1 day. The isomerization happened continuously during
this time and homogeneously throughout the polymer rather than in
isolated amorphous or crystalline regions. The crystallinity of
these samples is estimated to about 90{\%} and the amorphous
content did not increase with isomerization.

There are two methods for the synthesis of crystalline
polyacetylene. In the work of Sirakakawa and Ikeda \cite{SI79/80}
polyacetylene was grown in the form of randomly oriented fibrils.
Inside a particular fibril the single chains
of polyacetylene are directed along the axis of the fibril. In
order to obtain a material with parallelly oriented chains of
polyacetylene for further crystallographic and electronic
structure studies the grown polymer is stretched along some
direction. The corresponding electron micrographs can be found in
\cite{SI79/80}. The
characteristic diameter of fibrils obtained by this method ranges
from 200~{\AA} to 500~{\AA}. This technique was exploited in a
number of experimental works on polyacetylene, e.g.
\cite{Fincher79}, \cite{Tani80}, \cite{Fincher82}, \cite{Robin83},
\cite{Yannoni83} and \cite{Loendlund93}. The disadvantages of
samples produced this way are not perfectly aligned polyacetylene
chains and even the presence of a certain amount of randomly
oriented chains, and the fact that it is impossible to investigate
the electronic properties of tPA in the direction perpendicular to
the stretching due to the fibrillar morphology of the material.

The other method to produce crystalline tPA was developed at
Durham University (\cite{Edwards80} and \cite{Edwards84}) and
provides true bulk polyacetylene. The synthesis utilizes a
solution of a precursor polymer which is converted into
polyacetylene. Under usual conditions, both conformations are
present and the obtained polymer is amorphous. However, one can
get samples with perfectly oriented chains of {\it
trans}-polyacetylene by applying an appropriate stress at
temperatures up to 120~$^\circ$C. Then, during the conversion, an
elongation of the polymer film and simultaneous alignment of tPA
chains take place. As the result highly anisotropic free-standing
films of 100\% {\it trans}-polyacetylene are obtained. This
material exhibits a compact, non-fibrous morphology and smooth
surfaces appropriate for further optical study (\cite{Fink86},
\cite{Kahlert87}, \cite{Bradley87}, \cite{Uitz87} and
\cite{Leising88}).

\begin{figure}
\centerline{ \psfig{figure=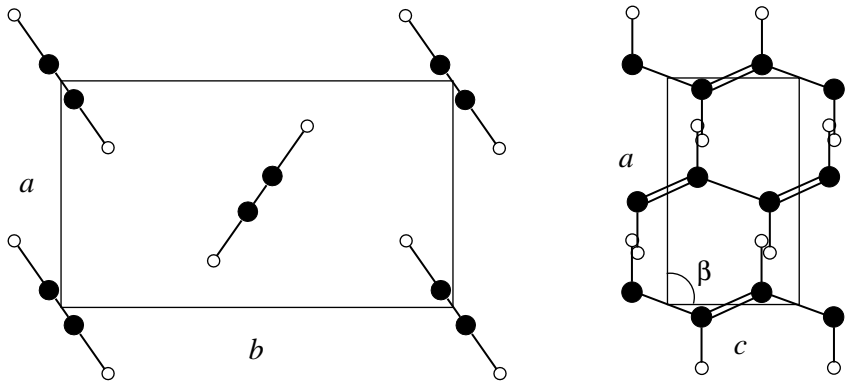,height=5cm} }
\caption{Two schematic projections of bulk {\it
trans}-polyacetylene. $a = 4.24 \; {\rm {\AA}}$, $b = 7.32 \; {\rm
{\AA}}$ and $c = 2.46 \; {\rm {\AA}}$ and $\beta = 91.5^\circ$
(from \cite{Fincher82}).} \label{fishbone}
\end{figure}

Thus, when we refer to {\it trans}-polyacetylene as a bulk polymer
we mean a crystal built by densely packed, flat and infinitely
long tPA single chains all aligned in one direction. The
crystalline structure of bulk tPA was studied in \cite{Fincher82}
by x-ray scattering experiments and a fishbone-like packing of
independent tPA chains was determined (Fig.~\ref{fishbone}). The
space group of tPA crystals is P2$_1$/n with a C$_4$H$_4$ unit
cell consisting of two C$_2$H$_2$ units on neighboring tPA single
chains. The parameters of the monoclinic structure are determined
with a unique angle $\beta = 91.5^\circ$ and lattice constants $a
= 4.24 \; {\rm {\AA}}$, $b = 7.32 \; {\rm {\AA}}$ and $c = 2.46 \;
{\rm {\AA}}$. The angle between a chain plane and the $b$ axis is
55$^\circ$.

As one can see in Fig.~\ref{fishbone}, for any atom of the tPA
crystal the first few closest atoms belong to the same tPA chain.
Therefore, the crystal consists of weakly coupled chains with
strong intrachain bonding. This is evident from the high
anisotropy of the electronic properties of bulk tPA: SCF
calculations give band widths between 5 and 10 eV for
crystal-momentum vectors $\bf k$ parallel to the chains and less
than 0.4 eV in perpendicular directions. 
These findings tell that the electrons in bulk tPA
are delocalized exclusively along the polymer chain and thus each
tPA single chain can indeed be considered as a
quasi-one-dimensional crystal. Therefore, one can focus on the
study of single 1D tPA chains to describe the electronic
properties of {\it trans}-polyacetylene neglecting for a moment
the weak interchain interactions.

\subsection{Geometry of tPA single chain}

\begin{figure}
\centerline{ \psfig{figure=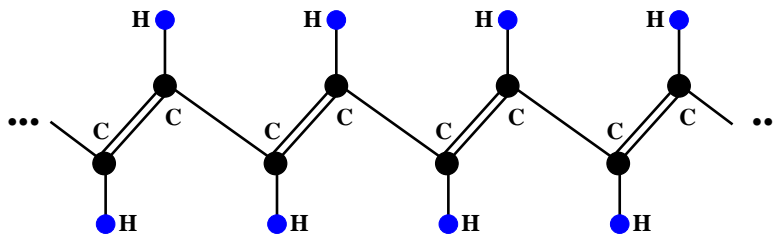,height=3cm} }
\caption{Investigated system: {\it trans}-polyacetylene single
chains.} \label{tPA}
\end{figure}

A single tPA chain consists of repeating C$_2$H$_2$ units with
alternating short and long C--C bonds (Fig.~\ref{tPA}). The
structure is produced by hybridization of the valence $sp^2$
electrons of carbon and the $s^1$ electron of hydrogen. Three out
of four valence carbon electrons take part in this formation of
$\sigma$ bonds (with charge density maximum in the molecule
plane), namely one long C--C bond, one (out of two) short C--C
bond and one C--H bond. The fourth electron occupies the $2p_z$
atomic orbital, whose lobes lie out of the plane, and together
with the $2p_z$ electron of the closest neighbor carbon atom form
the second short C--C bond which is of $\pi$ symmetry. The
alternation in the C--C bond lengths emerges from the Peierls
instability of 1D metals \cite{Peierls}. If there would be no bond
alternation (sometimes also called "dimerization") and all C--C
bond were equal the unit cell would consist of just one CH unit
with one unpaired electron on the $2p_z$ orbital per cell. In this
case tPA would be a metal with a half-filled conduction band as
can be seen from simple tight-binding-model considerations.
However, such a 1D metal can reduce its energy by spontaneous
breaking of the translational symmetry by moving two adjacent CH
groups towards each other and forming by this the short (or
double) C--C bond. By this the system gains electronic energy
whereas its elastic energy (i.e. the energy of the carbon
skeleton) increases. The minimum of the system energy as a
function of dimerization length defines the equilibrium C--C bond
lengths in a (C$_2$H$_2$)$_x$ chain. The reduction of translation
symmetry opens the band gap and tPA becomes a semiconductor with a
completely filled upper valence band and with a band gap of the
order of few electron-volts. The qualitative analysis of
dimerization effect in {\it trans}-polyacetylene is
comprehensively expounded in \cite{Fulde}, p.~179--183, where two
original papers \cite{Su79} and \cite{Su80} are summarized.

To perform single-point {\it ab initio} calculations we need to
know the positions of the atoms in the tPA chain. We take the
relevant experimental data on the structure of the carbon skeleton
of {\it trans}-polyacetylene from \cite{Kahlert87}: the long C--C
bond length is $r_1 = 1.45 \; {\rm {\AA}}$, the short C--C bond
length is $r_2 = 1.36 \; {\rm {\AA}}$ and the lattice constant is
$a = 2.455 \; {\rm {\AA}}$. The angle $\gamma$ between short and
long bonds is measured to be 122$^\circ$. The measured angle does
not perfectly agree with measured lengths and we set this angle to
121.7421$^\circ$ (which is well within the experimental error
bar).

The experimental data on the length and orientation of C--H bond
of tPA are not available in the literature. Therefore, we have
taken the C--H bond length from the paper \cite{Suhai92} where the
geometry of tPA was optimized on both HF and MP2 level using
different basis sets. The results for C--C bond lengths obtained
on the MP2 level with the largest basis set (of double-zeta
quality with polarization functions) agreed nicely with the
experimental data mentioned above ($r_1 = 1.4450 \; {\rm {\AA}}$
and $r_2 = 1.3634 \; {\rm {\AA}}$) however the angle between C--C
bonds was somewhat different ($\gamma = 123.75^\circ$). Therefore
from \cite{Suhai92} we have taken only the value of the C--H bond
length which is $r_3 = 1.0872 \; {\rm {\AA}}$.

The last undefined parameter of the geometry of tPA was the angle
$\vartheta$ between short C--C bond and C--H bond. To determine
this angle we used the geometry optimization option of the
GAUSSIAN98 program package \cite{GAUSSIAN}. We have optimized the
angle $\vartheta$ in a hydrogen-terminated C$_{12}$H$_{14}$
cluster of {\it trans}-polyacetylene taking $r_1 = 1.45 \; {\rm
{\AA}}$, $r_2 = 1.36 \; {\rm {\AA}}$, $r_3 = 1.0872 \; {\rm
{\AA}}$ and $\gamma = 121.7421^\circ$ as fixed parameters. The
angle between the two cluster-saturating C--H bonds and adjacent
short C--C bond was set to $\gamma$ and kept fixed as well. Such
an arrangement allows to reduce the influence of cluster edges on
the optimized angle $\vartheta$ in the center of the cluster where
its value is maximally close to that of the infinite chain. As
follows from \cite{Suhai92}, the value of the optimized angle
$\vartheta$ does almost not depend on the method (HF or MP2) and
the basis sets employed. Thus, we performed the optimization on
the HF level with cc-pVDZ basis sets for both hydrogen and carbon
atoms \cite{Dunning89}. We have tried several starting values $\vartheta_0$ in the
range from 118$^\circ$ to 122$^\circ$ and in all cases the
optimized angle $\vartheta$ for the chosen fixed parameters was
found to be 120.057$^\circ$.

To summarize, our test system is a plane infinitely long single
chain of {\it trans}-polyacetylene with alternating C--C bonds
(Fig.~\ref{tPA}). There are four $\sigma$ bonds and one $\pi$ bond
(on the short C--C bond) per unit cell C$_2$H$_2$. The lengths of
long C--C, short C--C and C--H bonds are 1.45~{\AA}, 1.36~{\AA}
and 1.087~{\AA} respectively. The angle between two neighbor C--C
bonds is 121.742$^\circ$ and the angle between C--H bond and
adjacent short C--C bond is 120.057$^\circ$.

\subsection{Basis sets}

Throughout all our further HF and correlation calculations we use valence
triple zeta (VTZ) atomic basis sets for carbon and hydrogen atoms
\cite{Dunning89} (see Table~\ref{basis}). These Gaussian-type-orbital
(GTO) basis sets are correlation consistent, have rather diffuse
exponents to describe properly the decay of electron wavefunctions
in the direction perpendicular to the chain, and are rich enough
to provide reasonable results for correlation calculations in case 
of the anionic ($N+1$)-electron states.

\begin{table}
\refstepcounter{table} \addtocounter{table}{-1} \label{basis}
\caption{Exponents and contraction coefficients of the cc-pVTZ
basis sets for carbon and hydrogen atoms used in all our
calculations.} \vspace{5mm}
\renewcommand{\baselinestretch}{1.2}\normalsize
\begin{tabular}{ccrrrcc}
\hline
atom & type & exponent & \multicolumn{4}{c}{contraction coefficients} \\
     &      &          & \;\;\; shell 1 \; & \;\;\; shell 2 \;
& \;\; shell 3 \;\; & \;\; shell 4 \;\; \\
\hline
C & & & \\
& $s$ & 8236.0000   & 0.000531 & -0.000878 &     &     \\
& &  1235.0000      & 0.004108 & -0.000878 &     &     \\
& &   280.8000      & 0.021087 & -0.004540 &     &     \\
& &    79.2700      & 0.081853 & -0.018133 &     &     \\
& &    25.5900      & 0.234817 & -0.055760 &     &     \\
& &     8.9970      & 0.434401 & -0.126895 &     &     \\
& &     3.3190      & 0.346129 & -0.170352 &     &     \\
& &     0.9059      & 0.039378 &  0.140382 & 1.0 &     \\
& &     0.3643     & -0.008983 &  0.598684 &     &     \\
& &     0.1285      & 0.002385 &  0.395389 &     & 1.0 \\
& $p$ & 18.7100     &          &  0.014031 &     &     \\
& &     4.1330      &          &  0.086866 &     &     \\
& &     1.2000      &          &  0.290216 &     &     \\
& &     0.3827      &          &  0.501008 & 1.0 &     \\
& &     0.1209      &          &  0.343406 &     & 1.0 \\
& $d$ & 1.0970      &          &           & 1.0 &     \\
& &     0.3180      &          &           &     & 1.0 \\
H & & & \\
& $s$ &   33.870000 & 0.006068 &          &     \\
& &        5.095000 & 0.045308 &          &     \\
& &        1.159000 & 0.202822 &          &     \\
& &        0.325800 & 0.503903 & 1.0 \;\; &     \\
& &        0.102700 & 0.383421 &          & 1.0 \\
& $p$ &    1.407000 &          & 1.0 \;\; &     \\
&     &    0.388000 &          &          & 1.0 \\
\hline
\end{tabular}
\end{table}
\renewcommand{\baselinestretch}{1}\normalsize

The Gaussian-type orbitals (or more precisely Cartesian Gaussians)
which were first proposed as an approximation to atomic orbitals
in \cite{Boys50} have the form

\begin{equation}
\label{GTO}
\chi_i ({\bf r}) = N_i x^l y^m z^n {\rm e}^{-\zeta_i r^2}
\end{equation}
where $N_i$ a normalizing prefactor and $l$, $m$ and $n$ are
integer numbers which are all zero for $s$-type orbitals, one of
them is equal to 1 such that they sum up to 1 for $p$-type
orbitals (e.g. for 2$p_x$, 3$p_x$, 4$p_x$ {\it etc.} one has $l=1$
and $m=n=0$), they sum up to two (e.g. 110 or 200) for $d$-type
orbitals and so on. To approximate better real atomic orbitals
fixed linear combinations of Gaussians are used:

\begin{equation}
\label{contractions} \chi_i ({\bf r}) = N_i x^l y^m z^n \biggl(
\sum_{j=1}^{L_i} c_{ij} {\rm e}^{-\zeta_{ij} r^2} \biggr)
\end{equation}
where $c_{ij}$ are the so-called contraction coefficients,
$\zeta_{ij}$ are the exponents of the so-called primitive
Gaussians and $L_i$ is the length of the contraction of the $i$-th
atomic-like orbital. The exponents and contraction coefficients of
the employed cc-pVTZ basis sets are listed in 
Table~\ref{basis}.

\subsection{Band structure of tPA chain on the HF level}

Now, that the geometry of our test system and the basis sets are
assessed, we can start the investigation. At first, we would like
to calculate the periodic HF solution for an infinite tPA single
chain, mean the ground-state energy (per unit cell), Bloch
orbitals and the corresponding orbital energies (the band
structure). This can be done by the program package CRYSTAL
\cite{CRYSTAL} which is able to perform HF calculations for
periodic systems and uses Gaussian-type orbitals. However, due to
the presence of rather diffuse Gaussians in the chosen basis set
(with exponents less than 0.2) one faces the problem of
"catastrophic" SCF results (non-converging SCF iterations) when
default values of the ITOL parameters which control the accuracy
of the calculation of the bielectronic Coulomb and exchange series
(see \cite{CRYSTAL}, p.~74 and 123-125) are used. The effect on
these parameters on the convergence of SCF calculations in CRYSTAL
is discussed in details in \cite{Pisani88}, p.~79-85 and here we
would only like to notice that in the case of {\it
trans}-polyacetylene single chains and the chosen basis sets
well-converged results are obtained with the ITOL parameters being
set to 8 8 8 14 24 (the defaults are 6 6 6 6 12). The SCF
ground-state energy per unit cell is -3.53887 eV and the iteration
is converged when the energy changes by less than 0.14 meV (5
$\mu$Hartree) from one iteration step to the next.

\begin{figure}
\centerline{
\psfig{figure=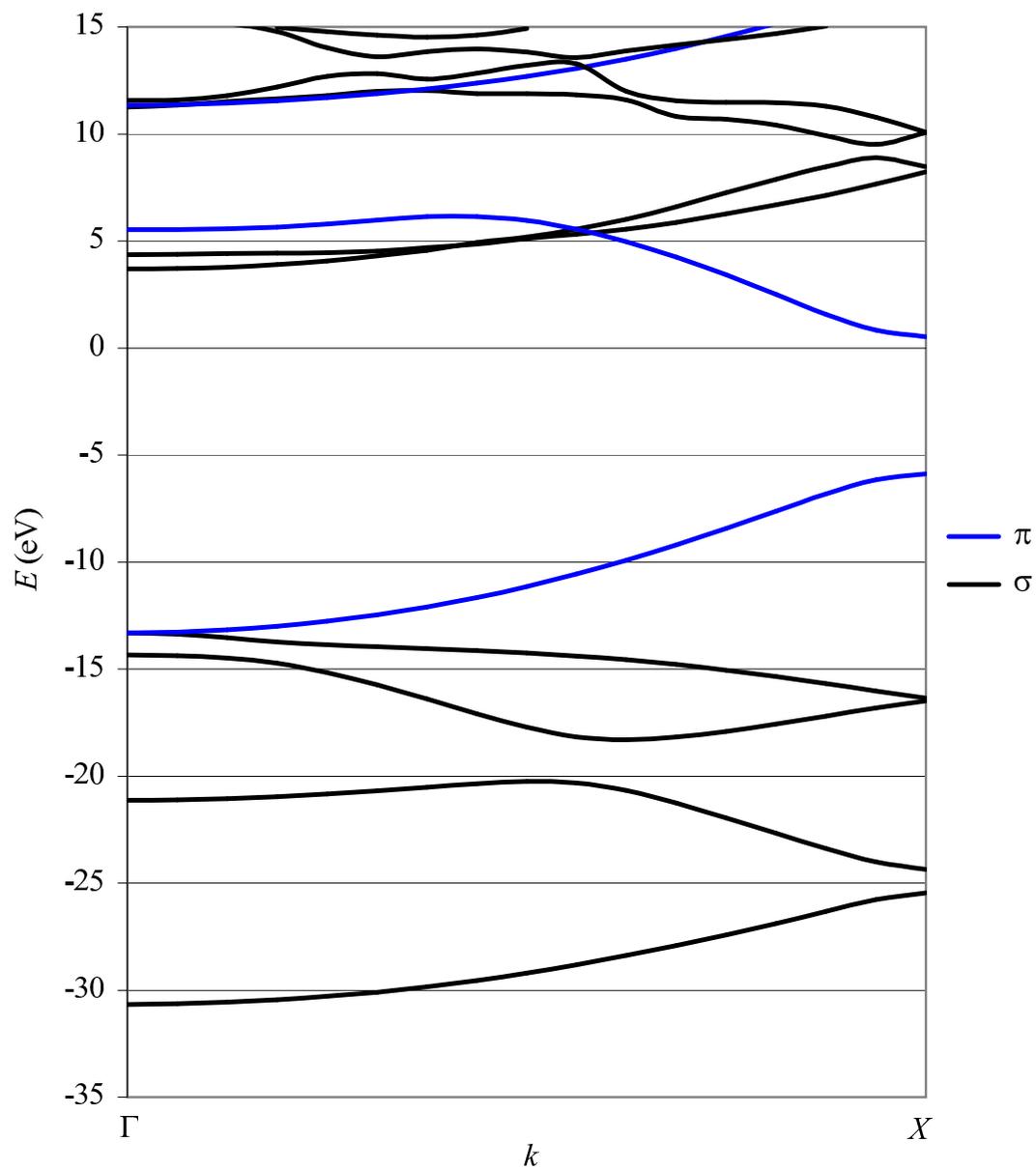,height=16cm}
}
\caption{The HF band structure of {\it trans}-polyacetylene.}
\label{SCFband}
\end{figure}

The resulting orbital energies as a function of the
one-dimensional crystal-momentum $k$ (the band structure) are
shown in Fig.~\ref{SCFband}. Blue lines correspond to Bloch states
of $\pi$ symmetry (single-electron wavefunction vanishes in the
molecule plane) and the rest are of $\sigma$ symmetry. Because of
the different symmetry blue and black lines may cross.

From Fig.~\ref{SCFband} one concludes that the ionization
potential and the electron affinity of an infinite tPA chain are
given by the top of $\pi$-type valence band and the bottom of
$\pi$-type conduction band in the $X$ point, respectively:

\begin{equation}
\label{IPtPA}
{\rm IP}= -\varepsilon_{\pi ({\rm val})}(k=\pi / a),
\end{equation}

\begin{equation}
\label{EAtPA}
{\rm EA}= -\varepsilon_{\pi ({\rm cond})}(k=\pi / a)
\end{equation}
On the HF level these quantities are IP$=5.90$~eV and
EA$=-0.52$~eV. The band gap, defined as the difference of these
two values, is equal to 6.42~eV. However, from a number of
experimental studies, {\it trans}-polyacetylene is known to have a
positive electron affinity \cite{Kaner89} and a band gap of
approximately 2~eV \cite{Fincher79}, \cite{Tani80},
\cite{Leising88}. Therefore, {\it trans}-polyacetylene is one of
many systems for which the HF approximation fails to give both
quantitatively and qualitatively correct results. By inclusion of
electron correlation effects, following the scheme sketched in
Chapter 2, we will approach the experimental data quite
noticeably.

In our correlation calculations we focus our attention on the five
valence bands and the three lowest conduction bands. Taking into
account general considerations \cite{Fulde} and earlier experience
of the application of wavefunction-based correlation methods to
covalent periodic systems \cite{Borrmann87}, \cite{Graef97} and
\cite{Albrecht00} we expect that the correlation effect will lead
to an upward shift of the HF valence bands and a downward shift of
the conduction bands and also to a flattening of the bands.

\section{Localized orbitals}

To set up the finite molecules for our correlation calculations
we have chosen the following hydrogen-saturated
clusters of {\it trans}-polyacetylene single chain:
C$_{6}$H$_{8}$, C$_{8}$H$_{10}$, C$_{10}$H$_{12}$ and
C$_{12}$H$_{14}$. All these clusters are terminated at single
C--C bonds on both edges.  There dangling C--C bonds are substituted by C--H
bonds having the same angle $\gamma$ with the adjacent double C--C bond
and a characteristic C--H bond length of 1.087~{\AA}. Such a termination on
long C--C bond exclusively preserves the flat geometry of clusters
and also minimizes the spurital impact of the saturating hydrogen
atoms on the cojugated $\pi$ system of tPA. Thus, we operate with
closed-shell molecules which perfectly reproduce some finite part of
our infinite system except for the two terminating C--H bonds.
Therefore we expect that the localized single-particle wavefunctions
(or molecular orbitals, LMOs) in these molecules
are similar to those in the infinite system: in particular, in 
the central part of the molecule where an LMO
has the same surrounding as in the infinite system they should
coincides with the localized crystalline (Wannier) orbital (WO). 
On the edges of the cluster on one side there is only a terminating
C--H bond which mimics the long C--C bond (and the remainder of the
infinite chain). As LMOs centered on some bond always also have some
small, but yet not negligible contribution of atomic orbitals of
atoms in the neighborhood of the bond, WOs can not be reproduced
correctly at the cluster edges. Thus, LMOs from edges of clusters 
will exhibit some deviations from the corresponding WOs (or 
equivalently from the corresponding LMOs in the central part of 
clusters). In accurate correlation calculations these edge LMOs 
are usually kept frozen or if not, the obtained results should be 
regarded as rough estimates only.

\subsection{The case of occupied orbitals}

To generate localized occupied spatial SCF orbitals (i.e. the bonding
orbitals) on the clusters we use the Foster-Boys localization scheme 
\cite{Boys} which is implemented in MOLPRO \cite{MOLPRO}. The idea 
of this scheme is to find that unitary matrix $u_{ij}$ in 
Eq.~(\ref{locorb}) which minimizes the sum of the spread of 
$N_{\rm orb}$ pre-selected spatial orbitals:

\begin{equation}
\label{FBfunctional} I = \sum^{N_{\rm orb}}_{i=1} \sigma^2_i
\end{equation}
where the orbital spread is defined as

\begin{equation}
\label{orb_spread}
\sigma^2_i = \int ({\bf r} - {\bf R}_i)^2
|\varphi_i({\bf r})|^2 {\rm d}V,
\end{equation}
and the center of $i$-th localized orbital $\varphi_i$ as

\begin{equation}
\label{el_position}
{\bf R}_i = \int {\bf r} |\varphi_i({\bf
r})|^2 {\rm d}V
\end{equation}

The question arises which canonical orbitals we have to
provide for the localization to get properly localized occupied 
LMOs, which correspond to the individual bonds. We have count the 
valence electrons in the molecule (i.e. the electrons on the
outer shells of each atoms which take part in the formation of bonds)
and half of this number defines the number of the
occupied canonical orbitals to be localized starting with the HOMO
(highest occupied molecular orbital) and continuing with the
successive energetically highest orbitals.
All the other occupied orbitals are core orbitals which do not
participate in bonding. In our case they are linear combinations
of 1$s$ orbitals of carbon atoms.

A special remark has to be made for flat molecules where orbitals
of two different symmetries ($\sigma$  and $\pi$) exist.
Canonical orbitals of $\pi$ type can be readily found in the set
of SCF orbitals of a flat molecule as they have zero value in the
molecule plane. Then the canonical $\sigma$ and $\pi$ orbitals can be
localized separately and LMOs of $\sigma$
and $\pi$ symmetries are obtained. By doing so one reduces the
number of LMEs to be treated on the correlation level as off-diagonal
LMEs between $\sigma$ and $\pi$ orbitals are zeros and the effort for
the correlation calculations, which are the most difficult and time
consuming part of the whole approach, is reduced substentially.

\begin{figure}
\centerline{ \psfig{figure=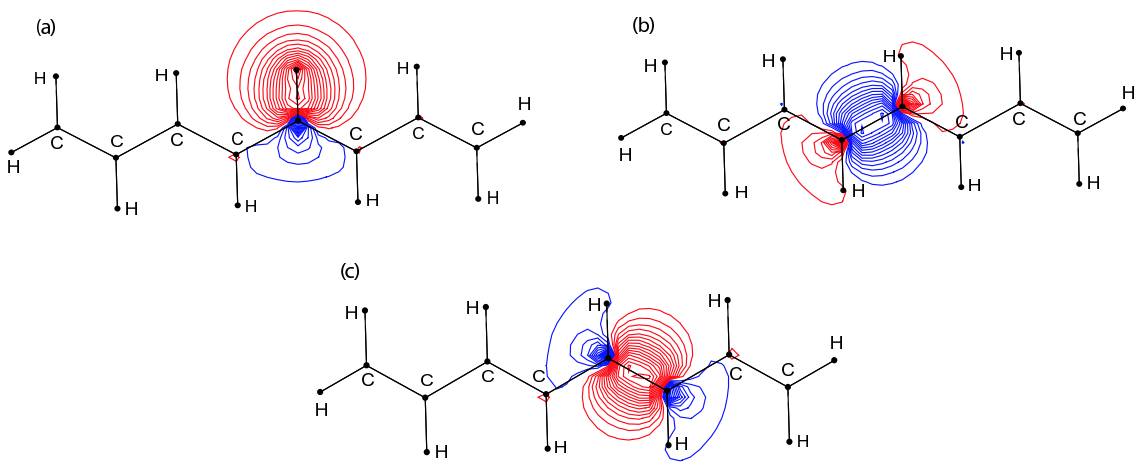,width=5.9in} }
\caption{Three different $\sigma$ bonds in {\it
trans}-polyacetylene clusters: (a) C--H bond, (b) long C--C bond
and (c) short C--C bond. The contour values increase in steps
of 0.025~a.u.}
\label{occ_sig}
\end{figure}

In Fig.~\ref{occ_sig} contour plots of localized
one-particle wavefunctions $\varphi_a({\bf r})$ for three
different $\sigma$ bonds of a C$_{8}$H$_{10}$ cluster are shown: (a)
C--H bond, (b) long C--C bond and (c) short C--C bond. Red color
denotes positive values of $\varphi_a({\bf r})$ and blue color
denotes negative ones.

\begin{figure}
\centerline{ \psfig{figure=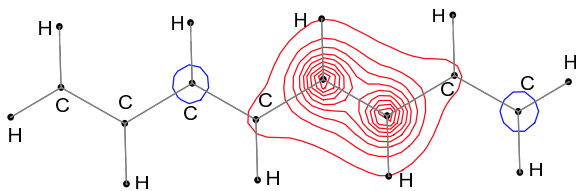,width=3in} }
\caption{One of the $\pi$ bonds of a C$_{8}$H$_{10}$ cluster of tPA.
Contours are in steps of 0.025 a.u. The plot
plane is parallel to the molecule plane at a hight of
$z=0.5 \; {\rm {\AA}}$.}
\label{occ_pi}
\end{figure}

In Fig.~\ref{occ_pi} a typical localized $\pi$ orbital is shown. As it
vanishes in the molecule plane ($\varphi({\bf r})_{|z=0}=0$) the
plot plane for the contour plot is shifted to $z=0.5 \;
{\rm {\AA}}$. All plots in Figs.~\ref{occ_sig} and
\ref{occ_pi} are obtained by the MOLDEN program package \cite{MOLDEN}.

Comparing Fig. \ref{occ_sig} and \ref{occ_pi} we see that the 
$\pi$-bonds are more diffuse than the $\sigma$-bonds. The former 
have also a quite noticeable contribution of atomic basis functions
from carbon atoms in the two nearest-neighbor unit cells. Therefore,
for an accurate treatment of the correlation corrections to LMEs
corresponding to $\pi$ bonds substantially larger clusters are needed 
than for the case of $\sigma$ bonds.

Having obtained the occupied LMOs we calculate the local SCF matrix
elements ${\rm IP}_{ab}$ as defined in (\ref{IPab2}) and identify
them with the corresponding crystalline local matrix elements 
IP$_{R,nn^\prime}$ which enter the truncated sum (\ref{IPn}). The
hopping matrix elements decay with the distance between two WOs.
Setting some threshold for the absolute values of LMEs
in (\ref{IPn}) we define the two most distant WOs and by this the 
minimal size of the cluster in which the LMEs are calculated.
Also, for choosing the proper cluster one takes into account that
the LMOs on the edges of clusters can not reproduce the corresponding 
WOs correctly. Thus, the size of the cluster on which the LMEs are
calculated is given by the two most distant WOs which give
hopping matrix elements above the threshold plus one additional unit
cell on both edges of the cluster. In the case of tPA the threshold
is set to 1 mHartree (0.027 eV), and the minimal-sized saturated cluster
arising from that is a C$_{10}$H$_{12}$ molecule.

\begin{figure}
\centerline{ \psfig{figure=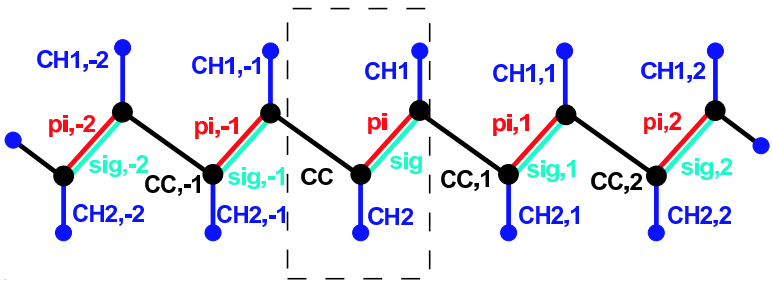,width=4in} }
\caption{C$_{10}$H$_{12}$ cluster of tPA with the names given to each bond.
The reference (central) unit cell is indicated by the dashed
frame.} 
\label{c10h12}
\end{figure}

This cluster is schematically shown in Fig.~\ref{c10h12}. For
convenience we give names to the different bonds in the unit cell:
on the double C--C bond we distinguish bonds of $\pi$ and $\sigma$
type and call them "pi" and "sig", respectively, the single C--C
bond is called "CC" and the two C--H bonds are labelled "CH1" and
"CH2". In the central unit cell which is marked by dashed
rectangle the bond names are taken as they are. Bonds in the other 
unit cells acquire an additional index which correspond to the 
ordering numbers $n_c$ of the cell they are located in starting 
from the central (reference) cell and being negative on the left- and
positive on the right-hand side. We build the finite matrix ${\rm
IP}_{R,nn^\prime}$ between the WOs $\varphi_{0n}$ and 
$\varphi_{Rn^\prime}$ defined in (\ref{IPnn}) with
$R=n_c a_l$. The SCF values of ${\rm IP}_{R,nn^\prime}$ matrix are
summarized in Table~\ref{SCF_IPs} for $\sigma$-type WOs and in
Table~\ref{SCF_IPp} for $\pi$-type WOs. The number of relevant
non-equivalent entries in this matrix is finite since we
omit hopping matrix elements which are below the threshold. The
matrix IP$_{R,nn^\prime}$ is clearly diagonal dominant and the 
matrix elements nicely decay with increasing distance of the
outers of the respective Wannier orbitals.

\begin{table}
\refstepcounter{table}
\addtocounter{table}{-1}
\label{SCF_IPs}
\caption{SCF values of the ${\rm IP}_{R,nn^\prime}$ matrix
for $\sigma$-type WOs (in eV).}
\vspace{5mm}
\centerline{
\renewcommand{\baselinestretch}{1.2}\normalsize
\begin{tabular}{cc|rrrr}
\hline
 $n^\prime$ & \;\;\; $R$ \;\;\; & IP$_{R,{\rm CH1} \; n^\prime}$ & IP$_{R,{\rm sig} \; n^\prime}$ &
 IP$_{R,{\rm CH2} \; n^\prime}$ & IP$_{R,{\rm CC} \; n^\prime}$ \\
\hline
 sig & 2$a_l$  &         & -0.049  &         &        \\
 CC  & 2$a_l$  & -0.153  &  0.202  &         &        \\
 CH1 & 1$a_l$  &  0.338  & -0.190  &         &        \\
 sig & 1$a_l$  &  0.634  & -0.696  & -0.190  &  0.202 \\
 CH2 & 1$a_l$  & -0.803  &  0.634  &  0.338  & -0.153 \\
 CC  & 1$a_l$  &  2.968  &  3.022  &  0.817  & -0.767 \\
 CH1 &    0    & 18.904  &  3.022  & -0.856  &  0.817 \\
 sig &    0    &  3.022  & 22.396  &  3.022  &  3.022 \\
 CH2 &    0    & -0.856  &  3.022  & 18.904  &  2.968 \\
 CC  &    0    &  0.817  &  3.022  &  2.968  & 21.230 \\
 CH1 & -1$a_l$ &  0.338  &  0.634  & -0.803  &  2.968 \\
 sig & -1$a_l$ & -0.190  & -0.696  &  0.634  &  3.022 \\
 CH2 & -1$a_l$ &         & -0.190  &  0.338  &  0.817 \\
 CC  & -1$a_l$ &         &  0.202  & -0.153  & -0.767 \\
 sig & -2$a_l$ &         & -0.049  &         &  0.202 \\
 CH1 & -2$a_l$ &         &         &         & -0.153 \\
\hline
\end{tabular}
}
\end{table}
\begin{table}
\refstepcounter{table}
\addtocounter{table}{-1}
\label{SCF_IPp}
\caption{SCF values of the ${\rm IP}_{R,nn^\prime}$ 
matrix for $\pi$-type WOs (in eV).}
\vspace{5mm}
\centerline{
\renewcommand{\baselinestretch}{1.2}\normalsize
\begin{tabular}{ccccc}
\hline
 IP$_{0,{\rm pi} \: {\rm pi}}$ &
 IP$_{\pm  a,{\rm pi} \: {\rm pi}}$ &
 IP$_{\pm 2a,{\rm pi} \: {\rm pi}}$ &
 IP$_{\pm 3a,{\rm pi} \: {\rm pi}}$ &
 IP$_{\pm 4a,{\rm pi} \: {\rm pi}}$ \\
\hline
 10.526 & 1.683 & -0.365 & 0.120 & -0.047 \\
\hline
\end{tabular}
}
\end{table}
\renewcommand{\baselinestretch}{1}\normalsize

Let us emphasize that the matrix ${\rm IP}_{R,nn^\prime}$ contains
orbital energies and hopping matrix elements for an infinite tPA
chain while its entries are calculated on the C$_{10}$H$_{12}$
molecule being ${\rm IP}_{ab}$. To build the matrix ${\rm
IP}_{R,nn^\prime}$ correctly we use both, translation and
inversion-center symmetry, of the infinite system and usually those matrix
elements of ${\rm IP}_{ab}$ which correspond to bonds close to the
center of the molecule. For example, all diagonal matrix elements ${\rm IP}_{0nn}$
are associated with ${\rm IP}_{aa}$ from the central unit cell. For
off-diagonal elements we have ${\rm
IP}^{\rm chain}_{2a,{\rm sig \; sig}}= {\rm IP}^{\rm
chain}_{-2a,{\rm sig \; sig}}={\rm IP}^{\rm mol}_{\rm (sig,-1), \;
(sig,1)}$ (use of translation symmetry) and ${\rm IP}^{\rm
chain}_{0,{\rm CC \; CH2}}= {\rm IP}^{\rm chain}_{-a,{\rm CC \;
CH1}}={\rm IP}^{\rm mol}_{\rm CC, \; CH2}$ (use of
inversion-center symmetry). Here, for the distinctness we temporarily use
superscripts "chain" for LMEs from the infinite chain and "mol"
for LMEs obtained from SCF calculations of molecules.

\begin{figure}
\centerline{ \psfig{figure=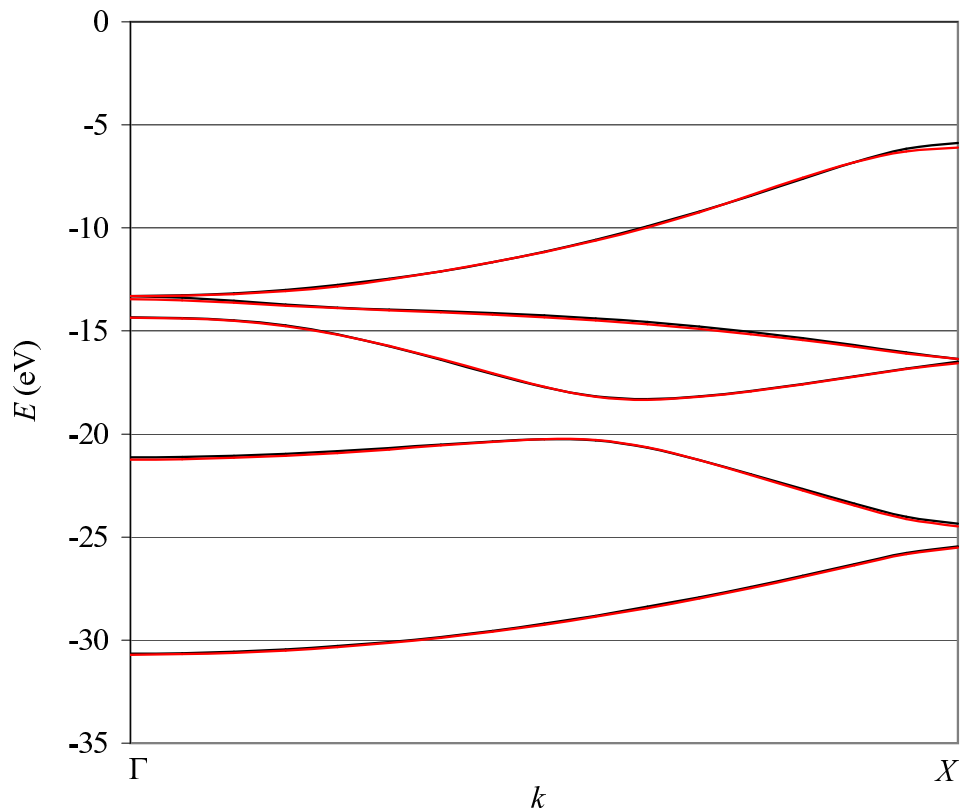,width=5in} }
\caption{Valence bands of {\it trans}-polyacetylene obtained on
the SCF level by LMEs calculated on clusters (red) and by a
CRYSTAL calculation on an infinite tPA single chain (black).}
\label{SCFvalband}
\end{figure}

Substituting the matrix elements from Tables~\ref{SCF_IPs} and 
\ref{SCF_IPp} in the truncated sum (\ref{IPn}) one obtains
$N_k$ matrices IP$_{nn^\prime}(k_i)$ on the chosen $k$-mesh of
the Brillouin zone ($i=1,\;
2,\ldots, \; N_k$). After diagonalization these matrices provide the
positions of the bands at each point $k_i$. By choosing a rather dense
grid one can easily produce a smooth band structure. In
Fig.~\ref{SCFvalband} the five valence bands of {\it trans}-polyacetylene
obtained this way are presented and compared
with the SCF bands of an infinite tPA single chain already shown in
Fig.~\ref{SCFband}. Comparison reveals that the LMEs reproduce the
SCF band structure very nicely and only small deviations (less than 0.2 eV)
from the band structure by CRYSTAL appear due to the truncation in
Eq.~(\ref{IPn}). This demonstrates that LMEs obtained in molecules are
practically equal to those comming directly for the infinite system
using Wannier orbitals.

\subsection{The case of virtual orbitals}

\begin{figure}
\centerline{ \psfig{figure=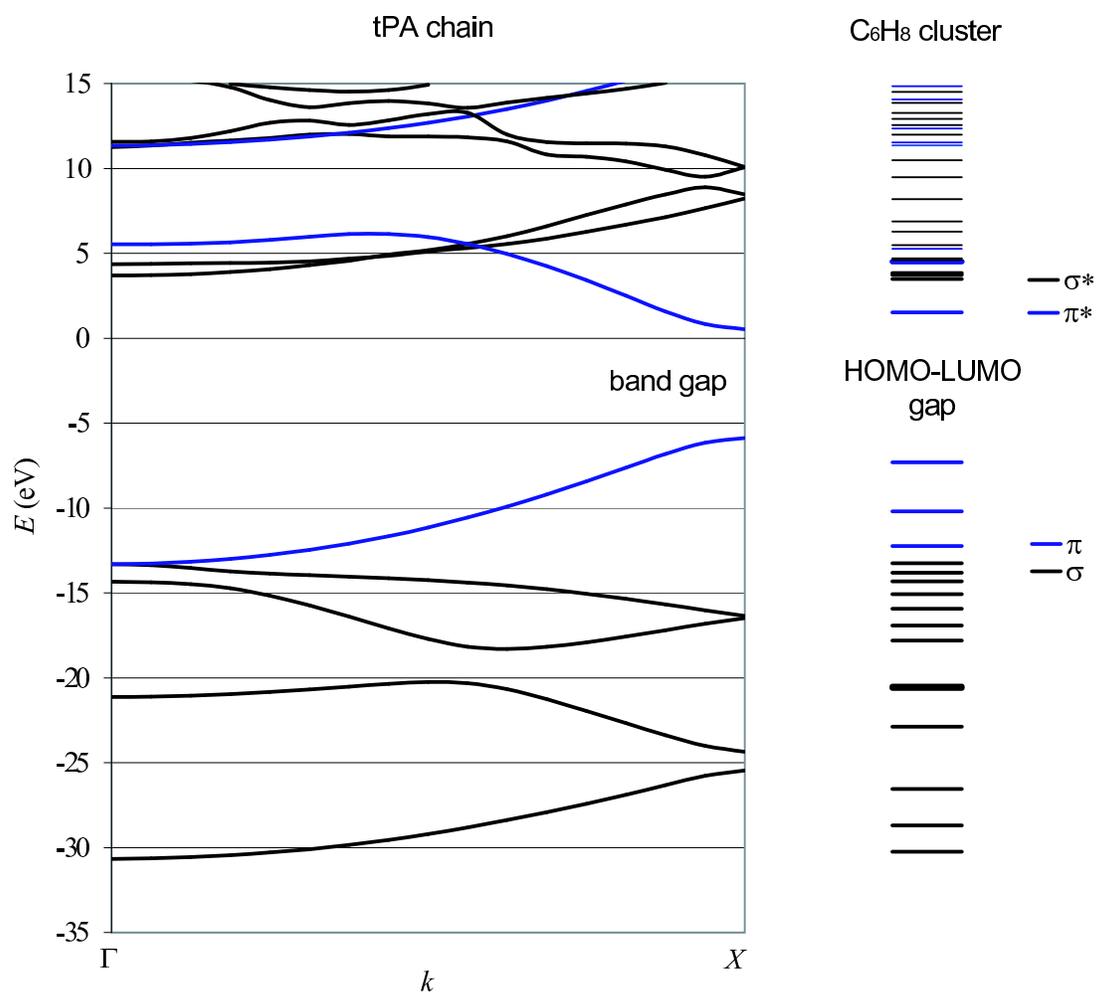,width=5.7in} }
\caption{SCF band structure of tPA (on the left) and the 
one-particle spectrum of a C$_{6}$H$_{8}$ cluster (on the right).}
\label{spectrum}
\end{figure}

The situation becomes more difficult when localized virtual
orbitals have to be generated. In general, one can not simply use the same
strategy as for the localization of the occupied orbitals. Occupied
LMOs are obtained by applying e.g. the Foster-Boys localization scheme
to the $N_{\rm bond}$ energetically highest occupied canonical spatial
orbitals where $N_{\rm bond}$ is the number of bonds in the
molecule. If one applies the localization procedure to the $N_{\rm
bond}$ energetically lowest unoccupied canonical spatial orbitals
one usually does {\it not} get any localized virtual orbitals (or
antibonds) which can be regarded as the unoccupied WOs corresponding
to the lowest few conduction bands. This difference in getting bonds but no
antibonds lies in the fact that the valence bands are well-separated from
the rest of the bandstructure by the band gap and that accordingly 
the occupied spectrum of the molecule representing a finite 
part of the crystal is clearly separated
from the virtual spectrum by the HOMO--LUMO gap
(see Fig.~\ref{spectrum}). In contrast, the lowest
conduction bands are usually entangled with higher conduction bands and
therefore the spectra of virtual states corresponding to lower and
higher conduction bands overlap. For instance, in the band
structure of tPA the five lowest conduction bands (which would
correspond to the five antibonds per unit cell) are not separated 
from the other conduction bands. In this case one can not select 
in a simple and controlled way some suitable $N_{\rm bond}$
unoccupied canonical orbitals which could be converted to
antibonds by the localization procedure. To illustrate this we
show on Fig.~\ref{spectrum} the spectrum of a C$_{6}$H$_{8}$ cluster
together with the SCF band structure of tPA.

Alternatively, one could localize {\it all} virtual canonical orbitals
of the molecule. However, even the energetically lowest localized virtual
orbitals obtained this way have too high orbital energies since they contain
contributions of both energetically low- and high-lying canonical
orbitals. In this case non of the virtual LMOs can be regarded
as a WO corresponding to the lowest few conduction bands. Only the
{\it whole} set of LMOs can be used to reproduce {\it all}
conduction bands.
Therefore LMOs obtained by localization of all virtual molecular
orbitals can not be used in our approach for correlation
correction to the {\it few} lowest conduction bands.

\begin{figure}
\centerline{ \psfig{figure=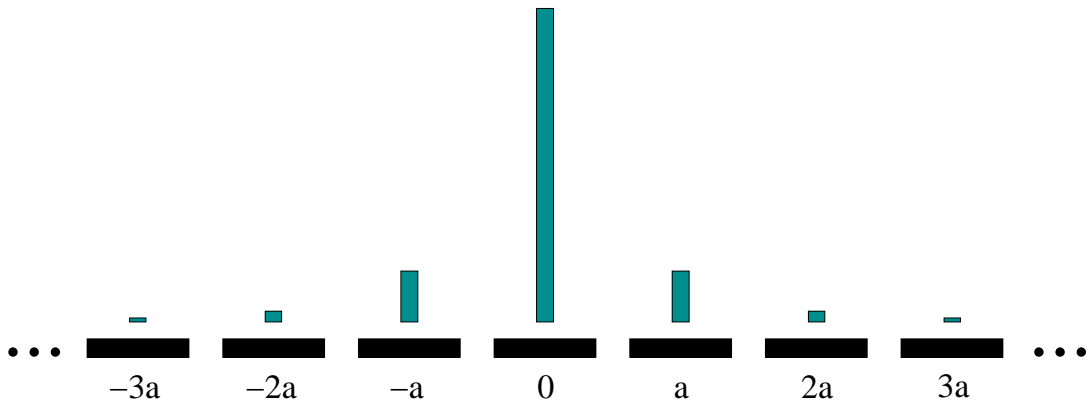,width=5.7in} }
\caption{Schematic diagram showing the contributions of the atomic orbitals from
different unit cells to a WO centered on the zero unit cell of a
1D crystal. Horizontal bars denote the unit cells and the hight of the
vertical bars is proportional to the squared norm of the
wavefunction contribution (or population) from the individual cells
$R = 0$,$\pm a$, $\pm 2a, \ldots$.}
\label{WOdecay}
\end{figure}

Thus, in general, antibonds can not be obtained on clusters by
localization of some set of virtual canonical molecular orbitals 
(except of the cases of well separated conduction bands). 
Hence, we propose a completely
different way of getting localized orbitals in clusters which can
reproduce the bands of our interest: Wannier orbitals obtained in an
infinite system are projected onto a cluster. If the Wannier
orbital can be obtained as a linear combination of atomic orbitals
with contributions of the latter rapidly decaying with the distance
from the orbital center (see Fig.~\ref{WOdecay}) such a WO can be
represented on the cluster when the contributions from those atoms which 
are not present in the cluster are negligibly small such that they can be omitted.
Then the projected WO can be used as the LMO. For instance,
the WO shown schematically in Fig.~\ref{WOdecay} can reasonably be 
represented in a cluster consisting of five unit cells or more.

So, we have first to get well-localized Wannier orbitals in the
infinite system which correspond to the conduction bands of interest.
The concept of maximally localized Wannier orbitals was introduced
for an isolated band or composite bands (an isolated group of 
bands) by Marzari and
Vanderbilt \cite{Marzari97}. There, the molecular localization
criteria of Foster and Boys were applied to minimize the
total spread of the WOs in the periodic system. In \cite{Marzari97}
Bloch functions $\psi_{n {\bf k}}({\bf r})$ represented on a discrete 
mesh of $k$-points ($n$ labels the band) are used and the Wannier
orbital in cell $\bf R$ associated with band $n$ is determined as
$w_n({\bf r-R}) = 1/N_k \sum_k {\rm e}^{-i {\bf k \cdot R}}
\psi_{n {\bf k}}({\bf r})$ ($N_k$ is the number of considered $k$
points in the Brillouin zone). The minimization of the total
spread functional is carried out in the space of unitary matrices
$U_{mn}^{({\bf k})}$ describing the rotation among the Bloch functions
at each $k$-point: $\psi_{n {\bf k}}^\prime = \sum_m U_{mn}^{({\bf
k})} \psi_{m {\bf k}}$.

Later this idea was generalized in \cite{Marzari01} for the case
of entangled bands to account for the case where no separate 
group of bands can be found.
There, one specifies an energy window with $N_{\rm tot}({\bf k})$
bands with the $N_b$ bands on interest being present together with
higher or/and lower lying bands (at each $k$-point $N_{\rm
tot}({\bf k}) \geq N_b$). At $k$-points where $N_{\rm tot}({\bf k})
> N_b$ an $N_b$-dimensional subspace of Bloch states which
correspond to the bands of interest is first established (by 
minimization of some functional) and then the localization
method from \cite{Marzari97} is applied to the selected $N_b$
states at each $k$-point. This way the maximally localized Wannier
orbitals associated with selected bands are obtained.

A variant of the method of \cite{Marzari97} for obtaining well-localized
Wannier orbitals corresponding to composite bands was recently
implemented in the CRYSTAL code (\cite{Baranek01} and \cite{Zicovich01}).
The Bloch orbital for band $n$ at point $k$ is represented in
CRYSTAL as the linear combination of atomic orbitals
$\chi_\mu({\bf r} -{\bf s}_\mu)$ (being contracted
Gaussian-type function) centered at an atomic site ${\bf s}_\mu$

\begin{equation}
\label{BlochFuncCRYSTAL}
\psi_{n {\bf k}}({\bf r}) = \sum_{\mu = 1}^M \alpha_{\mu}^n ({\bf k})
\sum_{\bf R} {\rm e}^{i {\bf k \cdot R}} \chi_\mu({\bf r} -{\bf s}_\mu - {\bf R})
\end{equation}
where $\mu$ labels atomic orbitals centered in the reference cell and 
${\bf R}$ denotes the lattice vector. The $N_b$ Wannier orbitals
associated with the $N_b$ composite bands at each unit cell
are then obtained as linear combinations of atomic orbitals

\begin{equation}
\label{WOCRYSTAL}
w_n({\bf r - R}) = \sum_{\mu = 1}^M \sum_{{\bf R}^\prime} 
c_{\mu, {\bf R}^\prime}^{n,{\bf R}}
\chi_\mu({\bf r} -{\bf s}_\mu - {\bf R}^\prime)
\end{equation}
with the orthonormality condition being fulfilled $\langle
w_n({\bf r - R}) | w_{n^\prime}({\bf r - R}^\prime) \rangle = 
\delta_{nn^\prime} \delta_{{\bf R, R}^\prime}$. 
This new feature of CRYSTAL was
explicitely used in our work to get well-localized virtual WOs in
{\it trans}-polyacetylene corresponding to the three lowest conduction
bands since they can be regarded as composite bands being
separated from other conduction bands by a gap in the whole
Brillouin zone (see Fig.~\ref{SCFband} or \ref{spectrum}). 
However, at this point we would like to note that
the idea of band disentanglement originally developed in
\cite{Marzari01} was very recently implemented in the CRYSTAL 
program package by our work group and WOs corresponding to the 
lowest conduction bands of bulk diamond and silicon were obtained
\cite{Izotov}. Thus, the limitation of the method reported in
\cite{Baranek01} and \cite{Zicovich01} for separated group of
bands is removed and WOs for any bands of interest can now be 
obtained for the general case.

Having obtained well-localized virtual WOs in periodic system as
linear combinations of atomic orbitals centered on atoms of a
finite region of the crystal (usually some few unit cells close to the
center of the WO) we need to project these orbitals onto the basis
set of a finite molecule being a hydrogen-saturated cluster of the
crystal. Then we could use them as virtual LMOs to get LMEs
in our approach and also for further correlation
calculations. For this purpose the program CRYREAD \cite{CRYREAD}
was recently designed in our group. 
It reads the WOs obtained in CRYSTAL as linear combinations
of atomic functions and projects them onto the basis set of
the user-specified cluster of the periodic system. The program
takes all those WOs with norms exceeding a user-specified threshold 
after the coefficients $c_{\mu, {\bf R}^\prime}^{n,{\bf R}}$ at 
atomic orbitals $\chi_\mu({\bf r} -{\bf s}_\mu - {\bf R}^\prime)$
centered outside the cluster are set to zero, makes them
othonormal and writes the corresponding matrix of coefficients $c_{\mu, {\bf
R}^\prime}^{n,{\bf R}}$ where vectors ${\bf R}^\prime$ and ${\bf R}$ are
restricted to the specified cluster. After some further
modification (contraction coefficients of atomic orbitals centered
on cluster-saturating hydrogen atoms are added and
set to zero) this matrix can be read by the program package
MOLPRO and virtual localized molecular orbitals are obtained
directly in the form of linear combinations of basis functions
of the molecule. Virtual LMOs obtained this way perfectly represent 
WOs of the periodic system.

To be more precise, in the case of {\it trans}-polyacetylene we
generate with CRYSTAL Wannier orbitals associated with core, valence and
the first three conduction bands, performing three separate calculations
for each subset of bands. The procedure for getting well-localized
crystalline orbitals in CRYSTAL (\cite{Baranek01} and
\cite{Zicovich01}) is an iterative one and requires (especially for 
virtual orbitals) an initial guess
for the Wannier orbitals to start with. As discussed in \cite{Zicovich01}
the choice of the initial guess is critical for systems with
fully covalent bonds (like diamond, silicon and also
{\it trans}-polyacetylene). Therefore, we have to design good starting
Wannier orbitals for each subset of bands. For core bands
1$s$ atomic orbitals of carbon atoms are
provided. In the case of valence bands we expect that the associated
WOs are similar to the molecular bonds of hydrogen-saturated clusters.
Therefore, we specify five bonding pairs of atomic orbitals per unit cell which
correspond to the five bonds of the covalent system. The choice of the
initial guess for three unoccupied WOs associated with the three
lowest conduction bands of tPA is more subtle. We suppose
that these three WOs correspond to three out of five antibonds
per unit cell. As one of the conduction bands is formed by states
of $\pi$ symmetry the corresponding WO should be the
$\pi$-antibond centered on the short C--C bond. For starting WOs
associated with the two $\sigma$-bands the C--H antibonds are
used. This choice is based on the fact that the C--H bonds in
hydrogen-saturated clusters of tPA are energetically higher than
the C--C $\sigma$ bonds and therefore the C--H antibonds are expected
to be energetically lower than the C--C $\sigma$ antibonds. Since we
already have occupied WOs $w_n({\bf r})$ (or bonds) the antibonds
of interest are obtained by a change of the sign of the coefficients
$c_{\mu, 0}^{n,0}$ from the atomic orbitals
$\chi_\mu({\bf r} -{\bf s}_\mu)$ centered at one of two atoms
of the corresponding bond. Providing antibonds constructed this
way we were able to generate well-localized unoccupied WOs associated with three
lowest conduction bands of tPA. The condition of orthonormality
is fulfilled for all Wannier orbitals: $\langle
w_i({\bf r - R}) | w_j({\bf r - R}^\prime) \rangle = \delta_{ij}
\delta_{{\bf R, R}^\prime}$ (indices $i$ and $j$ running over core,
valence and the first three conduction bands).

By the program CRYREAD we project the virtual WOs onto a 
C$_{12}$H$_{12}$ cluster (six unit cells, no saturating hydrogens)
of the infinite tPA chain. Thus we obtain 18 ($3\times 6$) WOs
centered on this cluster. The coefficient matrix produced by
CRYREAD is enlarged to be consistent with the set of basis
functions of an isolated C$_{12}$H$_{14}$ molecule
(hydrogen-saturated). The enlarged matrix is read by the 
MOLPRO program where the SCF calculation on the
C$_{12}$H$_{14}$ molecule has been performed and the set of
one-particle SCF wavefunctions have been obtained. Then, the
occupied and virtual spaces in C$_{12}$H$_{14}$ are constructed 
as follows. The
occupied space $\{ \psi_a \}$ is taken from the SCF calculation on
C$_{12}$H$_{14}$ and is not changed. Next, the WOs 
$w_n({\bf r - R}) = \tilde{\varphi}_r(\bf r)$ (obtained in the periodic
system and projected onto the C$_{12}$H$_{12}$ cluster) are provided and
are made orthogonal to the occupied space of the C$_{12}$H$_{14}$
molecule by Schmidt orthogonalization and re-orthonormalized thereafter
by L\"owdin (or symmetric) orthogonalization
(see \cite{Szabo}, p.~142-145). After this orthogonalization step the
obtained virtual LMOs $\varphi_r^\prime$ may slightly differ from
the projected WOs $\tilde{\varphi}_r$ as the occupied space of the
C$_{12}$H$_{14}$ molecule differs from the one of the embedded C$_{12}$H$_{12}$
cluster. The space of the remaining virtual one-particle orbitals of
C$_{12}$H$_{14}$ $\{\psi_r^\prime\}$ is constructed to be
orthogonal to the occupied space $\{ \psi_a \}$ and the specified
subspace of the 18 virtual LMOs $\{ \varphi_r^\prime \}$.

\begin{figure}
\centerline{ \psfig{figure=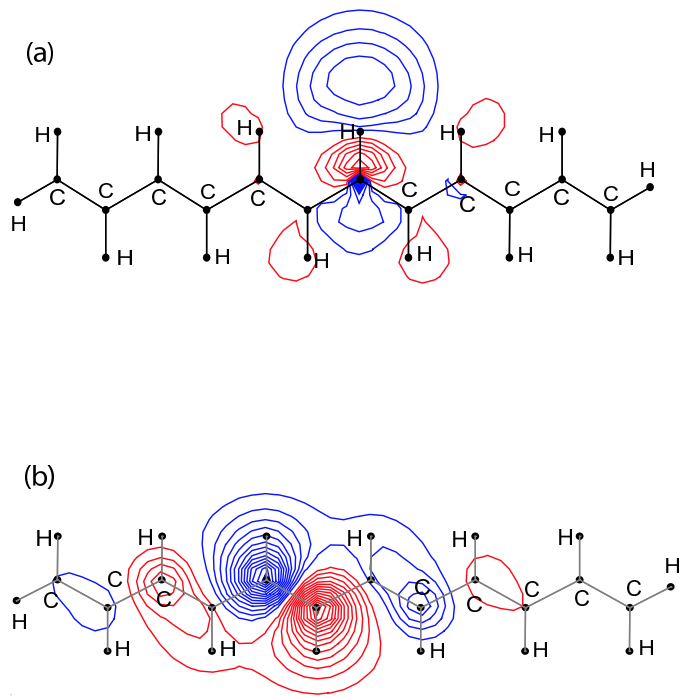,width=3.5in} }
\caption{One C--H (a) and one C--C $\pi$ (b) antibonds of a
C$_{12}$H$_{14}$ cluster of tPA: the plane for the contour plot
of the C--H antibond lies in the molecular plane, that
of the C--C $\pi$ antibond is lifted upwards to $z=0.5 \;
{\rm {\AA}}$. Red lines correspond to positive and
blue lines to negative vlues of $\varphi_r({\bf r})$.
The contour values increase in steps of 0.025~a.u.}
\label{virt}
\end{figure}

The 18 virtual LMOs constructed this way turned out to be 6 C--C antibonds
of $\pi$ type and 12 C--H antibonds of the C$_{12}$H$_{14}$ molecule.
Two of these antibonds (one C--H and one C--C) from the central
part of the molecule are presented in Fig.~\ref{virt}. Again
as $\pi$ orbitals have zero values in the molecule plane the plot
plane in the contour plot in Fig.~\ref{virt}(b) is lifted by
0.5~{\AA}. As one sees from these plots, the unoccupied LMOs
are substantially more diffuse than the corresponding occupied ones
(Fig.~\ref{occ_sig} and \ref{occ_pi}).

So, we got virtual LMOs of the molecule $\varphi_r^\prime$ in form
of linear combinations of basis functions $\chi_\mu$ centered on
atoms of this molecule. The LMOs represent WOs and can be used to
obtain LMEs on the SCF level and to reproduce the three conduction bands.
However, at this point we would like to turn back to the question
of getting virtual LMOs by applying the Foster--Boys localization
scheme to virtual canonical orbitals $\psi_r$. The problem was
that, in general, one can not simply select a subset of consecutive
virtual canonical orbitals. Now with the LMOs at hand we can project 
the virtual LMOs onto the space of the virtual canonical orbitals 
$\{ \psi_r \}$ of the C$_{12}$H$_{14}$ molecule (which altogether
consists of 359 orbitals):

\begin{equation}
\label{projection}
| \varphi_{s}^\prime \rangle = \sum_{r=1}^ {359} | \psi_r \rangle
\langle \psi_r |\varphi_s^\prime \rangle
\end{equation}
and select the required subset for the localization by chosing those 
with the largest projection coefficients
$\langle \psi_r |\varphi_s^\prime \rangle$. This procedure is justified
as long as all other coefficients are close to zero which was indeed
the case for all our clusters.
In case of the hydrogen-saturated C$_{12}$H$_{14}$
cluster of {\it trans}-polyacetylene the 20 energetically lowest virtual
canonical orbitals were selected by the projection (\ref{projection}).
After localization they give 6 C--C antibonds of $\pi$ symmetry and
14 C--H antibonds including the two C--H antibonds with saturating 
hydrogen atoms which are not present in the crystal. (Again
canonical orbitals of $\pi$ and $\sigma$ symmetry are localized
separately from each other.) The LMOs obtained this way will be
denoted by $\varphi_r$. We have to emphasize here that we could get
these LMOs $\varphi_r$ by Foster-Boys localization of suitably selected
virtual canonical orbitals only because of two unique properties of {\it
trans}-polyacetylene: i) it has three separated conduction bands
(see Fig.~\ref{SCFband}) which correspond to the C--C
$\pi$ antibond and two C--H antibonds (note that no separated
bands corresponding to two C--C $\sigma$ antibonds can be found in
the band structure of tPA) and ii) the two C--H bonds which connect
two carbon atoms on the edges of the cluster with the saturating
hydrogen atoms are similar to the characteristic C--H antibonds of the 
(C$_2$H$_2$)$_x$ 1D crystal and the energy of an electron in the
artificial antibond is approximately the same as the one in the
C--H antibond of the crystal. In general, localization of
virtual canonical orbitals fails to produce proper antibonds. For instance,
no suitable subset can be found in the virtual
spectrum of hydrogen-saturated clusters of diamond (or silicon) to
get the C--C (or Si--Si) antibonds by Foster--Boys localization. Thus,
in general, one has to use projected WOs as LMOs. We decide to use the 
LMOs $\varphi_r$ for {\it trans}-polyacetylene because
they are easier to handle than the projected WOs $\varphi_r^\prime$ but
we would like to analyze first the difference between $\varphi_r$
and $\varphi_r^\prime$.

For the C--C $\pi$ antibonds there is no noticeable difference between
the projected WOs and the LMOs obtained by localization of the first 6 
virtual canonical $\pi$ orbitals as there is no effect of the saturating 
C--H antibonds of $\sigma$ symmetry on the C--C $\pi$ antibonds. Therefore,
the LMEs corresponding to states with an attached electron in a projected
virtual WOs and an LMOs of $\pi$ symmetry are practically the same.
In Table~\ref{SCF_EAp} we list the LMEs EA$_{R,mm^\prime}$ defined in
(\ref{EAmm}) for ($N+1$)-electron states where the additional electron
is placed in a $\pi$ antibonds as obtained by Foster-Boys localization.
The names of antibonds coincide with those of the bonds but are marked
with an asterisk. Again a rapid decay of the matrics elements with
increasing bond distance is clearly discernable.

\begin{table}
\refstepcounter{table}
\addtocounter{table}{-1}
\label{SCF_EAp}
\caption{SCF values of the ${\rm EA}_{R,mm^\prime}$
matrix for $\pi$-type unoccupied WOs as obtained on a
C$_{12}$H$_{14}$ cluster (in eV).}
\vspace{5mm}
\centerline{
\renewcommand{\baselinestretch}{1.2}\normalsize
\begin{tabular}{ccccc}
\hline
 EA$_{0,{\rm pi}^\ast \: {\rm pi}^\ast}$ &
 EA$_{\pm  a,{\rm pi}^\ast \: {\rm pi}^\ast}$ &
 EA$_{\pm 2a,{\rm pi}^\ast \: {\rm pi}^\ast}$ &
 EA$_{\pm 3a,{\rm pi}^\ast \: {\rm pi}^\ast}$ &
 EA$_{\pm 4a,{\rm pi}^\ast \: {\rm pi}^\ast}$ \\
\hline
 -4.497 & -1.054 & 0.701 & -0.146 & 0.025 \\
\hline
\end{tabular}
}
\end{table}
\renewcommand{\baselinestretch}{1}\normalsize

When C--H antibonds are obtained by localization of the 14 energetically
lowest unoccupied canonical $\sigma$ orbitals of C$_{12}$H$_{14}$ with
the latter ones containing contibutions of the two C--H antibonds with 
the saturating hydrogen atoms. Since unoccupied LMOs are more diffuse 
than occupied ones (and spread over three unit cells of tPA as one sees
in Fig.~\ref{virt}) the C--H antibonds obtained in the molecule are 
affected by the presence of the saturating
atoms. As a consequence, the local SCF matrix elements
for attached electron states EA$_{R,mm^\prime}$ of $\sigma$ type 
calculated in the C$_{12}$H$_{14}$ molecule by means of the projected 
WOs as localized virtual orbitals differ slightly from those calculated
with the localized orbitals in the molecule. The two relevant matrices
are listed in Table~\ref{SCF_EAs}. The first two colunms of data
show the values of the LMEs obtained when
projected virtual WOs of $\sigma$ type are used as C--H antibonds.
The last two columns represent the same matrix elements obtained
with localized virtual orbitals from a Foster-Boys localization
of the canonical $\sigma$ orbitals of C$_{12}$H$_{14}$. They differ
by up to 0.12 eV which is small but not negligible for the calculation
of the SCF bandstructure. The difference between the matrix elements
EA$_{R,mm^\prime}$ is caused by the presense of
the two saturating hydrogen atoms in C$_{12}$H$_{14}$ molecule.

\begin{table}
\refstepcounter{table}
\addtocounter{table}{-1}
\label{SCF_EAs}
\caption{SCF values of the ${\rm EA}_{R,mm^\prime}$ matrix for C--H 
antibonds in C$_{12}$H$_{14}$ cluster (in eV).}
\vspace{5mm}
\centerline{
\renewcommand{\baselinestretch}{1.2}\normalsize
\begin{tabular}{cc|rr|rr}
\hline
  &  & \multicolumn{2}{c|}{Projected WOs} &  \multicolumn{2}{c}{LMOs by FB localization}
  \\
 $m^\prime$ & \;\;\; $R$ \;\;\; & EA$_{R,{\rm CH1^\ast} \: m^\prime}$ & EA$_{R,{\rm CH2^\ast} \: m^\prime}$ &
 EA$_{R,{\rm CH1^\ast} \: m^\prime}$ & EA$_{R,{\rm CH2^\ast} \: m^\prime}$ \\
\hline
 CH2$^\ast$  & 4$a_l$  & -0.063  &         & -0.068  &        \\
 CH1$^\ast$  & 3$a_l$  &  0.120  & -0.083  &  0.092  & -0.054 \\
 CH2$^\ast$  & 3$a_l$  &  0.097  &  0.120  &  0.107  &  0.092 \\
 CH1$^\ast$  & 2$a_l$  & -0.289  &  0.126  & -0.244  &  0.078 \\
 CH2$^\ast$  & 2$a_l$  & -0.189  & -0.289  & -0.177  & -0.244 \\
 CH1$^\ast$  & 1$a_l$  &  1.012  & -0.236  &  0.914  & -0.194 \\
 CH2$^\ast$  & 1$a_l$  & -0.031  &  1.012  & -0.025  &  0.914 \\
 CH1$^\ast$  &      0  & -5.603  & -0.013  & -5.487  & -0.042 \\
 CH2$^\ast$  &      0  & -0.013  & -5.603  & -0.042  & -5.487 \\
 CH1$^\ast$  & -1$a_l$ &  1.012  & -0.031  &  0.914  & -0.025 \\
 CH2$^\ast$  & -1$a_l$ & -0.236  &  1.012  & -0.194  &  0.914 \\
 CH1$^\ast$  & -2$a_l$ & -0.289  & -0.189  & -0.244  & -0.177 \\
 CH2$^\ast$  & -2$a_l$ &  0.126  & -0.289  &  0.078  & -0.244 \\
 CH1$^\ast$  & -3$a_l$ &  0.120  &  0.097  &  0.092  &  0.107 \\
 CH2$^\ast$  & -3$a_l$ & -0.083  &  0.120  & -0.054  &  0.092 \\
 CH1$^\ast$  & -4$a_l$ &         & -0.063  &         & -0.068 \\
\hline
\end{tabular}
}
\end{table}
\renewcommand{\baselinestretch}{1}\normalsize

\begin{figure}
\centerline{ \psfig{figure=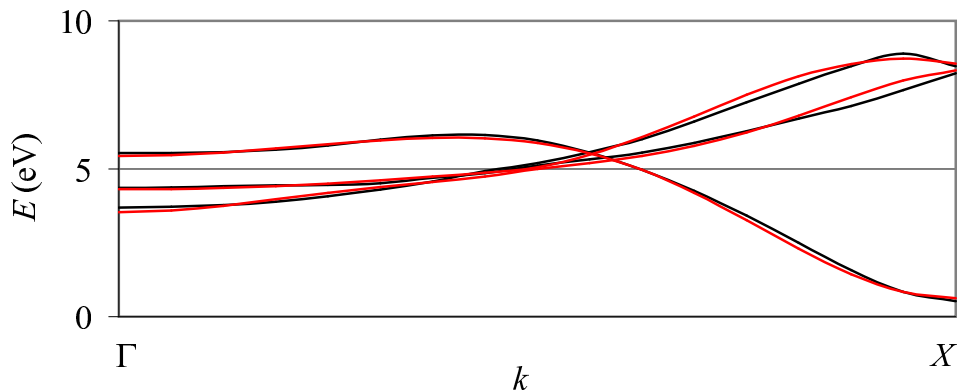,width=5in} }
\caption{Conduction bands of {\it trans}-polyacetylene obtained
on the SCF level by LMEs calculated in a C$_{12}$H$_{14}$ molecule
with projected unoccupied Wannier orbitals (red) and by CRYSTAL
calculation on the infinite tPA single chain (black).}
\label{SCFcondband}
\end{figure}

To demostrate how the corresponding virtual LMEs reproduce the
conduction bands we substitute the data from Table~\ref{SCF_EAp}
and the first two columns of Table~\ref{SCF_EAs} into the
truncated sum (\ref{EAm}) and compare in Fig.~\ref{SCFcondband} 
the bands obtained this way (red lines) with the three first
conduction  bands of infinite tPA chains as obtained by CRYSTAL
(black lines). The deviation of bands
by LMEs from those by CRYSTAL does not exceed 0.2 eV for all
$k$-points like in the case of the valence bands and is mainly due to
the truncation of the infinite summation (\ref{EAm}).

It has to be noticed that one needs hopping matrix elements
between more distant virtual LMOs (up to 4-th nearest neighbor cell) 
than in the case of the occupied LMOs
to reproduce the $\sigma$ conduction bands with the same accuracy as
valence bands (where 2nd nearest neighbor cells were enough). 
Comparing Tables~\ref{SCF_EAs} and \ref{SCF_IPs} we
see that the decay of the SCF hopping matrix elements with the distance
between the localized virtual orbitals is slower than in
the case of the occupied orbitals since the former are more diffuse (cf.
Fig.~\ref{occ_sig}(a) and \ref{virt}(a)). This implies that one needs
larger clusters for getting the EA$_{R,mm^\prime}$ matrix elements
than the IP$_{R,nn^\prime}$ matrix elements if the same value for the 
truncation threshold in (\ref{EAm}) and (\ref{IPn}) is used.

\begin{figure}
\centerline{ \psfig{figure=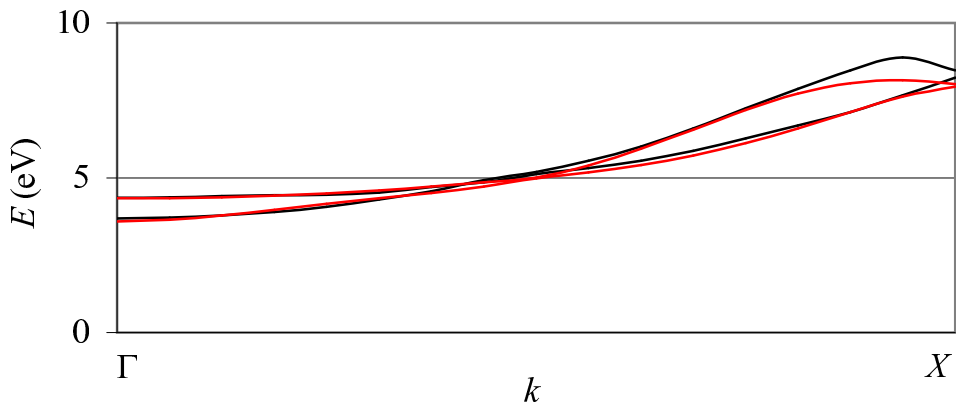,width=5in} }
\caption{Two lowest conduction bands of $\sigma$-symmetry obtained
on the SCF level by LMEs calculated fully in C$_{12}$H$_{14}$ molecule
(red lines) as compared to $\sigma$ bands from CRYSTAL calculation on
the infinite tPA single chain (black lines).}
\label{ch_cond}
\end{figure}

In the case of {\it trans}-polyacetylene we have the unique possibility to
compare the local matrix elements EA$_{R,mm^\prime}$ obtained with projected
unoccupied Wannier orbitals $\varphi^\prime_r$ and
antibonds $\varphi_r$ obtained on a cluster by Foster-Boys localization of
canonical virtual orbitals $\psi_r$ of this cluster. If we also produce the
lowest $\sigma$ conduction bands by substitution the matrix elements from the
two last columns of Table~\ref{SCF_EAs} into (\ref{EAm}) and compare them to
the corresponding $\sigma$ bands obtained by CRYSTAL for the infinite
tPA chain (see Fig.~\ref{ch_cond}) we see rather big deviation for $k$-points 
close to the $X$ point. This is most likely due to the onset of an avoided 
crossing in the upper $\sigma$ conduction band close to the $X$ point 
(see Fig.~\ref{SCFband}). Therefore, for the SCF values of the
EA$_{R,mm^\prime}$ matrix elements we take those which are obtained with the
projected unoccupied WOs.

\section{Correlation corrections to the local matrix elements}

The aim of our approach is to obtain the local matrix elements as
defined in (\ref{IP_corr}) and (\ref{EA_corr}) with the correlation effects
taken into account. This is equivalent to finding corrections
$\delta H_{ij}^{\rm corr}$ and $\delta E_{0}^{\rm corr}$ to the 
matrix elements $H_{ij}$ from Eq.~(\ref{H_ij_SCF}) and
the ground-state energy $E_0$ obtained in the framework of
the HF approximation, respectively:

\begin{equation}
\label{Hcorrcorr}
\delta H_{ij}^{\rm corr} =  H_{ij}^{\rm corr} - H_{ij}
\equiv  H_{ij}^{\rm eff} - H_{ij}
\end{equation}
and

\begin{equation}
\label{Ecorrcorr}
\delta E_0^{\rm corr} =  E_0^{\rm corr} - E_0.
\end{equation}
Then, the local matrix elements including the correction effects are
obtained by adding these corrections to the corresponding SCF
matrix elements as follows from Eqs.~(\ref{IP_corr}), (\ref{EA_corr}),
(\ref{Hcorrcorr}) and (\ref{Ecorrcorr}):

\begin{equation}
\label{IP_correlated}
{\rm IP}_{ab}^{\rm corr} = {\rm IP}_{ab}
+\delta {\rm IP}_{ab}^{\rm corr}
= H_{ab} + \delta H_{ab}^{\rm corr}
- \delta_{ab} (E_0 + \delta E_0^{\rm corr})
\end{equation}

and

\begin{equation}
\label{EA_correlated}
{\rm EA}_{rs}^{\rm corr} = {\rm EA}_{rs}
+\delta {\rm EA}_{rs}^{\rm corr}
= - (H_{rs} + \delta H_{rs}^{\rm corr})
+ \delta_{rs} (E_0 + \delta E_0^{\rm corr})
\end{equation}
The correlation corrections to the local SCF matrix elements

\begin{equation}
\label{IPcorrcorr}
\delta {\rm IP}_{ab}^{\rm corr} = \delta H_{ab}^{\rm corr}
- \delta_{ab} \delta E_0^{\rm corr}
\end{equation}

and

\begin{equation}
\label{EAcorrcorr}
\delta {\rm EA}_{rs}^{\rm corr} = -\delta H_{rs}^{\rm corr}
+ \delta_{rs} \delta E_0^{\rm corr}
\end{equation}
appear as the consequence of electron correlation effects in the
charged and the neutral system (which are two different effects
as explained in chapter~2.4). Formally,
$\delta {\rm IP}_{ab}^{\rm corr}$ and
$\delta {\rm EA}_{rs}^{\rm corr}$ are non-zero because of the
CI expansion of the wavefunctions according to Eqs.~(\ref{Psi_i^corr}) and
(\ref{Phi0_corrCI}). Without this expansion (i.e. when SCF
many-particle wavefunctions are used) we get the SCF local
matrix elements IP$_{ab}$ and EA$_{rs}$ which are listed in
Tables~\ref{SCF_IPs}--\ref{SCF_EAs}.

In principle, correlation corrections to all matrix elements
listed in these tables can be calculated. However, in order to get
correlated band structure with sufficiant accuracy one needs to
account for correlation effects only on the diagonal and the few
largest off-diagonal matrix elements. As known from earlier works
\cite{Graef93}, \cite{Graef97}, \cite{Albrecht98} and
\cite{Albrecht00} (and as confirmed by the present work) the
correlation corrections $\delta {\rm IP}_{ab}^{\rm corr}$ and
$\delta {\rm EA}_{rs}^{\rm corr}$ are small fractions of the
corresponding SCF matrix elements IP$_{ab}$ and EA$_{rs}$. Also
these corrections are obtained with final accuracy (10\% say).
Therefore one can conclude that for a given set of SCF local
matrix elements reproducing some set of bands on the SCF level
there is no sense to calculate correlation corrections to LMEs
which are of the order of (or smaller than) 10\% of the largest
matrix element in the given
set since the obtained correction to such small LMEs i) has no
noticeable effect on the band structure and ii) themselves are of
the order of the error bar for the largest matrix element.
Thus we can restrict the number of matrix
elements which need to be corrected to {\it all} diagonal elements
and all those off-diagonal elements whose absolute values exceeds
10\% of the largest diagonal matrix element in the given set
(see Tables~\ref{SCF_IPs}--\ref{SCF_EAs}).

Practically, these off-diagonal elements are those where the two 
involved localized orbitals have the largest overlap. So, for $\pi$
orbitals (occupied and virtual) the hopping matrix element for two
$\pi$ bonds (or antibonds) from two adjasent unit cells are 
considered. In the case of the occupied $\sigma$ bonds we calculate
correlation corrections to off-diagonal elements $\delta {\rm
IP}_{ab}^{\rm corr}$ for all pairs of bonds sharing one atom. In
tPA there are three such non-equivalent hopping matrix elements between
adjacent $\sigma$ bonds: IP$_{\rm CH2,sig}$, IP$_{\rm CH2,CC}$ and
IP$_{\rm CC,sig}$ (see Fig.~\ref{c10h12}). The C--H antibond is
extended along the chain and therefore has the largest overlap
with its translational copy from the nearest-neighbor unit cell
(see Fig.~\ref{virt}a). Thus, taking into account the inversion
and translation symmetry of tPA we only need to care for one off-diagonal
matrix element, EA$_{\rm CH2^\ast,CH2,1^\ast}$, to correct the
$\sigma$ conduction bands.

As discussed in chapter 2.4, electron correlation is mainly a
local effect. It takes into account the possibility of electrons
in two neighbor localized orbitals to avoid each other and, by 
this, to reduce the energy of electron-electron repulsion. Also
the rearrangement of electrons due to the presence of an
extra charge happens mainly in the close vicinity of the charge.
Therefore, the major contribution to the corrections $\delta {\rm
IP}_{ab}^{\rm corr}$ and $\delta {\rm EA}_{rs}^{\rm corr}$ to the
local SCF matrix elements IP$_{ab}$ and EA$_{rs}$ is produced by
correlation of electrons in localized orbitals which are located
in unit cells close to orbitals $a$ and $b$ (or $r$ and $s$
respectively). Thus, for each matrix element IP$_{ab}$ (or
EA$_{rs}$) one can explicitly define a cluster in which electron
correlation gives the major contribution to $\delta {\rm
IP}_{ab}^{\rm corr}$ ($\delta {\rm EA}_{rs}^{\rm corr}$). For
this cluster correlation should be treated with high accuracy
(that is why the multireference CI method is used in our work).

To account for correlation effect on IP$_{ab}$ matrix
elements coming from electrons in $\sigma$ bonds a C$_6$H$_8$
cluster is enough. However, in order to treat
accurately correlation between electrons in $\pi$ bonds
one needs C$_8$H$_{10}$ and C$_{10}$H$_{12}$ molecules
because the occupied $\pi$ bonds are more diffuse than the
$\sigma$ bonds. As unoccupied localized orbitals are even
more diffuse than the occupied ones, correlation calculations
in C$_{12}$H$_{14}$ molecule have to be performed to properly account for
corrections to the EA$_{rs}$ matrix elements. The details of the
application of the multireference single and double
configuration interaction method to get the corrections
$\delta {\rm IP}_{ab}^{\rm corr}$ and $\delta {\rm EA}_{rs}^{\rm corr}$
are outlined in the next section.

\section{The method of local increments}

Having chosen the molecule (a hydrogen-saturated cluster of the
crystal) for explicit correlation calculations we have to treat
all electrons of the ($N-1$), $N$ or ($N+1$)-electron system on
the correlation level except of the electrons in the core orbitals
and in the bonds with the saturating hydrogen atoms. Let us say
that in some cluster we have to handle electrons from $n_{\rm
occ}$ occupied bonds. Then, to get the correlation correction to
the ground-state energy all singly and doubly excited
configurations with electrons removed from these $n_{\rm occ}$
bonds are included into the expansion (\ref{Phi0_corrCI}). For
hole states, $n_{\rm occ}$ reference configurations $\Psi_i^{\rm
corr}$ are formed as linear combinations of $n_{\rm occ}$
one-particle configurations $\Phi_a$ with the hole in one of the
$n_{\rm occ}$ bonds. Then, excited configurations are produced by
taking one or two out of ($2 \: n_{\rm occ}-1$) electrons from the
occupied bonds and placing them into some virtual orbitals as
defined in Eq.~(\ref{Psi_i^corr}), and the energies of $n_{\rm
occ}$ hole states have to be calculated. In the case of an
attached electron in $n_{\rm virt}$ antibonds of the cluster we
have $n_{\rm virt}$ reference states with the extra electron being
delocalized over these antibonds. Due to the presence of the extra
electron the electrons in the $n_{\rm occ}$ occupied bonds
rearrange and in order to account for the correlation effect in
this case all excited configurations with one or two electrons
removed from either one of the $n_{\rm occ}$ occupied or virtual
orbital which carries the extra electron and placed in some free
virtual orbitals have to be taken into account in
Eq.~(\ref{Psi_i^corr}). Then, the energies of $n_{\rm virt}$
states are calculated. In all these three cases the number of
excited configurations which are taken into account is huge even
for small molecules and robust CI calculations as described above
are not feasible (an enormous number of coefficients $\alpha$ from
Eqs.~(\ref{Phi0_corrCI}) or (\ref{Psi_i^corr}) have to be
optimized simultaneously). To proceed one needs an approximate
scheme which consists of a series of calculations with the number
of excited configurations being drastically reduced but which
still accounts for electron correlation in the cluster with the
desired accuracy.

\subsection{Formulation in terms of bonds}

A simple but powerful scheme to estimate correlation correction to
SCF ground-state energy of clusters and solids was proposed by
Stoll in Ref.~\cite{Stoll92a} and \cite{Stoll92b}. This scheme is
based on a clear physical idea and systematically sums
contributions to $\delta E_0^{\rm corr}$ where each term accounts
for correlation of a very limited number of electrons in localized
orbitals. The method is as follows. First, corrections to the
ground-state energy from electrons in each bond $a$ individually
are calculated: $\epsilon_a = E_0^{\rm corr}(a) - E_0$, the energy
$E_0^{\rm corr}(a)$ is obtained from Eq.~(\ref{E_corr}) where in
the CI expansion for the correlated wavefunction
(\ref{Phi0_corrCI}) {\it only} electrons in bond $a$ are used to
construct the excited configurations $\Phi_a^r$ and
$\Phi_{a\bar{a}}^{rs}$. Here $\bar{a}$ stands for a spin orbital
with the same spatial orbital as $a$ but with opposite
spin-orientation with respect to the spin orbital $a$. Let us
notice here that a drastic reduction of the number of excited
configurations in (\ref{Phi0_corrCI}) takes place when excitations
of two electrons only are considered. Summing up the contributions
$\epsilon_a$ from all bonds one gets a first estimate $\delta
E_{0}^{(1)}$ for the correction to the HF ground-state energy. To
account for the non-additive corrections $\epsilon_{a,b} =
E_0^{\rm corr}(a,b) - (\epsilon_a + \epsilon_b) - E_0$,
correlation calculations with only the pair $(a,b)$ of bonds being
open for producing excited configurations are performed. Adding
all pair corrections $\epsilon_{a,b}$ to $\delta E_0^{(1)}$ we
reach the next order of approximation $\delta E_0^{(2)}$. The
third order is obtained when we account for correlation
corrections of electrons in bond triples $(a,b,c)$ in the cluster:
$\delta E_0^{(3)} = \delta E_0^{(2)} + \sum_{a<b<c}
\epsilon_{a,b,c}$; the non-additivity corrections to the
second-order approximation are $\epsilon_{a,b,c} = E_0^{\rm
corr}(a,b,c) -(\epsilon_a + \epsilon_b + \epsilon_c)
-(\epsilon_{a,b}+ \epsilon_{a,c}+ \epsilon_{b,c})- E_0$.
Continuing this way one builds order by order a converging series

\begin{equation}
\label{Eincr}
\delta E_0^{\rm corr} =  \sum_{a} \epsilon_{a}
+ \sum_{a<b} \epsilon_{a,b} + \sum_{a<b<c} \epsilon_{a,b,c}
+ \sum_{a<b<c<d} \epsilon_{a,b,c,d} +\ldots
\end{equation}
which in its limit gives the exact value of $\delta E_0^{\rm
corr}$. However, as was discussed in a number of papers (e.g. in
Ref.~\cite{Stoll92a}, \cite{Stoll92b}, \cite{Paulus95},
\cite{Paulus96}, \cite{Kaldova97}, \cite{Doll97} and
\cite{Fulde02}) one obtains a very accurate estimate of $\delta
E_0^{\rm corr}$ when the series (\ref{Eincr}) is truncated after
the third term. Thus, in each separate correlation calculation at
most six electrons are treated simultaneously. Moreover, as
correlation effect decays rapidly with the distance between
localized orbitals, the two- and three-bond increments
$\epsilon_{a,b}$ and $\epsilon_{a,b,c}$ involving distant bonds
can be neglected and therefore the effort for evaluating the
correlation correction to the ground-state energy of a big
molecules is substantially reduced. In fact, in the case of a
periodic system only a finite number of energy increments has to
be considered.

An extension of this approach to the correlation corrections to
local SCF matrix elements of {\it hole} states was proposed in the
paper \cite{Graef93} and developed further in \cite{Graef97},
\cite{Albrecht00} and \cite{Albrecht02}. It was shown that the
corrections to the matrix elements $\delta {\rm IP}_{ab}^{\rm
corr}$ given by Eq.~(\ref{IPcorrcorr}) can be also obtained with
any desired accuracy by means of local increments. In
Eq.~(\ref{IPcorrcorr}) the term $\delta H_{ab}^{\rm corr}$
accounts for correlation in the ($N-1$)-electron system where
electrons react on the presence of the hole while $\delta E_0^{\rm
corr}$ accounts for electron correlation in the neutral system. As
the difference of these two quantities define $\delta {\rm
IP}_{ab}^{\rm corr}$ they must be evaluated consistently in the
incremental scheme.

By analogy to Eq.~(\ref{Eincr}) for the correlation correction to
the ground-state energy the incremental approach to $\delta {\rm
IP}_{ab}^{\rm corr}$ can be written as

\begin{equation}
\label{IPincr}
\delta {\rm IP}_{ab}^{\rm corr} = \Delta {\rm
IP}_{ab}^{\rm corr}() + \sum_{c\neq a,b} \Delta {\rm IP}_{ab}^{\rm
corr}(c) + \sum_{c<d \neq a,b} \Delta {\rm IP}_{ab}^{\rm
corr}(c,d) +\ldots
\end{equation}
where in parentheses we denote those additional bonds which are
"open" for correlation together with the matrix element defining
bonds $a$ and $b$. Below we explain the performance of the
approach separately for diagonal and off-diagonal elements.

The first order of approximation to the diagonal element $\delta
{\rm IP}_{aa}^{\rm corr}$ is obtained when only bond $a$ is open
for correlation. This means that in the ($N-1$)-electron system
one electron in bond $a$ is destroyed and the second electron from
this bond can be excited to the virtual orbitals thus forming
singly excited configurations in the expansion (\ref{Psi_i^corr})
of the reference wavefunction:

\begin{equation}
\label{Psi_a^corr} \Psi_a^{\rm corr} = \alpha_a(a) \Phi_a +
{\sum_{r}} \alpha_{a \bar{a}}^{\; \; r} (a) \Phi_{a\bar{a}}^{\; \;
r}
\end{equation}
with energy $E_a^{\rm corr}$. In this case the effective
Hamiltonian is a $1\times 1$ matrix whose element is equal to
$E_a^{\rm corr}$. The corresponding energy of the neutral (i.e.
$N$-electron) system is $E_0^{\rm corr}(a)$. The first-order
approximation to $\delta {\rm IP}_{aa}^{\rm corr}$ thus is

\begin{equation}
\label{IPincr1} \Delta {\rm IP}_{aa}^{\rm corr}() = \delta {\rm
IP}_{aa}^{\rm corr}() = H_{aa}^{\rm eff}() - H_{aa} -\bigl[
E_0^{\rm corr}(a) - E_0 \bigr] = {\rm IP}_{aa}^{\rm corr}() - {\rm
IP}_{aa}.
\end{equation}

The second-order approximation is obtained when the contributions
to $\delta {\rm IP}_{aa}^{\rm corr}$ from each of the remaining
($n_{\rm occ}-1$) bonds individually are summed up, where the
corrections are defined relative to $\Delta {\rm IP}_{aa}^{\rm
corr}()$,

\begin{equation}
\label{D_IPincr2}
\Delta {\rm IP}_{aa}^{\rm corr}(c) = \delta
{\rm IP}_{aa}^{\rm corr}(c) - \Delta {\rm IP}_{aa}^{\rm corr}()
\end{equation}
The first term in the right-hand side of Eq.~(\ref{D_IPincr2}) is

\begin{equation}
\label{d_IPincr2}
\delta {\rm IP}_{aa}^{\rm corr}(c) =
H_{aa}^{\rm eff}(c) - H_{aa} -\bigl[ E_0^{\rm corr}(a,c) -
E_0 \bigr]  = {\rm IP}_{aa}^{\rm corr}(c) - {\rm IP}_{aa}.
\end{equation}
In Eq.~(\ref{d_IPincr2}) the effective Hamiltonian element
$H_{aa}^{\rm eff}(c)$ stems from a $2 \times 2$ matrix defined in
the configurations $\Phi_a$ and $\Phi_c$. At this point we would
like to notice that due to the decay of the correlation effect
with the distance between the correlated electrons one finds a
rapid decrease of the increments $\Delta {\rm IP}_{aa}^{\rm
corr}(c)$ when the distance between the orbitals $a$ and $c$
increases. This ensures that one can indeed use finite clusters to
evaluate the correlation corrections to the local matrix elements
defined in a crystal.

It is obvious that summing over all second-order increments does
not yet give a correct estimate of $\delta {\rm IP}_{aa}^{\rm
corr}$ as correlation between electrons in different additional
bonds ($c$ and $d$) is missing. The next order of approximation is
obtained when we account for the corresponding non-additivity
corrections

\begin{equation}
\label{D_IPincr3}
\Delta {\rm IP}_{aa}^{\rm corr}(c,d) = \delta
{\rm IP}_{aa}^{\rm corr}(c,d) - \Delta {\rm IP}_{aa}^{\rm corr}()
- \bigl[ \Delta {\rm IP}_{aa}^{\rm corr}(c) + \Delta {\rm IP}_{aa}^{\rm
corr}(d) \bigr].
\end{equation}
Here the correlation correction $\delta {\rm IP}_{aa}^{\rm
corr}(c,d)$ to the diagonal matrix element IP$_{aa}$ is calculated
with three bonds ($a$, $c$ and $d$) being open:

\begin{equation}
\label{d_IPincr3}
\delta {\rm IP}_{aa}^{\rm corr}(c,d) =
H_{aa}^{\rm eff}(c,d) - H_{aa} -\bigl[ E_0^{\rm corr}(a,c,d) -
E_0 \bigr]  = {\rm IP}_{aa}^{\rm corr}(c,d) - {\rm IP}_{aa}.
\end{equation}
The hole is delocalized within these three bonds and thus three
correlated ($N-1$)-electron states are obtained. This procedure
can be continued further order by order. However, as increments
are expected to decrease rapidly with increasing order (see
Ref.~\cite{Graef97}, \cite{Albrecht00} and \cite{Albrecht02}), the
series (\ref{IPincr}) can be truncated after the third term.

The correlation corrections to off-diagonal (or hopping) matrix
elements ${\rm IP}_{ab}^{\rm corr}$ are evaluated the same way,
however, the procedure is started with the two bond $a$ and $b$
and no ground-state energies are present:

\begin{equation}
\label{IPoff_incr1} \Delta {\rm IP}_{ab}^{\rm corr}() = \delta
{\rm IP}_{ab}() = H_{ab}^{\rm eff}() - H_{ab} = {\rm IP}_{ab}^{\rm
corr}() - {\rm IP}_{ab}.
\end{equation}
Here the starting effective Hamiltonian matrix elements
$H_{ab}^{\rm eff}()$ already originate from a $2 \times 2$ matrix
defined in configurations $\Phi_a$ and $\Phi_b$. Next we compute
the correlation contribution of an individual additional orbital:

\begin{equation}
\label{D_IPoff_incr2}
\Delta {\rm IP}_{ab}^{\rm corr}(c) = \delta
{\rm IP}_{ab}^{\rm corr}(c) - \Delta {\rm IP}_{ab}^{\rm corr}()
\end{equation}
where

\begin{equation}
\label{d_IPoff_incr2}
\delta {\rm IP}_{ab}^{\rm corr}(c) =
H_{ab}^{\rm eff}(c) - H_{ab}  = {\rm
IP}_{ab}^{\rm corr}(c) - {\rm IP}_{ab}.
\end{equation}
In this case the hole is delocalized over a triple of bonds ($a$,
$b$ and $c$). Summing over $c$ we get the second term in the
series (\ref{IPincr}). The third-order increments are

\begin{equation}
\label{D_IPoff_incr3}
\Delta {\rm IP}_{ab}^{\rm corr}(c,d) = \delta
{\rm IP}_{ab}^{\rm corr}(c,d) - \Delta {\rm IP}_{ab}^{\rm corr}()
- \bigl[ \Delta {\rm IP}_{ab}^{\rm corr}(c) + \Delta {\rm IP}_{ab}^{\rm
corr}(d) \bigr]
\end{equation}
where correlated matrix elements $\delta {\rm IP}_{ab}^{\rm
corr}(c,d)$ are obtained when four bonds are open for correlations
always including the bonds $a$ and $b$:

\begin{equation}
\label{d_IPoff_incr3}
\delta {\rm IP}_{ab}^{\rm corr}(c,d) =
H_{ab}^{\rm eff}(c,d) - H_{ab}  =
{\rm IP}_{ab}^{\rm corr}(c,d) - {\rm IP}_{ab}.
\end{equation}
Again, the series (\ref{IPincr}) can be truncated after the third
term.

In the present work this approach is used to obtain good estimates
for the correlation corrections of the local matrix elements
IP$_{ab}$. For the $\sigma$ bonds a C$_6$H$_8$ molecule is used.
Due to the truncation of the series (\ref{IPincr}) in each CI
calculation we have at most four bonds open for correlation which
keeps the calculations reasonably cheap. However, even in such a
small molecule as C$_6$H$_8$ the number of all third-order
increments is 78 (all pair combinations among 13 bonds) for the
diagonal elements IP$_{aa}$ and 66 for the off-diagonal elements
(2 out of 12 combinations). Dozens of input files have to be
prepared for the correlation calculations and the results of these
calculations have to be gathered and summed up properly. As
discussed in the section 3.3 we need much larger molecules to
account for the correlation corrections to local matrix elements
IP$_{ab}$ with $a$ and $b$ being $\pi$ bonds and also for selected
EA$_{rs}$ matrix elements. The number of increments increases
quadratically with the size of the cluster (i.e. with the number
of bonds). Therefore, the incremental scheme introduced above in
terms of bond contributions becomes unfeasible because of the huge
number of increments though each increment individually can be
evaluated by the CI method at low cost.

\subsection{Formulation in terms of bond groups}

To proceed in the case of big clusters we propose to reformulate
the incremental scheme in terms of {\it groups} of orbitals. In
this approach to the correlation corrections of the matrix element
IP$_{ab}$ we subdivide all bonds except of $a$ and $b$ in groups
$g_1$, $g_2$, $\ldots$ consisting of a couple of bonds each
($n_{g_i} \geq 1$, $n_{g_i}$ being the number of bonds in the
$i$-th group $g_i$). It is natural that the bonds in a particular
group $g_i$ belong to the same unit cell (or two adjacent unit
cells) and that the groups do not overlap. Then, in
Eq.~(\ref{IPincr}) we replace the single orbitals $c$ and $d$
in parentheses by entire orbital groups $g_i$ and $g_j$ and sum
over group indices $i$ and $j$:

\begin{equation}
\label{IPincr_group}
\delta {\rm IP}_{ab}^{\rm corr} =
\Delta {\rm IP}_{ab}^{\rm corr}() +
\sum_{i} \Delta {\rm IP}_{ab}^{\rm corr}(g_i) +
\sum_{i<j} \Delta {\rm IP}_{ab}^{\rm corr}(g_i,g_j) +\ldots
\end{equation}
In the case of a diagonal element, IP$_{aa}$ say, the other
orbital $b$ forms its own group and has to be included in the
summation over the groups $g_i$ in Eq.~(\ref{IPincr_group}).
Consequently, we replace the bonds $c$ and $d$ in
Eqs.~(\ref{D_IPincr2})--(\ref{d_IPincr3}) and
(\ref{D_IPoff_incr2})--(\ref{d_IPoff_incr3}) by groups $g_i$ and
$g_j$. Thus, the increments in the series (\ref{IPincr_group})
become

\begin{equation}
\label{D_IP_incr2_group}
\Delta {\rm IP}_{ab}^{\rm corr}(g_i) =
\delta {\rm IP}_{ab}^{\rm corr}(g_i) -
\Delta {\rm IP}_{ab}^{\rm corr}()
\end{equation}
and

\begin{equation}
\label{D_IP_incr3_group}
\Delta {\rm IP}_{ab}^{\rm corr}(g_i,g_j) = \delta
{\rm IP}_{ab}^{\rm corr}(g_i,g_j) - \Delta {\rm IP}_{ab}^{\rm corr}()
- \bigl[ \Delta {\rm IP}_{ab}^{\rm corr}(g_i) + \Delta {\rm IP}_{ab}^{\rm
corr}(g_j) \bigr]
\end{equation}
The substitution of a group $g_i$ instead of a single bond $c$ in
(\ref{D_IPincr2}) physically means that we account for all the
contribution to the correlation correction $\delta {\rm
IP}_{ab}^{\rm corr}$ which are associated with electrons from the
{\it whole} group of bonds $g_i$ instead of a single electron from
bond $c$:

\begin{equation}
\label{d_IP_incr2_group}
\delta {\rm IP}_{ab}^{\rm corr}(g_i) =
H_{ab}^{\rm eff}(g_i) - H_{ab} -
\delta_{ab} \bigl[ E_0^{\rm corr}(a,g_i) - E_0 \bigr] =
{\rm IP}_{ab}^{\rm corr}(g_i) - {\rm IP}_{ab}.
\end{equation}
To get the matrix element $H_{ab}^{\rm eff}(g_i)$ of the effective
Hamiltonian, ($n_{g_i}+2-\delta_{ab}$) correlated hole states have
to be calculated explicitly. Also for diagonal elements the
correlation correction to the ground-state energy is obtained when
all bonds of the group $g_i$ together with the bond $a$ are open
for correlation simultaneously and the rest of the bonds of the
cluster are frozen. Summing up the correlation contributions
associated with the individual {\it groups} one gets a first
estimate of the response of the cluster to the presence of a hole.
Note that this first-order estimate $\delta {\rm IP}_{ab}^{(1)}$
contains much more correlation contributions than the sum over all
{\it single bond} increments because intragroup correlation is
fully included. This result is improved when the non-additivity
corrections $\Delta {\rm IP}_{ab}^{\rm corr}(g_i,g_j)$ for all
{\it pairs of groups} are summed up. The scheme can be continued
by adding increments of higher orders. However, as group
increments also decrease rapidly with the order, the series
(\ref{IPincr_group}) can be terminated after the third term to get
rather accurate estimate for the correlation correction to the LME
IP$_{ab}$. The incremental scheme in terms of bond groups, as it
is introduced in this subsection, was used in our approach to
evaluate the correlation correction to the matrix element IP$_{\pm
a,{\rm pi} \: {\rm pi}}$ from Table~\ref{SCF_IPp}. In that case
groups of $\sigma$ bonds were composed. The corresponding data on
the increments are given in chapter~4.1.

Grouping the bonds substantially reduces the numbers of increments
in (\ref{IPincr_group}) as compared to (\ref{IPincr}) however to
the price of more expensive CI calculations for each increment in
(\ref{IPincr_group}). In the latter case up to ($n_{g_i} + n_{g_j}
+2$) bonds can be open for correlation simultaneously and the same
number of hole states have to be calculated to get the matrix
element $H_{ab}^{\rm eff}(g_i,g_j)$ for a particular increment
$\Delta {\rm IP}_{ab}^{\rm corr}(g_i,g_j)$.

\subsection{Incremental scheme for attached-electron states}

The method of local increments can be also used for the evaluation
of the correlation corrections to the matrix elements EA$_{rs}$.
In this case ($2n_{\rm bond}+1$) electrons have to be correlated
and the additional electron is delocalized over all antibonds of
the chosen molecule. To reduce the computational efforts for the
MRCI calculations, one can additionally freeze the majority of
antibonds when calculating a particular increment forbidding
electrons to occupy the frozen antibonds. This way one reduces i)
the number of calculated ($N+1$)-electron states and ii) the
external space. At the same time only electrons from the bonds
belonging to one or two groups are open for correlation
simultaneously. (Here we need to group the bonds again, because
large clusters are required to account for correlation effects in
the ($N+1$)-electron systems.) In the incremental scheme the
extension of the external space (i.e. the opening of antibonds)
and the opening of the bonds must be done consistently. Otherwise
the increment series will converge poorly. Thus, we find it
reasonable to group the virtual localized orbitals centered on one
bond together with the occupied orbitals from this bond (or
adjacent bonds). The general formula for correlation correction to
matrix element EA$_{rs}$ by means of local increments is analogous
to Eq.~(\ref{IPincr_group}):

\begin{equation}
\label{EAincr_group} \delta {\rm EA}_{rs}^{\rm corr} = \Delta {\rm
EA}_{rs}^{\rm corr}() + \sum_{i} \Delta {\rm EA}_{rs}^{\rm
corr}(g_i) + \sum_{i<j} \Delta {\rm EA}_{rs}^{\rm corr}(g_i,g_j)
+\ldots
\end{equation}
where

\begin{equation}
\label{D_EA_incr2_group}
\Delta {\rm EA}_{rs}^{\rm corr}(g_i) =
\delta {\rm EA}_{rs}^{\rm corr}(g_i) - \Delta {\rm EA}_{rs}^{\rm
corr}()
\end{equation}
and

\begin{equation}
\label{D_EA_incr3_group}
\Delta {\rm EA}_{rs}^{\rm corr}(g_i,g_j)
= \delta {\rm EA}_{rs}^{\rm corr}(g_i,g_j) - \Delta {\rm
EA}_{rs}^{\rm corr}() - \bigl[ \Delta {\rm EA}_{rs}^{\rm
corr}(g_i) + \Delta {\rm EA}_{rs}^{\rm corr}(g_j) \bigr].
\end{equation}
Here indices $r$ and $s$ denote two groups $g_r$ and $g_s$ each
containing the relevant antibond plus corresponding bond (or few
adjacent bonds). Again for the diagonal element, EA$_{rr}$ say,
the summation over the groups in Eq.~(\ref{EAincr_group}) includes
the group $g_s$ of the other antibond.

The first-order increment

\begin{equation}
\label{EAincr1} \Delta {\rm EA}_{rs}^{\rm corr}() = -\bigl[
H_{rs}^{\rm eff}() - H_{rs} \bigr] + \delta_{rs} \bigl[ E_0^{\rm
corr}(g_r) - E_0 \bigr] = {\rm EA}_{rs}^{\rm corr}() - {\rm
EA}_{rs}
\end{equation}
contains the effective Hamiltonian defined in configurations
$\Phi_r$ and $\Phi_s$, all occupied orbitals from the groups $g_r$
and $g_s$ are open for constructing the excited configurations in
the MRCI ansatz for the correlated wavefunctions
(\ref{Psi_k^corr}) and all antibonds except of $r$ and $s$ are
forbidden for the excited electrons. For the diagonal element
($s=r$) only one group is considered in Eq.~(\ref{EAincr1}). The
correlated energy of the $N$-electron state $E_0^{\rm corr}(g_r)$
is calculated with only bonds of the group $g_r$ being open.

When we add individual groups for the second-order increments and
pairs of groups for the third-order increment we simultaneously
open antibonds and the corresponding bonds for correlation. This
way we approach in a systematic way the value of $\delta {\rm
EA}_{rs}^{\rm corr}$ for {\it all} bonds and {\it all} antibonds
being open simultaneously, however, in each separate MRCI
calculation the number of configurations (and the number of
coefficients $\alpha$ in Eq.~(\ref{Phi0_corrCI}) and
(\ref{Psi_k^corr}) to be optimized) is considerably restricted.

\subsection{Additional features of the incremental scheme}

The incremental scheme described above has five important
additional features. The calculation of individual second-order
increments $\Delta {\rm IP}_{ab}^{\rm corr}(g_i)$ and $\Delta {\rm
EA}_{rs}^{\rm corr}(g_i)$ gives information on the decay of the
correlation effect with the distance between the bonds of a group
$g_i$ and the localized charge in bonds $a$ and $b$ (or antibonds
$r$ and $s$). This information allows one to estimate whether a chosen
cluster is big enough to account for the dominant part of the
correlation corrections.

The second feature is that one uses the symmetry of the crystal to
reduce the number of increments which have to be calculated: the
contribution of symmetry-equivalent increments to the correlation
correction is the value of one of them times the number of such
equivalent increments. This reduces the computational efforts and
works especially well for highly symmetric crystals with simple
unit cells like diamond, silicon and germanium (see
Ref.~\cite{Graef93}, \cite{Graef97}, \cite{Albrecht00} and
\cite{Albrecht02}).

The third feature: the main computational effort is directed to
get the effective Hamiltonian from which we only need one of its
matrix element for a particular increment. However, it is easy to
see that some other matrix elements of a given effective
Hamiltonian can be used for other increments without any extra
cost. For instance, the diagonal element $H_{aa}^{\rm eff}(b,c)$
and the off-diagonal element $H_{ab}^{\rm eff}(c)$ are obtained
from one and the same MRCI calculation with $\Phi_a$, $\Phi_b$ and
$\Phi_c$ being the model configurations.

The fourth feature is the possibility to use the translational
symmetry of the crystal and reduce the so-called {\it
cluster-edge} error by this. Such errors arise when one of the
bonds open for correlation is close to the edge of a cluster and
this localized orbital is affected by the lack of supporting atoms
(i.e. the energy of the electron in this orbital differs
noticeably from the energy of the electron in a
translation-equivalent orbital in the central part of the
cluster). Then, for the evaluation of this particular increment
one can open a translational-equivalent set of orbital groups for
correlation which does not contain orbitals at the edge of the
cluster if the size of the cluster allows one to do so. Let's say
that we want to get the contribution to the correlation correction
to the diagonal matrix element IP$_{\rm pi,pi}$ associated with
the additional bond "pi,-2" in the molecule C$_{10}$H$_{12}$ which
is shown in Fig.~\ref{c10h12}. We know that orbital "pi,-2" is not
equivalent to orbital "pi,-1" as one more unit cell from the left
side is needed to reproduce properly the occupied Wannier $\pi$
orbital centered on the bond "pi,-2" (see Fig.~\ref{occ_pi}). Also
we expect that the increment $\Delta {\rm IP}_{\rm pi,pi}^{\rm
corr}({\rm pi},-2)$ is still relatively large as orbitals "pi" and
"pi,-2" overlap as can be understood from Fig.~\ref{occ_pi}.
Therefore, this increment has to be evaluated when two equivalent
bonds with proper crystalline surrounding are open for
correlations and in the series (\ref{IPincr}) we substitute it by
the translation-equivalent increment $\Delta {\rm IP}_{\rm
pi1,pi1}^{\rm corr}({\rm pi},-1)$. This way we reduce the
influence of the edge of the cluster on the correlation
corrections to the local matrix elements defined for the periodic
infinite systems.

The fifth feature becomes important when one uses a correlation
method which suffers from size-consistency problems. This is
particularly the case when truncated CI methods are used. When the
second- and third-order increments are calculated, correlation
corrections obtained with different numbers of open bonds are
subtracted from each other. The difference itself may easily be of
the order of the size-consistency error (better to say
size-extensivity error) for each term entering this difference.
Then, by the incremental scheme, we would sum up these errors
which would accumulate and finally lead to wrong results.
Therefore, one needs an accurate correction to minimize the
size-extensivity error when the MRCI(SD) method is used for the
evaluation of the increments. Such a correction is the topic of
the next section.

\section{Size-extensivity correction}

The occurrence of a size-extensivity error in truncated CI
calculations can easily be understood from the example of two
H$_2$ molecules at large distance from each other. Each molecule
has one bond (an orbital obtained in the Hartree--Fock
approximation) occupied by two electrons. Let us evaluate the
correlation correction $\delta E_0^{\rm corr}(2H_2)$ to the HF
ground-state energy $E_0(2H_2)$ of this system by the single and
double configuration interaction method (for shortness we will call
here this correction as {\it correlation energy}). We can do this in two
ways. Firstly, we can calculate the correlation energies for each
H$_2$ molecule separately, $\delta E_0^{\rm corr}(H_2)$, and add
them. This approximation gives exact result since on big distances
between the H$_2$ molecules excited configurations with electrons
being destroyed in the bond of one molecule and created in some
virtual orbital of the second H$_2$ molecule give no contribution
to the correlation energy (dissociation limit). Alternatively, we
can directly calculate the correlation energy of the whole system
of two H$_2$ molecules. Due to the truncation of the CI expansion
(\ref{Phi_corrCI}) we have more configurations in the first case
than in the second one: for instance, double excitations on both
molecules simultaneously are neglected in the second case while
they are taken into account in the first case. Therefore, the
correlation energy obtained in the whole system $\delta E_0^{\rm
corr}(2H_2)$ is higher than the sum of the two correlation
energies of the separate H$_2$ molecule $2\delta E_0^{\rm
corr}(H_2)$. The error in the CI(SD) correlation energy of the
system consisting of $n$ H$_2$ molecules increases with increasing
$n$ (see the third column in Table~\ref{correctionE0}). Thus, when
one investigates correlation energy of systems with different
number of bonds open for the correlation one has to correct for
the size-extensivity error.

Explicitly this situation arises when one uses the incremental
scheme to evaluate correlation energies. There, correlation
energies of the system with different numbers of correlated bonds
are subtracted from each other. Therefore, all the energy values
defining a particular increment must be consistent.

The general strategy to obtain size-extensive correlation energy
is to estimate the value $n \times \delta E_0^{\rm corr}(H_2)$ by
means of information on the wavefunction and the correlation
energy $\delta E_0^{\rm corr}(nH_2)$ of the whole system. There
exist two well-known size-extensivity corrections to the
correlation energy of the ground state calculated by the CI method
which are widely used in quantum chemistry.

The first is the Davidson correction \cite{Davidson74} which is
the simplest estimate of the size-extensivity error in CI(SD)
calculations:

\begin{equation}
\label{Davidson} \Delta E_{\rm D}^{\rm SE} = \delta E_0^{\rm
corr}(1-\alpha_0^2).
\end{equation}
Here $\alpha_0$ is the coefficient of the SCF ground-state
configuration $\Phi$ in (\ref{Phi0_corrCI}) when the correlated
wavefunction $\Phi^{\rm corr}$ is {\it normalized}. The Davidson
correction does not depend explicitly on the number of correlated
bonds. It gives a rather accurate estimate of correlation energy
for $n$ ranging from 4 to 10 (the residual error is within 1\% as
pointed out in \cite{Szabo}, p.268) however the error is larger
for $n=2$ or 3 (see Table~\ref{correctionE0}) and it diverges for
large numbers of open bonds. Since in our incremental calculations
we have dozens of increments the error may accumulate up to 100\%
of the correlation energy. Thus, for our purposes we need a more
accurate size-extensivity correction.

The Pople correction \cite{Pople77} estimates the size-extensivity
error much better than the Davidson correction (see
Table~\ref{correctionE0}) giving perfect size-extensive
correlation energy in the case of any number of equivalent
non-interacting two-level systems:

\begin{equation}
\label{Pople} \Delta E_{\rm P}^{\rm SE} = \delta E_0^{\rm corr}
\Biggl( \frac{\sqrt{n^2 + n \tan^2 2\theta} - n}{\sec 2\theta -1}
-1 \Biggr)
\end{equation}
where $\theta = \arccos \alpha_0$.

\begin{figure}
\centerline{ \psfig{figure=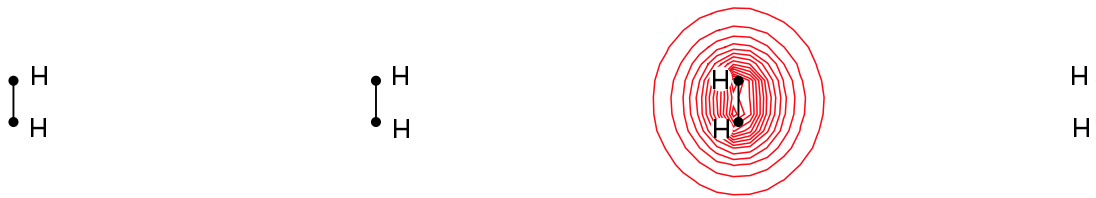,width=5.8in} }
\caption{Four H$_2$ molecules at large distances. The contour plot
of the bonding HF orbital of one H$_2$ molecule is shown.}
\label{4H2}
\end{figure}

We have investigated the reduction of the size-extensivity
error of the CI(SD) method in evaluating the correlation
correction to the ground-state energy when the two error
corrections are used. As a test system we have chosen four H$_2$
molecules in the arrangement shown in Fig.~\ref{4H2}. In the
dissociation limit the distance between nearest-neighbor molecules
(20~{\AA}) is much larger than the distance between the two H
atoms in a molecule (the characteristic bond length in H$_2$
molecule is 0.7414~{\AA} which is taken from
Ref.~{\cite{Holleman}, p.~52). The basis set for hydrogen used
here is the same as in Table~\ref{basis}. The correlation energy
of the system with $n$ correlated H$_2$ molecules ($n=2$, 3 and 4)
obtained by the CI(SD) method is compared in
Table~\ref{correctionE0} to the size-extensive one which is equal
to $n$ times the correlation energy of a single H$_2$ molecule.

\begin{table}
\refstepcounter{table} \addtocounter{table}{-1}
\label{correctionE0} \caption{Comparison of the size-extensivity
errors of the CI(SD) method with and without error corrections.
The correct size-extensive correlation energy for different
numbers $n$ of correlated H$_2$ molecules is presented in the
second column (in eV). The correlation energy of three different
methods (pure CI(SD), CI(SD) with Davidson correction and CI(SD)
with Pople correction) are presented in the three next columns as
deviation from the correct correlation energy.} \vspace{5mm}
\centerline{
\renewcommand{\baselinestretch}{1.2}\normalsize
\begin{tabular}{ccccc}
\hline
 $n$ &
 $n\; \delta E_0^{\rm corr}$(H$_2$) &
 $\delta E_0^{\rm corr}(n{\rm H}_2)$ &
 $\delta E_0^{\rm corr}(n{\rm H}_2) + \Delta E_{\rm D}^{\rm SE}$ &
 $\delta E_0^{\rm corr}(n{\rm H}_2) + \Delta E_{\rm P}^{\rm SE}$ \\
\hline
 2 & -2.063 & 1.721\% & -1.706\% & 0.008\% \\
 3 & -3.094 & 3.318\% & -1.533\% & 0.031\% \\
 4 & -4.126 & 4.806\% & -1.318\% & 0.064\% \\
\hline
\end{tabular}
}
\end{table}
\renewcommand{\baselinestretch}{1}\normalsize

The Pople correction reduces the size-extensivity error in average
by two order of magnitudes as compared to pure CI(SD) or Davidson
corrected CI(SD) which both yield error of more than 1\% (see
Table~\ref{correctionE0}). Thus, we use the Pople correction to
CI(SD) correlation energies to get size-extensive results for the
neutral closed-shell systems. (More detailed study of the
performance of these two corrections for different systems can be
found in e.g. Ref.~\cite{Pople77}, \cite{Langhoff87},
\cite{Gdanitz88} and \cite{Duch94}). Unfortunately, being applied
to the correlation energies of open-shell systems (as the ($N-1$)-
and ($N+1$)-particle systems we need to consider) the two
size-extensivity corrections do not yield satisfactorily accurate
results (see Table~\ref{correctionEa}) to be used in the method of
local increments. Therefore, we need a size-extensivity error
correction which gives accurate estimates of the correlation
energy in the case of open-shell systems (systems with a hole or
with captured electron) like the Pople correction does for
closed-shell systems.

Let us consider again the system consisting of $n$ equivalent
H$_2$ molecules at large distances and let us consider a ($2 \:
n-1$)-electron state with one electron annihilated at some
molecule $a$. We can get the correlation energy of this state on
the MRCI(SD) level but we would like to approach the correct
size-extensive correlation energy

\begin{equation}
\label{EcorrH+} \delta E_{a}^{\rm SE} = (n-1)\; \delta E_0^{\rm
corr} ({\rm H}_2) + \delta E_a^{\rm corr} ({\rm H}_2^+)
\end{equation}
having information on the correlation energy and the wavefunction
from the MRCI(SD) calculation of the whole system. In
Eq.~(\ref{EcorrH+}) $\delta E_0^{\rm corr} ({\rm H}_2)$ denote the
correlation energy of a neutral H$_2$ molecule and $\delta
E_a^{\rm corr} ({\rm H}_2^+)$ the one of the molecule carrying the
hole. The latter is in fact just the orbital relaxation energy.

Like in the derivation of the Pople correction \cite{Pople77} we
switch to a minimal basis description such that the H$_2$ molecule
becomes a two-level system. The spin-adapted reference
configurations for the MRCI ansatz for our system consists of all
hole configurations $\{ \Phi_a \}_{a=1, \ldots , n}$ with $a=1,
\ldots , n$ being the occupied spin-down orbitals on the different
H$_2$ molecules. In the dissociation limit the only 2-hole
1-particle configurations $\Phi_{bc}^{\; \; r}$ which couple to a
given model configuration $\Phi_a$, i.e. for which the CI matrix
element $\langle \Phi_{a} | H | \Phi_{bc}^{\; \; r} \rangle$ is
non-zero, are the "vertical" single excitations $\Phi_{ab}^{\; \;
b^\ast}$ where $b^\ast$ labels the unoccupied counter part of the spin
orbital $\varphi_b$. Because of Brillouin's theorem $\langle \Phi_{a} | H
| \Phi_{ab}^{\; \; b^\ast} \rangle$ vanishes if the spin orbital
$\varphi_b$ is located on a different molecule than $a$. Hence, only the
configuration $\Phi_{aa}^{\; \; a^\ast}$ couples to $\Phi_{a}$.

Similarly, the only 3-hole 2-particle configurations
$\Phi_{b \: cd}^{\; \; rs}$ which couple to a given model
configuration $\Phi_{a}$ are the "vertical" double excitations
$\Phi_{a \: b\bar{b}}^{\; \: b^\ast \bar{b}^\ast}$ on sites different
than the one of the spin orbital $\varphi_a$. Furthermore, the
configuration space $\{ \Phi_a \}_{a=1, \ldots , n} \oplus \{
\Phi_{a\bar{a}}^{\; \; \bar{a}^\ast} \}_{a=1, \ldots , n} \oplus
\{ \Phi_{a \: b\bar{b}}^{\; \: b^\ast \bar{b}^\ast} \}_{a \neq b =1,
\ldots , n}$ is closed in the sense that {\it no} other 2-hole
1-particle or 3-hole 2-particle configuration couple to any of the
configurations in that space. Hence, (in the dissociation limit)
we can write the MRCI wavefunction of our system as

\begin{equation}
\label{Psi_corr_SE_A} \Psi_i^{\rm corr} = \sum_{a=1}^n \alpha_a(i)
\Phi_a + \sum_{a=1}^n \alpha_{a \bar{a}}^{\; \; \bar{a}^\ast}(i)
\Phi_{a \bar{a}}^{\; \; \bar{a}^\ast}(i) + \sum_{a \neq b =1}^n
\alpha_{ab \bar{b}}^{ \; b^\ast \bar{b}^\ast}(i) \Phi_{ab
\bar{b}}^{ \; b^\ast \bar{b}^\ast}(i).
\end{equation}

For the sake of simplicity let us denote $\Phi_a$ by $\Phi_a^0$,
$\Phi_{a \bar{a}}^{\; \; \bar{a}^\ast}$ by $\Phi_a^1$,
$\Phi_{a \: b \bar{b}}^{\: \; b^\ast \bar{b}^\ast}$ by 
$\Phi_a^2,\ldots$. Then,
the CI matrix elements for the MRCI ansatz (\ref{Psi_corr_SE_A})
can be written as

\begin{equation}
C_{ab}^{xy} = \langle \Phi_{a}^x | H | \Phi_{b}^{y} \rangle.
\end{equation}
In the dissociation limit the CI matrix becomes block-diagonal,
i.e.

\begin{equation}
C_{ab}^{xy} = \delta_{ab}C_{aa}^{xy} \nonumber
\end{equation}
with all subblocks $\{ C_{aa}^{xy} \}_{x,y=0, \ldots , n}$ being
of the form

\begin{equation}
\left(
\begin{matrix}
E_{1h} & \beta  & \gamma & \cdots & \gamma \\
\beta  & E_{2h} & 0      & \cdots & 0      \\
\gamma & 0      & E_{3h} & \cdots & 0      \\
\vdots & \vdots & \vdots & \ddots & \vdots \\
\gamma & 0      & 0      & \cdots & E_{3h} \\
\end{matrix}
\right).
\end{equation}
The entries $E_{1h}$, $E_{2h}$, $E_{3h}$, $\beta$ and $\gamma$ do
not depend on the position of the hole in the H$_2$ chain which
implies that our ($2n-1$)-electron system of $n$H$_2$ molecule has
$n$ degenerate lowest hole states each of them residing on one of
the H$_2$ molecules. Delocalization of the hole over the chain
does not lead to a gain in energy.

The matrix elements $\gamma = \langle \Phi_{a} | H | \Phi_{a \: b
\bar{b}}^{\: \; b^\ast \bar{b}^\ast} \rangle = \langle \varphi_b
\bar{\varphi}_b | v_{12} | \varphi_b^\ast \bar{\varphi}_b^\ast
\rangle$ are identical for all $b \neq a$ since all the neutral
H$_2$ molecules are equivalent (in the dissociation limit). For
that reason the CI coefficients $\alpha_{a \: b \bar{b}}^{\: \; b^\ast
\bar{b}^\ast}(i)$ must all be equal, and the wavefunction
for a hole state on the $a$-th H$_2$ molecule (with energy
$E_a^{\rm corr}$) can be written as

\begin{equation}
\label{Psi_corr_SE2} \Psi_a^{\rm corr} = \Phi_a + c_1 \Phi_a^1 +
c_2 \sum_{x=2}^n \Phi_b^x.
\end{equation}
As a consequence the secular equation to determine the coefficients $c_1$ and
$c_2$ reduces to

\begin{equation}
\label{matrix_equation}
\left(
\begin{array}{ccc}
0      & \beta     & (n-1)\gamma \\
\beta  & 2\Delta_1 & 0           \\
\gamma & 0         & 2\Delta_2   \\
\end{array}
\right)
\left(
\begin{array}{c}
1\\ c_1 \\c_2
\end{array}
\right)
=\delta E_a^{\rm corr}
\left(
\begin{array}{c}
1\\ c_1 \\c_2
\end{array}
\right)
\end{equation}
where $\delta E_a^{\rm corr}= E_a^{\rm corr}-E_a$ is the
correlation energy, and $2 \Delta_1 = E_{2h} - E_a$ and $2
\Delta_2 = E_{3h} - E_a$ are the energies of the 2-hole 1-particle
and the 3-hole 2-particle configurations $\Phi_{a \bar{a}}^{\; \;
\bar{a}^\ast}$ and $\Phi_{a \: b \bar{b}}^{\: \; b^\ast \bar{b}^\ast}$,
respectively, relative to the Hartree--Fock energy $E_{1h} \equiv
E_a$ of the model configuration $\Phi_a$.
From the first row one concludes that the correlation energy
consists of two parts: $\delta E_a^{\rm corr}= c_1 \beta + (n-1)
c_2 \gamma = E_s + E_d$ where the part $E_s$ ($E_d$) is associated
with singly (doubly) excited configurations only. The quantities
$E_s$ and $E_d$ (and also $c_s \equiv c_1$  and $c_d \equiv
\sqrt{n-1}\; c_2$) can be obtained from the output of an
MRCI(SD) calculation with the MOLPRO program packadge \cite{MOLPRO}, 
\cite{Werner88}, \cite{Knowles88}, \cite{Knowles92}
and our aim is to
express the size-extensive correlation energy in terms of these
parameters.

As can be seen from Eq.~(\ref{EcorrH+}) the size-extensive 
correlation energy of the considered system
is equal to the correlation energy the of charged H$_2^+$ molecule
plus ($n-1)$ times the correlation energy of a separated neutral
H$_2$ molecule. To find the required energies 
$\delta E_a^{\rm corr}({\rm H}_2^+)$ and
$\delta E_0^{\rm corr}({\rm H}_2)$
of the two isolated molecules
(charged and neutral) we note, that one gets the following CI matrices 
when the isolated molecules are treated analogously to the
whole system:

\begin{equation}
\label{matrix_equationH2+}
\left(
\begin{array}{cc}
0     & \beta     \\
\beta & 2\Delta_1 \\
\end{array}
\right)
\left(
\begin{array}{c}
1\\ c_1^\prime
\end{array}
\right)
= \delta E_a^{\rm corr}({\rm H}_2^+)
\left(
\begin{array}{c}
1\\ c_1^\prime
\end{array}
\right)
\end{equation}
and

\begin{equation}
\label{matrix_equationH2}
\left(
\begin{array}{cc}
0      & \gamma    \\
\gamma & 2\Delta_2 \\
\end{array}
\right)
\left(
\begin{array}{c}
1\\ c_2^\prime
\end{array}
\right)
=\delta E_0^{\rm corr}({\rm H}_2)
\left(
\begin{array}{c}
1\\ c_2^\prime
\end{array}
\right)
\end{equation}
with the entries $\beta$, $\Delta_1$, $\gamma$ and $\Delta_2$ 
being the {\it same} as in Eq.~(\ref{matrix_equation}).
From these two equations the correlation energies
of the charged and neutral H$_2$ molecules can
easily be derived yielding:

\begin{equation}
\label{dE_aH2+}
\delta E_a^{\rm corr}({\rm H}_2^+) =
\Delta_1 -\sqrt{\Delta_1^2 + \beta^2}
\end{equation}
and

\begin{equation}
\label{dE_0H2}
\delta E_0^{\rm corr}({\rm H}_2) =
\Delta_2 -\sqrt{\Delta_2^2 + \gamma^2}.
\end{equation}
Substitution of (\ref{dE_aH2+}) and (\ref{dE_0H2}) into 
Eq.~(\ref{EcorrH+}) gives the size-extensive correlation energy
of the lowest $n$ hole states in our system of
$n$ H$_2$ molecules in terms of the parameters entering the
secular equation~(\ref{matrix_equation}):

\begin{equation}
\label{SE_a2}
\delta E_a^{\rm SE} = \Delta_1 -\sqrt{\Delta_1^2 + \beta^2} +
(n-1) \biggl( \Delta_2 -\sqrt{\Delta_2^2 + \gamma^2} \biggr).
\end{equation}
With the help of Eq.~(\ref{matrix_equation}) the parameters $\beta$, $\gamma$,
$\Delta_1$ and $\Delta_2$ can be linked to the output
quantities $\delta E_a^{\rm corr}$, $E_s$, $E_d$, $c_s$ and
$c_d$ which are available
from the MRCI(SD) calculations for the $n \times {\rm H}_2$
molecules as a whole: 

\begin{eqnarray}
\label{SE_parameters}
\beta = E_s / c_s \:, \quad \qquad & \quad & 
2\Delta_1 = \delta E_a^{\rm corr} - E_s / c_s^2
\nonumber \\
\gamma = E_d / \sqrt{n-1} \: c_s \:, & \quad &
2\Delta_2 = \delta E_a^{\rm corr} - E_d / c_d^2 \;.
\end{eqnarray}
Then, the final formula for the
size-extensive energy of the hole state reads

\begin{eqnarray}
\label{SE_a_final}
\delta E_a^{\rm SE} & = & \frac{1}{2} \Biggl[ n \; \delta E_a^{\rm corr} -
\frac{E_s}{c_s^2} - (n-1) \frac{E_d}{c_d^2} -
\sqrt{\biggl( \delta E_a^{\rm corr} - \frac{E_s}{c_s^2} \biggr)^2 +
4 \frac{E_s^2}{c_s^2}} - \nonumber \\
& & (n-1)\sqrt{\biggl( \delta E_a^{\rm corr} - \frac{E_d}{c_d^2} \biggr)^2 +
4 \frac{E_d^2}{(n-1)c_d^2}}   \; \;   \Biggr].
\end{eqnarray}
This correlation energy effectively takes into account the missing
configurations in the truncated CI expansion of the wavefunction of 
an open-shell system as a function of the number of bonds open for 
correlation. For a system more complicated than a chain of isolated
two-level systems the quantities $E_s$ ($E_d$) and $c_s$ ($c_d$) are
associated with the contribution from all single (double) excitations
to the correlation energy $E_i^{\rm corr}$ and the squared norm of
the (normalized) wavefunction $\Psi_i^{\rm corr}$, respectively, as 
is also done usually for the Davidson and Pople corrections to the MRCI
correlation energies in quantum chemistry.

\begin{table}
\refstepcounter{table}
\addtocounter{table}{-1}
\label{correctionEa}
\caption{Comparison of the size-extensivity errors of the CI(SD) method
with and without error correction (three different methods) for a
hole states. The correct size-extensive correlation energy for
different numbers $n$ of correlated H$_2$ molecules is presented in the
second column (in eV). The relative deviation of the correlation energy
of four different methods (the same as in Table~\ref{correctionE0} plus
our own correction $\delta E_a^{\rm SE}$) are presented in the four next
columns.}
\vspace{5mm}
\centerline{
\renewcommand{\baselinestretch}{1.2}\normalsize
\begin{tabular}{cccccc}
\hline
 $n$ &
 $\delta E_a^{\rm exact}$ &
 $\delta E_a^{\rm corr}(n{\rm H}_2)$ &
 $\delta E_a^{\rm corr}(n{\rm H}_2) + \Delta E_{\rm D}^{\rm SE}$ &
 $\delta E_a^{\rm corr}(n{\rm H}_2) + \Delta E_{\rm P}^{\rm SE}$ &
 $\delta E_a^{\rm SE}$ \\
\hline
 2 & -1.857 & 2.342\% & -2.176\% & 2.342\% & 0.002\% \\
 3 & -2.888 & 4.050\% & -1.749\% & 1.153\% & 0.017\% \\
 4 & -3.919 & 5.553\% & -1.400\% & 0.807\% & 0.045\% \\
\hline
\end{tabular}
}
\end{table}
\renewcommand{\baselinestretch}{1}\normalsize

To test this new size-extensivity correction for the MRCI(SD) method
we perform correlation calculations for the hole
states in our $4 \times {\rm H}_2$ system. The correlation energies
of this system size-corrected by different methods (Davidson
and Pople correction as realized in MOLPRO \cite{MOLPRO}, 
\cite{Werner88}, \cite{Knowles88}, \cite{Knowles92} and
$\delta E_a^{\rm SE}$ from Eq.~(\ref{SE_a_final})) are compared to
the correct size-extensive correlation energy
computed for a separate charged H$_2^+$ molecule and ($n-1$) neutral 
H$_2$ molecules (denoted as $\delta E_a^{\rm exact}$ for
shortness). They are presented in Table~\ref{correctionEa} for
different number of open bonds ($n=2$, 3 and 4).
The distance between nearest-neighbor H$_2$ molecules in these
calculations is set to 50~{\AA} to reach
the dissociation limit in our charged system.

From Table~\ref{correctionEa} one concludes that the new size-extensivity
correction for open-shell systems proposed in this section gives
results which are by several orders of magnitude more accurate than the convensional
corrections to the MRCI(SD) method. The accuracy obtained with
our size-extensivity correction is the same as that of the Pople
correction to correlation energy of a neutral closed-shell system. The use of
these two corrections in all our calculations for
individual increments (the first one for open-shell
and the second for closed-shell calculations) finally allowed us to reduce
the estimated error for the correlation corrections to the
local matrix elements (the target quantities of our approach)
down to a few percents of their values.

\section{Satellite states problem}

In the last section of this chapter we would like
to draw our attantion to the problem of satellite
states which may ruin the convergence of the MRCI
calculations. This problem appears when one (or more) of
model configurations $\Phi_i$ which
define the effective Hamiltonian $H_{ij}^{\rm eff}$
for some particular increment has an energy higher
than the energy $E_{jk}^{\;\; r}$ of a single
excitation $\Phi_{jk}^{\;\; r}$ of another model 
configuration $\Phi_{j}$.
This situation may happen
in systems with relatively broad bands and
a narrow band gap. For instance, in the case of
{\it trans}-polyacetylene the HF energies of hole
states range from approximately -31 to -6~eV
and the band gap is 6.42~eV. The position and
dispersion of the lowest valence band of tPA
(see Fig.~\ref{SCFband}) is mainly
defined by states with a hole in the carbon sceleton. 
In a C$_6$H$_8$ molecule the lowest localized valence
orbital is a short
C--C $\sigma$ bond and the energy of taking out
an electron from this bond is 22.3~eV.
The position and the dispersion of the highest
valence band (a $\pi$ band) in turn is
determined by states with a hole in the $\pi$ system
of tPA. In the C$_6$H$_8$ molecule the highest
localized valence orbital is a $\pi$ bond with
an orbital energy of 10.4~eV. The energy of a state
with two electrons destroyed in that $\pi$ bond
and one electron created in the lowest
unoccupied molecular orbitals (LUMO) is found to be
19.8~eV in a C$_6$H$_8$ molecule, and is
lower than the energy of the $\sigma$ hole. A
schematic sketch of this three configurations is
shown in Fig.~\ref{satellite}.

If in such a case we would perform MRCI(SD) calculations 
to determine the correlation contributions to a matrix
element IP$_{\sigma \sigma}(\pi)$, say, we would
construct the model space of two one-particle configurations
$\Phi_\pi$ and $\Phi_\sigma$ and calculate the two lowest 
correlated $(N-1)$-electron states with wavefunctions defined 
by Eq.~(\ref{Psi_i^corr}). During the CI iteration a
decrease of the CI coefficient $\alpha_\sigma (2)$ for the
model configuration
$\Phi_\sigma$ and an increase of the CI coefficient 
$\alpha_{\pi \bar{\pi}}^{\;\; \bar{\pi}^\ast} (2)$ for the
satellite configuration $\Phi_{\pi \bar{\pi}}^{\;\; \bar{\pi}^\ast}$
(which is present in the CI expansion (\ref{Psi_i^corr}))
would lead to a gain in energy and after few iterations
we would arrive at a state which is dominated by the satellite 
configuration
$\Phi_{\pi \bar{\pi}}^{\;\; \bar{\pi}^\ast}$ instead of the model
configuration $\Phi_\sigma$. The obtained correlated wavefunction does
not obey the requirement of the MRCI approximation that
reference configurations (the model configurations here) must be 
dominant in the CI
expansion of the wavefunctions. At this point the program
stops the iteration with an error message and no data for the
two correlated states are obtained. To proceed in this
situation we include the satellite configuration in the
reference space and calculate the first {\it three}
correlated states. Two hole and one satellite state will
emerge from this calculation.

\begin{figure}
\centerline{
\psfig{figure=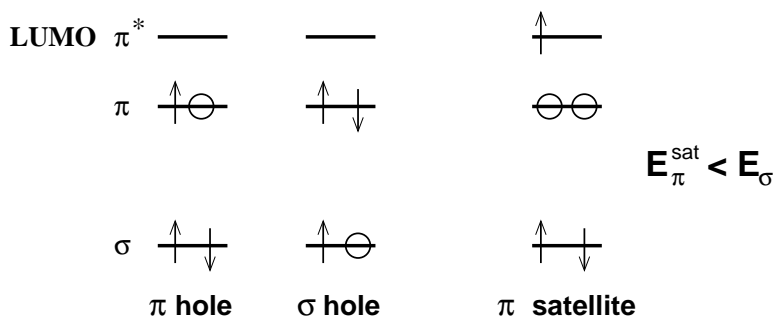,width=4in}
}
\caption{Three possible configurations for an $(N-1)$-electron
states of tPA: single hole in iether a $\pi$ or a
$\sigma$ bonds and a satellite state of $\pi$ symmetry.}
\label{satellite}
\end{figure}

The hole states can easily be separated from the satellite
state by checking the square of the norms $(\alpha_a(i))^2$ of
their projection onto the subspace spanned by the one-particle 
model configurations. In all our calculations where
satellite states show up the squares of the norms of the projected
$\sigma$ hole states are higher than 0.70 such that the 
distinction between satellite states and single hole states is 
clear. Thus, the requirement that we have to find those
wavefunctions $\Psi_i^{\rm corr}$ which are dominated by
model configurations is fulfilled.

In Table~\ref{sat1} we present the correlation data corresponding to
the case described above. In the second column the
squares of the norm of the correlated wavefunctions
projected onto the subspace spanned by model configurations
$\Phi_\pi$ and $\Phi_\sigma$ are shown. The wavefunction
of the satellite state has essentially zero projection
and is clearly more stable than the deep-lying $\sigma$
hole state.

\begin{table}
\refstepcounter{table}
\addtocounter{table}{-1}
\label{sat1}
\caption{Correlation data for two hole and one
satellite state in a C$_6$H$_8$ molecule.
Correlation energies are given with respect to the HF ground-state
energy of the molecule.}
\vspace{5mm}
\centerline{
\renewcommand{\baselinestretch}{1.2}\normalsize
\begin{tabular}{p{3cm}cc}
\hline
state &  \;  $|\Psi_i|^2$ projected  \; & $E_i^{\rm corr}$ (eV) \\
\hline
$\pi$ hole        &   0.96    &     8.63    \\
$\pi$ satellite   &   0.00    &    16.15    \\
$\sigma$ hole     &   0.86    &    18.54    \\
\hline
\end{tabular}
}
\end{table}
\renewcommand{\baselinestretch}{1}\normalsize

In the present case, the inclusion of singly excited configurations into the
reference space helps to solve the problem of competing states
in MRCI calculations and allows us to get the desired target matrix
elements of the effective Hamiltonian. However, this approach
has limitations. When many one-particle configurations
with relatively large energy dispersion are included in the
model space or bigger clusters are considered (which implies
more low-lying unoccupied orbitals) there may be a lot of
satellite states competing energetically with the
one-particle states. Therefore, many states have to
be optimized simultaneously in the MRCI iterations and such
calculations may become quite expensive (or even unfeasible). 
To proceed in this case
one needs an improved version of the MRCI program which could
sort out the relevant states at each step of the iteration (again by the norm
of the projection) and converge only states with a dominant
contribution of one-particle configurations in the
CI wavefunctions. Such a treatment would completely solve the
problem of satellite states in the MRCI method as long as the 
satellites do not strongly mix with the single hole states.

\chapter{Results for {\it trans}--polyacetylene}

In this chapter we present and discuss the numerical results for
{\it trans}--polyacetylene. In section~4.1 the individual
increments obtained in clusters are given and the convergence
and accuracy of the incremental scheme is explored. The
correlation corrections to the local matrix elements are obtained
and the change of the LMEs due to electron correlation is
discussed. In the next section we present the {\it correlated} band structure
of {\it trans}--polyacetylene and compare it to that on
the HF level. In section~4.3 we estimate the error for the
individual local matrix elements, the band gap and the band widths.
In section~4.4 we discuss how results obtained for the
one-dimensional system can be compared to experimental data
on bulk {\it trans}--polyacetylene.

\section{From single increments to final local matrix elements}

We start the discussion of the numerical results with the correlation
corrections to the local matrix elements associated with a hole in
$\sigma$ bonds. In this case, the convergence of the correlation effect
with the distance from the localized hole is most rapid as
$\sigma$ bonds are the most compact localized orbitals. Therefore,
we only need a relatively small C$_6$H$_8$ cluster to account for
the correlation effects.

\begin{figure}
\centerline{ \psfig{figure=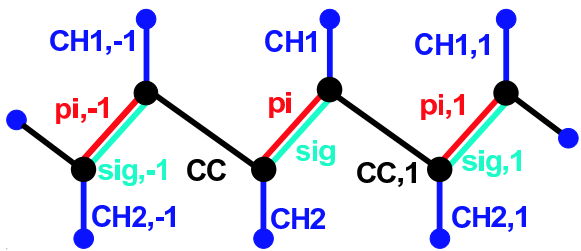,width=3in} }
\caption{C$_{6}$H$_{8}$ cluster of tPA with bond labels.}
\label{c6h8}
\end{figure}

First, we would like to discuss the effect of truncating the 
incremental scheme after the third order. For this purpose we
perform a correlation calculation for nine hole states when 
the most inner nine
bonds of C$_6$H$_8$ are open simultaneously ("CH1,-1",
"CC", "CH2", "sig", "CH1", "CC,1", "CH2,1", "sig,1" and "CH1,1").
The names of the bonds are indicated in Fig.~\ref{c6h8}.
We do not include $\pi$ bonds here to avoid the appearance of
satellite states. From this calculation we get the correlation
corrections to the local matrix elements which are associated 
with these nine bonds (e.g. one diagonal element
$\delta^\prime {\rm IP_{CH1,CH1}^{corr}}=-1.077$ eV and one
off-diagonal element $\delta^\prime 
{\rm IP_{CH1,CC,1}^{corr}}=-0.407$ eV).

\begin{table}
\refstepcounter{table}
\addtocounter{table}{-1}
\label{CHincr}
\caption{Increments for the correlation correction to
the diagonal matrix element IP$_{\rm CH1,CH1}$ of a C$_6$H$_8$ 
molecule (in eV) when only the
$\sigma$ bonds are open for correlation.
The reference value $\delta^\prime {\rm IP_{CH1,CH1}^{corr}}$
where the electrons from all nine $\sigma$ bonds are correlated
is given as well.}
\vspace{5mm}
\centerline{
\renewcommand{\baselinestretch}{1.1}\normalsize
\begin{tabular}{llp{25mm}llr}
\hline
 $\Delta {\rm IP_{CH1,CH1}^{corr}}()$ &
 \multicolumn{2}{c}{$\Delta {\rm IP_{CH1,CH1}^{corr}}(c)$ \; \; \;} &
 \multicolumn{3}{c}{$\Delta {\rm IP_{CH1,CH1}^{corr}}(c,d)$ } \\[2mm]
& \quad $c$ & $\Delta$IP & $c$ & $d$ & $\Delta$IP \; \; \\
\hline
\quad 0.188 & \quad sig    & -0.336 & sig    & CC,1    &  0.013 \; \\
      & \quad CC,1   & -0.393 & CH2    & sig     & -0.064 \; \\
      & \quad CH1,-1 & -0.069 & CC     & sig     & -0.012 \; \\
      & \quad CH1,1  & -0.073 & CH1,-l & sig     & -0.000 \; \\
      & \quad CH2    & -0.123 & CC,1   & CH2,1   & -0.070 \; \\
      & \quad CH2,1  & -0.114 & CC,1   & sig,1   & -0.015 \; \\
      & \quad CC     & -0.092 & CC,1   & CH1,1   &  0.003 \; \\
      & \quad sig,1  & -0.080 & CH2    & CC,1    &  0.013 \; \\
      &              &        & CC     & CC,1    &  0.008 \; \\
      &              &        & CH1,-1 & CC,1    &  0.009 \; \\
      &              &        & CH1,-1 & CC      &  0.003 \; \\
      &              &        & CH1,-1 & CH2     &  0.003 \; \\
      &              &        & CH1,-1 & CH2,1   &  0.006 \; \\
      &              &        & CH1,-1 & sig,1   &  0.003 \; \\
      &              &        & CH1,-1 & CH1,1   &  0.005 \; \\
      &              &        & CC     & CH2     & -0.017 \; \\
      &              &        & CC     & CH2,1   &  0.007 \; \\
      &              &        & CC     & sig,1   &  0.003 \; \\
      &              &        & CC     & CH1,1   &  0.005 \; \\
      &              &        & CH2    & CH2,1   &  0.011 \; \\
      &              &        & CH2    & sig,1   &  0.006 \; \\
      &              &        & CH2    & CH1,1   &  0.006 \; \\
      &              &        & sig    & sig,1   &  0.005 \; \\
      &              &        & sig    & CH2,1   &  0.008 \; \\
      &              &        & sig    & CH1,1   &  0.006 \; \\
      &              &        & CH2,1  & sig,1   & -0.010 \; \\
      &              &        & CH2,1  & CH1,1   &  0.002 \; \\
      &              &        & sig,1  & CH1,1   &  0.007 \; \\
\hline
\quad {\bf 0.188} &  & {\bf -1.280} & &          & {\bf -0.057} \; \\
\hline
 \multicolumn{5}{l}{} &  \\[-2mm]
 \multicolumn{5}{l}{$\sum_{i=1}^3 \Delta {\rm IP}^{(i)}_{\rm CH1,CH1}$} & {\bf -1.149} \; \\[2mm]
 \multicolumn{5}{l}{$\delta^\prime {\rm IP_{CH1,CH1}^{corr}}$} & {\bf -1.077} \; \\[2mm]
\hline
\end{tabular}
}
\end{table}
\renewcommand{\baselinestretch}{1}\normalsize
\begin{table}
\refstepcounter{table}
\addtocounter{table}{-1}
\label{CHCCincr}
\caption{Increments for the correlation correction to
the off-diagonal matrix element IP$_{\rm CH1,CC,1}$ of a C$_6$H$_8$
molecule (in eV) when only the $\sigma$ bonds are open for correlation.
The reference value $\delta^\prime {\rm IP_{CH1,CC,1}^{corr}}$
where the electrons from all nine $\sigma$ bonds are correlated
is given as well.}
\vspace{5mm}
\centerline{
\renewcommand{\baselinestretch}{1.2}\normalsize
\begin{tabular}{clrcllr}
\hline
 $\Delta {\rm IP_{CH1,CC,1}^{corr}}()$ &
 \multicolumn{2}{c}{$\Delta {\rm IP_{CH1,CC,1}^{corr}}(c)$ } & \qquad \quad &
 \multicolumn{3}{c}{$\Delta {\rm IP_{CH1,CC,1}^{corr}}(c,d)$ } \\[2mm]
& \quad $c$ & $\Delta$IP \; & & $c$ & $d$ & $\Delta$IP \; \; \\
\hline
-0.213 & \quad sig   & -0.118 & & sig    & CH1,-1 &  0.001 \; \\
       & \quad CH2,1 & -0.029 & & sig    & CC     &  0.004 \; \\
       & \quad sig,1 & -0.024 & & sig    & CH2    &  0.004 \; \\
       & \quad CH1,1 & -0.014 & & sig    & CH2,1  & -0.001 \; \\
       & \quad CH2   & -0.002 & & sig    & sig,1  &  0.001 \; \\
       & \quad CH1,-1&  0.001 & & sig    & CH1,1  &  0.004 \; \\
       & \quad CC    &  0.000 & & CH1,-1 & CC     & -0.002 \; \\
       &             &        & & CH1,-1 & CH2    &  0.000 \; \\
       &             &        & & CH1,-1 & CH2,1  &  0.000 \; \\
       &             &        & & CH1,-1 & sig,1  &  0.003 \; \\
       &             &        & & CH1,-1 & CH1,1  &  0.001 \; \\
       &             &        & & CC     & CH2    &  0.000 \; \\
       &             &        & & CC     & CH2,1  &  0.003 \; \\
       &             &        & & CC     & sig,1  &  0.001 \; \\
       &             &        & & CC     & CH1,1  &  0.001 \; \\
       &             &        & & CH2    & CH2,1  &  0.002 \; \\
       &             &        & & CH2    & sig,1  &  0.002 \; \\
       &             &        & & CH2    & CH1,1  &  0.000 \; \\
       &             &        & & CH2,1  & sig,1  & -0.004 \; \\
       &             &        & & CH2,1  & CH1,1  & -0.003 \; \\
       &             &        & & sig,1  & CH1,1  & -0.005 \; \\
\hline
\quad {\bf -0.213} & & {\bf -0.186}  & &     &   & {\bf 0.010} \; \\
\hline
 \multicolumn{6}{l}{} &  \\[-2mm]
 \multicolumn{6}{l}{$\sum_{i=1}^3 \Delta {\rm IP}^{(i)}_{\rm CH1,CC,1}$} & {\bf -0.389} \; \\[2mm]
 \multicolumn{6}{l}{$\delta^\prime {\rm IP_{CH1,CC,1}^{corr}}$} & {\bf -0.407} \; \\[2mm]
\hline
\end{tabular}
}
\end{table}
\renewcommand{\baselinestretch}{1}\normalsize

We can get the same corrections by the method of local
increments. The corresponding data on the individual increments are
presented in Tables~\ref{CHincr} and \ref{CHCCincr} where
increments of different orders are placed in different columns.
Numbers at the bottom of these tables are the sums over all
increments $\Delta {\rm IP}_{ab}^{(n)}$ of the given order $n$.
These numbers demonstrate the
convergence of the incremental scheme with the order of
increments: the third term in Eq.~(\ref{IPincr}) drops with respect
to the second one by more than an order of magnitude that insures
the possibility to truncate the series (\ref{IPincr}) after the
third term. Actually, non of the third-order increments of the
{\it off-diagonal} metrix elements is larger than 5 meV, so that
these contributions might be neglected as well. Summing up 
the increments of 1st, 2nd and 3rd order one gets
the correlation correction estimated by the incremental scheme:
$\delta^{\prime \prime} {\rm IP_{CH1,CH1}^{corr}}=-1.149$ eV and
$\delta^{\prime \prime} {\rm IP_{CH1,CC,1}^{corr}}=-0.389$ eV. The
differences between the exact and the estimated values of
correlation corrections are $-72$~meV (or 6.3\%) and
$-18$~meV (or 4.6\%) for the diagonal and off-diagonal matrix
elements, respectively. Thus, we can argue that the incremental scheme
estimates the correlation corrections with an error of a few
percent.

From Tables~\ref{CHincr} and \ref{CHCCincr} one also nicely sees the
convergence of the second-order increments with the distance to
the added $\sigma$ bond $c$. To study this decay
of correlation effect with the distance from the localized 
extra charge more extensively we provide the
second-order increments of the matrix elements
IP$_{\rm CC,CC}$ (diagonal) and IP$_{\rm CC,sig}$ (off-diagonal)
in Table~\ref{conver_sig} ordered by the distance of the added 
bond from the reference bonds. The largest value occur for the
added bond adjacent to the one of reference bonds (or even to both
in the case of off-diagonal matrix elements).
Values of increments drop by more than an order of magnitude when
the added bond is moved outwards by one unit cell. The contribution
to the correlation correction coming from farther bonds is negligibly
small and will not be treated explicitly by our correlation calculations.
Rather the infinite-size extrapolation of the correlation correction
obtained in a cluster can be estimated in a continuum approximation
(see section 4.3). The data presented in Table~\ref{conver_sig}
justifies the use of the C$_6$H$_8$ molecule as the smallest cluster
of {\it trans}-polyacetylene to calculate the correlation corrections
to all local matrix elements corresponding to $\sigma$ hole states.

\begin{table}
\refstepcounter{table}
\addtocounter{table}{-1}
\label{conver_sig}
\caption{Convergence of the second-order increments of the
matrix elements IP$_{\rm CC,CC}$ and IP$_{\rm CC,sig}$ with the
distance to the added $\sigma$ bond (in eV).}
\vspace{5mm}
\centerline{
\renewcommand{\baselinestretch}{1.2}\normalsize
\begin{tabular}{p{2cm}p{45mm}p{2cm}p{25mm}}
\hline
c & $\Delta {\rm IP_{CC,CC}^{corr}(c)}$ & c & $\Delta {\rm IP_{CC,sig}^{corr}(c)}$ \\
\hline
CH2   & -0.576 & CH2   & -0.129 \\
sig   & -0.380 & CH1   & -0.064 \\
CH1   & -0.132 & CC,1  & -0.031 \\
CH2,1 & -0.025 & CH2,1 & -0.007 \\
sig,1 & -0.022 & sig,1 & -0.003 \\
\hline
\end{tabular}
}
\end{table}
\renewcommand{\baselinestretch}{1}\normalsize

Of course, the $\pi$ orbitals also contribute to the correlation
corrections of the local matrix elements of $\sigma$-type.
Together with the pure $\sigma$- and mixed $\pi$-$\sigma$-type 
contributions they sum up to the total
correlation corrections summarized in Table~\ref{IPcorrections_sig}
where the results for all diagonal and first-nearest-neighbor
off-diagonal matrix elements are given. They are obtained by the
incremental scheme when electrons from {\it all} bonds
of the C$_6$H$_8$ molecule are correlated: the $\sigma$ bonds
discussed above but also the $\pi$ bonds.
Only the  bonds with saturating hydrogen atoms are excluded.
To show the convergence of the incremental scheme
we provide the data for each increment order of the series
(\ref{IPincr}) in different columns of this table. As a rule,
the major contributions are coming from the second-order
increments $\Delta {\rm IP}^{\rm corr}_{ab}(c)$. The third-order
increments $\Delta {\rm IP}^{\rm corr}_{ab}(c,d)$ are usually more
than an order of magnitude smaller than the second-order ones,
and even in the worst case (IP$_{\rm CC,CC}$) they are by a
factor of 4 smaller.

\begin{table}
\refstepcounter{table}
\addtocounter{table}{-1}
\label{IPcorrections_sig}
\caption{Correlation corrections to local matrix elements
corresponding to $\sigma$ bonds (in eV). The last column gives
the total sum of the incremental contributions.}
\vspace{5mm}
\centerline{
\renewcommand{\baselinestretch}{1.2}\normalsize
\begin{tabular}{lcccc}
\hline
matrix element & $\Delta {\rm IP_{a,b}^{corr}}()$ &
$\sum_c \Delta {\rm IP_{a,b}^{corr}}(c)$ &
$\sum_{c<d}\Delta {\rm IP_{a,b}^{corr}}(c,d)$ &
${\bf \delta {\rm IP_{a,b}^{corr}}}$ \\
\hline
IP$_{\rm CH1,CH1}$  &  0.188 & -3.274 &  0.468 & {\bf -2.804} \\
IP$_{\rm CC,CC}$    &  0.166 & -5.422 &  1.554 & {\bf -3.702} \\
IP$_{\rm sig,sig}$  &  0.064 & -3.834 & -0.098 & {\bf -3.867} \\
IP$_{\rm CH1,CC,1}$ & -0.213 & -0.514 &  0.011 & {\bf -0.718} \\
IP$_{\rm CH1,sig}$  & -0.210 & -0.816 & -0.014 & {\bf -1.040} \\
IP$_{\rm CC,sig}$   & -0.165 & -0.756 & -0.014 & {\bf -0.936} \\
\hline
\end{tabular}
}
\end{table}
\renewcommand{\baselinestretch}{1}\normalsize

The same cluster is used to estimate the contribution from the
$\sigma$ bonds to the correlation correction of the diagonal matrix
element ${\rm IP_{pi,pi}^{corr}}$. However, one needs
C$_{8}$H$_{10}$ molecule to get useful values for the correlation
correction to the {\it off-diagonal} matrix element
${\rm IP_{pi,pi,1}^{corr}}$ as the two $\pi$ bonds of
{\it trans}-polyacetylene are spread over four
unit cells. In this case the adjacent $\sigma$ bonds
were grouped in pairs as indicated in Fig.~\ref{c8h10groups}.

\begin{figure}
\centerline{ \psfig{figure=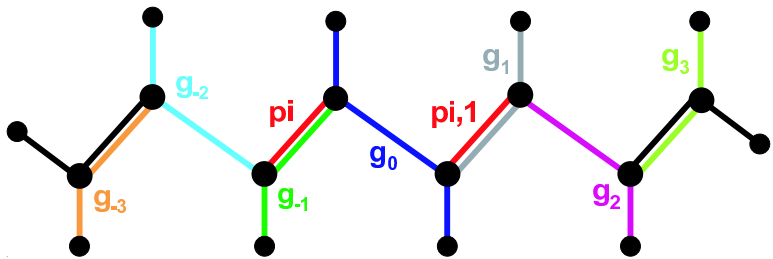,width=4in} }
\caption{Grouping of $\sigma$ bonds in the C$_{8}$H$_{10}$ cluster 
of tPA for the calculation of the correlation correction of
IP$_{pi,pi,1}$ matrix element.}
\label{c8h10groups}
\end{figure}

Nevertheless even this cluster turned out to be not big enough for
the evaluation of the contributions from the $\pi$ bonds of tPA to
$\delta {\rm IP_{pi,pi}^{corr}}$ and
$\delta {\rm IP_{pi,pi,1}^{corr}}$. The localized $\pi$ orbitals are
more diffuse than the $\sigma$ ones (see Fig.~\ref{occ_pi}) and
therefore the correlation effects decay noticeably slowler
with the distance from the $\pi$ hole. Hence, the calculation of the
increments with added $\pi$ bonds is performed in a
C$_{10}$H$_{12}$ cluster which contains five bonds of $\pi$ symmetry
(see Fig~\ref{c10h12}).

\begin{table}
\refstepcounter{table}
\addtocounter{table}{-1}
\label{pi_incr}
\caption{Increments for the correlation correction to
the diagonal matrix element IP$_{\rm pi,pi}$ of a
C$_{10}$H$_{12}$ molecule (in eV) when only the $\pi$ bonds 
are open for correlation.
The reference value $\delta^\prime {\rm IP_{pi,pi}^{corr}}$
where the electrons from all five $\pi$ bonds are correlated
is given as well.}
\vspace{5mm}
\centerline{
\renewcommand{\baselinestretch}{1.2}\normalsize
\begin{tabular}{clp{25mm}lp{10mm}r}
\hline
 $\Delta {\rm IP_{pi,pi}^{corr}}()$ &
 \multicolumn{2}{c}{$\Delta {\rm IP_{pi,pi}^{corr}}(c)$ } \; \; \; \; \; &
 \multicolumn{3}{c}{$\Delta {\rm IP_{pi,pi}^{corr}}(c,d)$ } \\[1mm]
& \; $c$ & $\Delta$IP & \; $c$ & $d$ & $\Delta$IP \: \\
\hline
0.445 & \; pi,1 \;  & -0.604 & \; pi,2  & pi,1    & -0.291  \\
      & \; pi,2     & -0.162 & \; pi,-2 & pi,-1   & -0.291  \\
      & \; pi,-1    & -0.604 & \; pi,-2 & pi,1    &  0.061  \\
      & \; pi,-2 \; & -0.162 & \; pi,2  & pi,-1   &  0.061  \\
      &             &        & \; pi,2  & pi,-2   &  0.036  \\
      &             &        & \; pi,1  & pi,-1   &  0.154  \\
\hline
{\bf 0.445} & & {\bf -1.532} & & & {\bf -0.271}  \\
\hline
 \multicolumn{5}{l}{} &  \\[-4mm]
 \multicolumn{5}{l}{$\sum_{i=1}^3 \Delta {\rm IP}^{(i)}_{\rm pi,pi}$} & {\bf -1.358} \\[1mm]
 \multicolumn{5}{l}{$\delta^\prime {\rm IP_{pi,pi}^{corr}}$} & {\bf -1.145} \\[2mm]
\hline
\end{tabular}
}
\end{table}
\renewcommand{\baselinestretch}{1}\normalsize

We are going to show again the accuracy of the incremental scheme in 
the case of $\pi$ bonds by first calculating the correlation corrections to the matrix
elements ${\rm IP_{pi,pi}^{corr}}$ and ${\rm IP_{pi,pi,1}^{corr}}$
when all five $\pi$ bonds of C$_{10}$H$_{12}$ are correlated simultaneously
($\delta^\prime {\rm IP_{pi,pi}^{corr}}=-1.145$ eV and
$\delta^\prime {\rm IP_{pi,pi,1}^{corr}}=-0.573$ eV )
and comparing them to the values obtained by the incremental scheme. 
The data on the increments are presented
in Tables~\ref{pi_incr} and \ref{pi_pi1_incr}. The convergence
of the incremental scheme with the order is clearly discernable. 
Yet, there are also some third-order contributions which are still
in the order of the second-order terms ($\sim$0.3 eV). Those are
{\it not} increments $\Delta {\rm IP_{pi,pi}}(c,d)$ with the
orbitals $c$ and $d$ being arranged symmetrically around the
reference $\pi$ bond (as one would intuitively expect) but a linear
configuration of the three bonds with the two added bonds $c$ and $d$
on the {\it same} side, i.e. $\Delta {\rm IP_{pi,pi}(pi,1;pi,2)}$.
Summing up the individual contributions in each
table one gets the total correlation corrections for the five 
$\pi$-bonds as estimated by the incremental scheme:
$\delta^{\prime \prime} {\rm IP_{pi,pi}^{corr}}=-1.358$ eV and
$\delta^{\prime \prime} {\rm IP_{pi,pi,1}^{corr}}=-0.615$ eV. The
differences between the exact and the estimated values of
correlation corrections are $213$~meV (or 15.7\%) and
$42$~meV (or 6.8\%) for the diagonal and off-diagonal matrix
elements, respectively. We see that the performance of the incremental
scheme is still good even in the case of rather diffuse orbitals and in the
worst case it produces an error of approximately 0.2 eV.

\begin{table}
\refstepcounter{table}
\addtocounter{table}{-1}
\label{pi_pi1_incr}
\caption{Individual increments for the correlation correction to
the off-diagonal matrix element IP$_{\rm pi,pi,1}$ of a 
C$_{10}$H$_{12}$ molecule (in eV) obtained when only the $\pi$
bonds are open for correlation.}
\vspace{5mm}
\centerline{
\renewcommand{\baselinestretch}{1.2}\normalsize
\begin{tabular}{clp{25mm}lp{10mm}r}
\hline
 $\Delta {\rm IP_{pi,pi,1}^{corr}}()$ &
 \multicolumn{2}{c}{$\Delta {\rm IP_{pi,pi,1}^{corr}}(c)$ } \; \; \; \; \; &
 \multicolumn{3}{c}{$\Delta {\rm IP_{pi,pi,1}^{corr}}(c,d)$ } \\[1mm]
& \; $c$ & $\Delta$IP & \; $c$ & $d$ & $\Delta$IP \: \\
\hline
-0.460 & \; pi,2  \; & -0.082 & \; pi,-1 & pi,-2 & -0.043  \\
       & \; pi,-1 \; & -0.079 & \; pi,-1 & pi,2  &  0.031  \\
       & \; pi,-2 \; & -0.005 & \; pi,-2 & pi,2  &  0.015  \\
\hline
{\bf -0.460} &  & {\bf -0.157} & & & {\bf 0.003}  \\
\hline
 \multicolumn{5}{l}{} &  \\[-4mm]
 \multicolumn{5}{l}{$\sum_{i=1}^3 \Delta {\rm IP}^{(i)}_{\rm pi,pi,1}$} & {\bf -0.614} \\[1mm]
 \multicolumn{5}{l}{$\delta^\prime {\rm IP_{pi,pi,1}^{corr}}$} & {\bf -0.573} \\[2mm]
\hline
\end{tabular}
}
\end{table}
\renewcommand{\baselinestretch}{1}\normalsize

To illustrate the decay of the correlation effects with the distance
from the localized $\pi$ hole a representavive selection of the 
second-order increments
obtained in the C$_{10}$H$_{12}$ molecule when only $\pi$ bonds are
added are given in Table~\ref{conver_pi}. The values of the increments
of the diagonal matrix element drop by an order of
magnitude when distance of the added bond increases by two lattice constants. The
decay of the increments for the off-diagonal matrix element is even more
rapid. The dash in Table~\ref{conver_pi} (and also in
Table~\ref{conver_virt}) means that the absolute value of the 
corresponding increment is below the estimated size-extensivity error for this
increment.

\begin{table}
\refstepcounter{table}
\addtocounter{table}{-1}
\label{conver_pi}
\caption{Convergence of the diagonal (left) and first nearest-neighbor
off-diagonal increments $\Delta {\rm IP}_{b,a}^{\rm corr}(c)$ with the
distance between the additionally correlated bond (in eV). The dash 
means that the increment is too small to be reliable due to
size-extensivity problem of MRCI.}
\vspace{5mm}
\centerline{
\renewcommand{\baselinestretch}{1.2}\normalsize
\begin{tabular}{p{17mm}p{2cm}p{4cm}p{17mm}p{2cm}c}
\hline
$a,a$ & $c$ & $\Delta$IP &
$a,b$ & $c$ & $\Delta$IP \\
\hline
pi,pi     & pi,-1 & -0.604 &
pi,pi,1   & pi,-1 & -0.079 \\
pi,pi     & pi,-2 & -0.162 &
pi,pi,1   & pi,-2 & -0.005 \\
pi,1,pi,1 & pi,-2 & -0.053 &
pi,1,pi,2 & pi,-2 & ---    \\
pi,2,pi,2 & pi,-2 & -0.015 &  &  \\
\hline
\end{tabular}
}
\end{table}
\renewcommand{\baselinestretch}{1}\normalsize

The increments for the correlation corrections to the matrix elements
${\rm IP_{pi,pi}^{corr}}$ and ${\rm IP_{pi,pi,1}^{corr}}$
obtained on the C$_6$H$_8$ and C$_{10}$H$_{12}$ clusters
are summed up for both, $\sigma$ and $\pi$ orbitals, and presented in Table~\ref{IPcorrections_pi}.
The contribution of the part of the crystal outside  the
C$_{10}$H$_{12}$ cluster will be estimated in section~4.3 using
the data from Table~\ref{conver_pi}. There we show that even
in the case of the rather slow decay of the correlation effect with
the distance the polarization of the remaining part of the infinite
one-dimensional crystal outside the C$_{10}$H$_{12}$ cluster
amounts to 1\% of the correlation correction
$\delta {\rm IP_{pi,pi}^{corr}}$ only that is well-below the error bar
of the incremental scheme itself.

\begin{table}
\refstepcounter{table}
\addtocounter{table}{-1}
\label{IPcorrections_pi}
\caption{Correlation corrections to the local matrix elements
${\rm IP_{pi,pi}^{corr}}$ and ${\rm IP_{pi,pi,1}^{corr}}$
(in eV). The total values are presented in the last column.}
\vspace{5mm}
\centerline{
\renewcommand{\baselinestretch}{1.2}\normalsize
\begin{tabular}{lcccc}
\hline
matrix element & $\Delta {\rm IP_{a,b}^{corr}}()$ &
$\sum_c \Delta {\rm IP_{a,b}^{corr}}(c)$ &
$\sum_{c<d}\Delta {\rm IP_{a,b}^{corr}}(c,d)$ &
${\bf \delta {\rm IP_{a,b}^{corr}}}$ \\
\hline
IP$_{\rm pi,pi}$    &  0.445 & -2.006 & -0.446 & {\bf -2.007} \\
IP$_{\rm pi,pi,1}$  & -0.460 &  0.087 &  0.057 & {\bf -0.316} \\
\hline
\end{tabular}
}
\end{table}
\renewcommand{\baselinestretch}{1}\normalsize

Let us turn to the electron affinities now. To calculate the 
correlation corrections to the matrix elements corresponding
to attached-electron states we proceed as described in
section~3.4.3. In the case of the diagonal matrix elements
${\rm EA_{pi^\ast,pi^\ast}^{corr}}$ and
${\rm EA_{CH2^\ast,CH2^\ast}^{corr}}$ the molecule C$_{10}$H$_{12}$
may serve as a minimal-size cluster for the correlation calculations.
The decay of the second-order increments with the distance to the
added $\pi$ bond-antibond pair of $\pi$ symmetry (the most diffuse ones,
referred to by the $\pi^\ast$ orbital only for simplicity)
is presented for both matrix elements in
Table~\ref{conver_virt}. The convergence with the distance looks 
even better than in the case of $\pi$ hole states, however, that
is most probably simply because only a limited number of the
antibonds are open for excitations from the occupied bonds in
the incremental scheme for attached electrons which leads to a
partial reduction of the virtual space, the effect which is more
pronounced for increments $\Delta {\rm EA}_{rs}(t)$ with separated
bond groups $r$, $s$ and $t$.

\begin{table}
\refstepcounter{table}
\addtocounter{table}{-1}
\label{conver_virt}
\caption{Convergence of increments with the distance between
correlated bonds (in eV). Dashes mean that the
increments are too small to be reliable due to
size-extensivity problem of MRCI.}
\vspace{5mm}
\centerline{
\renewcommand{\baselinestretch}{1.2}\normalsize
\begin{tabular}{p{5cm}c|p{4cm}c}
\hline
$\Delta {\rm EA_{CH2^\ast,CH2^\ast}^{corr}(pi^\ast)}$  & 0.181 &
$\Delta {\rm EA_{pi^\ast,pi^\ast}^{corr}(pi^\ast,-1)}$  & 0.314 \\
$\Delta {\rm EA_{CH2^\ast,CH2^\ast}^{corr}(pi^\ast,-1)}$   & 0.034 &
$\Delta {\rm EA_{pi^\ast,pi^\ast}^{corr}(pi^\ast,-2)}$  & 0.088 \\
$\Delta {\rm EA_{CH2^\ast,CH2^\ast}^{corr}(pi^\ast,-2)}$   & --- &
$\Delta {\rm EA_{pi^\ast,1,pi^\ast,1}^{corr}(pi^\ast,-2)}$  & --- \\
\hline
\end{tabular}
}
\end{table}
\renewcommand{\baselinestretch}{1}\normalsize

The convergence of incremental series (\ref{EAincr_group}) with
order is shown in Table~\ref{EAcorrections}; the {\it total} values of
the correlation corrections to the diagonal matrix elements are presented
in the last column of this Table. As in the case of the hole states
the major contribution to the local matrix elements EA$_{rs}$ comes
from the second-order increments $\Delta {\rm EA}_{rs}(t)$ while
the third-order contributions $\Delta {\rm EA}_{rs}(t,u)$ are at least
by a factor of 4 smaller than the second-order terms.

\begin{table}
\refstepcounter{table}
\addtocounter{table}{-1}
\label{EAcorrections}
\caption{Correlation corrections to diagonal local matrix elements
corresponding to antibonds obtained in the cluster C$_{10}$H$_{12}$ 
(in eV).}
\vspace{5mm}
\centerline{
\renewcommand{\baselinestretch}{1.2}\normalsize
\begin{tabular}{lcccc}
\hline
matrix element & $\Delta {\rm EA_{a,b}^{corr}}()$ &
$\sum_c \Delta {\rm EA_{a,b}^{corr}}(c)$ &
$\sum_{c<d}\Delta {\rm EA_{a,b}^{corr}}(c,d)$ &
${\bf \delta {\rm EA_{a,b}^{corr}}}$ \\
\hline
EA$_{\rm CH2^\ast,CH2^\ast}$    & 0.191 & 0.816 & -0.161 & {\bf 0.846} \\
EA$_{\rm pi^\ast,pi^\ast}$      & 0.237 & 1.815 & -0.430 & {\bf 1.623} \\
\hline
\end{tabular}
}
\end{table}
\renewcommand{\baselinestretch}{1}\normalsize

\begin{table}
\refstepcounter{table}
\addtocounter{table}{-1}
\label{LMEresults}
\caption{Change of the local matrix elements ${\rm IP}_{R,nn^\prime}$ and
${\rm EA}_{R,mm^\prime}$ of the periodic system due to correlation effects.}
\vspace{5mm}
\centerline{
\renewcommand{\baselinestretch}{1.2}\normalsize
\begin{tabular}{p{3cm}rrrr}
\hline
matrix element & IP$_{0,\rm pi\: pi}$ &  IP$_{0,\rm CH1\: CH1}$ & IP$_{0,\rm CC\: CC}$  &  IP$_{0,\rm sig\: sig}$ \\
\hline
SCF value      & 10.526 eV &  18.904 eV & 21.230 eV &  22.396 eV \\
     reduction & -2.007 eV &  -2.804 eV & -3.702 eV &  -3.867 eV \\
               & 19.1 \%   & 14.8 \%    & 17.4 \%   &  17.2 \%   \\
total value    & 8.519 eV  & 16.100 eV  & 17.528 eV &  18.529 eV \\
\hline
\\[5mm]
\hline
matrix element &  IP$_{\pm a,\rm pi\: pi}$ & IP$_{a,\rm CC\: CH1}$ & IP$_{0,\rm CH1\: sig}$  & IP$_{0,\rm CC\: sig}$ \\
\hline
SCF value      &  1.683 eV          &  2.968 eV       &   3.022 eV      &    3.022 eV        \\
     reduction & -0.316 eV          & -0.718 eV       &  -1.040 eV      &   -0.936 eV        \\
               & 18.7 \%            & 24.2 \%         &  34.4 \%        &   31.0 \%          \\
total value    &  1.367 eV          &  2.250 eV       &   1.982 eV      &    2.087 eV        \\
\hline
\\[5mm]
\hline
matrix element & EA$_{0,\rm pi^\ast\: pi^\ast}$ &  EA$_{0,\rm CH1^\ast\: CH1^\ast}$  &
EA$_{\pm a, \rm pi^\ast\: pi^\ast}$ & EA$_{\pm a, \rm CH1^\ast\: CH1^\ast}$ \\
\hline
SCF value      &  -4.497 eV          &   -5.603 eV      &   -1.054 eV      &    1.012 eV        \\
     reduction &   1.623 eV          &   0.846 eV       &    0.365 eV      &   -0.035 eV        \\
               &  36.1\%             &  15.1 \%         &   34.6 \%        &    3.5 \%          \\
total value    &  -2.941 eV          &  -4.764 eV       &   -0.689 eV      &    0.976 eV        \\
\hline
\end{tabular}
}
\end{table}
\renewcommand{\baselinestretch}{1}\normalsize

To evaluate the correlation corrections to the off-diagonal matrix
elements ${\rm IP_{pi^\ast,pi^\ast,1}^{corr}}$ and
${\rm EA_{CH2^\ast,CH2^\ast,1}^{corr}}$ we had to use even larger,
the C$_{12}$H$_{14}$
molecule. The convergence of the increments with the order and the
distance between the correlated electrons was checked and turned out
to be satisfactory. Summing up these
increments we get the following values for the corrections:
${\bf \delta {\rm EA_{\rm CH2^\ast,CH2^\ast,1}^{corr}}}=-0.035$ eV
and ${\bf \delta {\rm EA_{\rm pi^\ast,pi^\ast,1}^{corr}}}=0.365$
eV.

To summarize our results on the correlation corrections to the
local matrix elements we present in Table~\ref{LMEresults} the SCF matrix elements
together with the absolute corrections (and their percentage
relative the SCF value) and the {\it total correlated} matrix
elements IP$^{\rm corr}$ and EA$^{\rm corr}$. Note
that we also switched back to the notation for the matrix elements 
in the infinite
periodic system as done in Tables~\ref{SCF_IPs}--\ref{SCF_EAs}.
though the correlation contributions are actually calculated in 
molecules representing finite parts of the crystal. In the
one-dimensional crystal the polarization outside the
specified clusters only gives negligibly small effect as compared
to the correlation effect in the close vicinity of the localized
extra charge (see section~4.3) and therefore the correlation correction to a local
matrix element obtained in the molecules are equal to the corresponding one
in the crystal within the accuracy of the incremental
scheme.

From Table~\ref{LMEresults} one can see that the change of the
SCF matrix elements due to correlation effects is quite pronounced, 
ranging from 15\% to 30\%. One also sees
that in all cases the absolute values of the SCF matrix elements
are reduced, i.e. overall the ionization potentials decrease while
the electron affinities increase. This indicates that the reduction 
of the total energy due to electron correlation is always stronger 
in the charged systems than in the neutral one.

\section{Correlated band structure of {\it trans}-polyacetylene}

Having obtained the local matrix elements including all correlation
effects (the "total values" in Table~\ref{LMEresults}) we are
able to compile the band structure of {\it trans}-polyacetylene
which takes into account electron correlation, the ultimate
goal of our approach. For this purpose we replace
the LMEs in the truncated series (\ref{IPn}) and (\ref{EAm}) 
obtained on the SCF level
by the correlated ones and diagonalize the
obtained matrices at each $k$-point analogously to the case
of the SCF bands. The new band structure (we call it
{\it correlated} band structure) will exhibit small deviations
from the one obtained by an infinite summation in Eqs.~(\ref{IPn}) and
(\ref{EAm}) like the SCF band structure (see
Figs.~\ref{SCFvalband} and \ref{SCFcondband}). To correct
for this (minor) deficiency we
simply add at each $k$-point the difference between
the band energies obtained by the local SCF matrix
elements and those obtained by the CRYSTAL code for an infinite periodic tPA
single chain. The resulting correlated band structure of
{\it trans}-polyacetylene is presented in Fig.~\ref{corrbands} 
(red lines) and compared to the SCF band structure by the CRYSTAL
code (black lines).

\begin{figure}
\centerline{ \psfig{figure=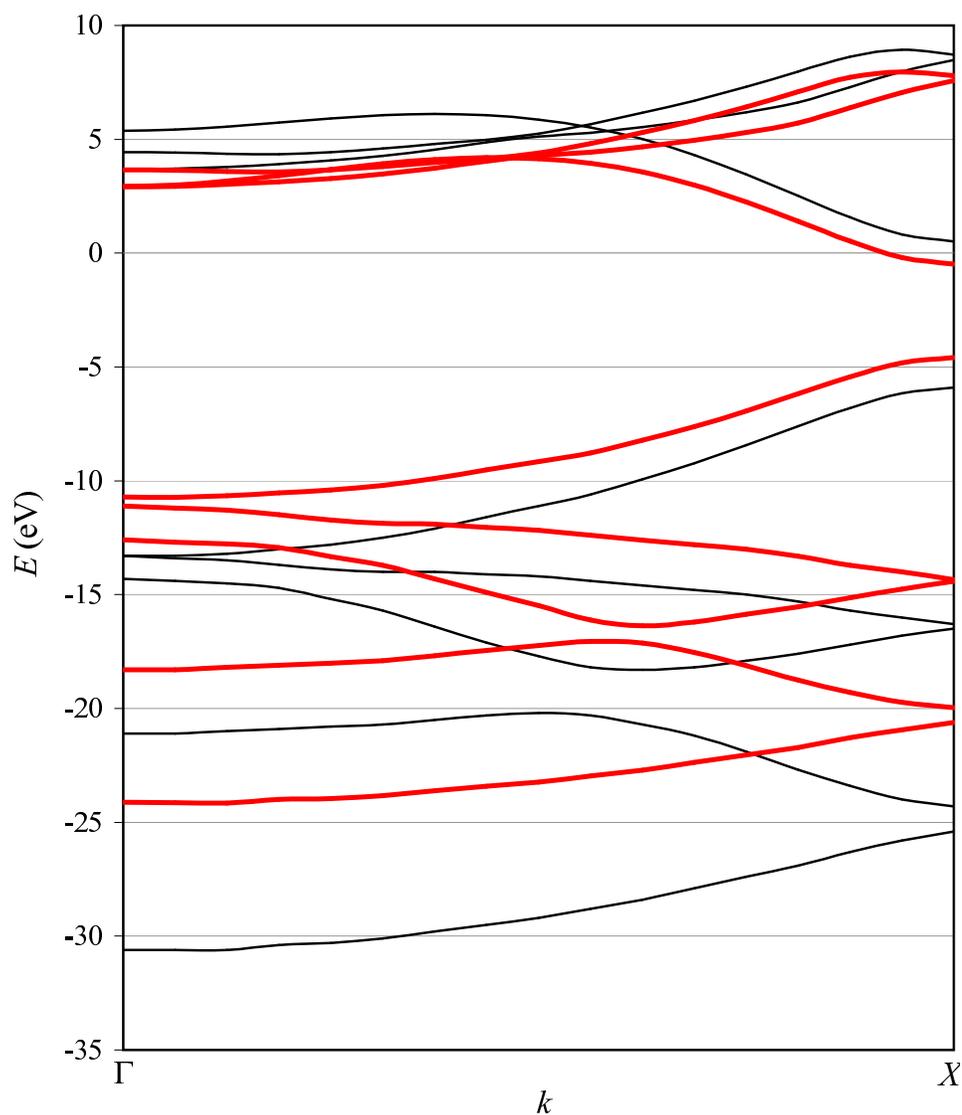,width=5in} }
\caption{Correlated band structure of the infinite
{\it trans}-polyacetylene single chain (red lines) compared
to the SCF one (black lines).}
\label{corrbands}
\end{figure}

From this Figure we see that the band structure changes
dramatically when electron correlation is taken into account.
Two effects are clearly seen: i) a shift of the "center-of-mass"
positions of all bands (valence bands are shifted upwards and
conduction bands downwards) and ii) a flattening of all bands
(significantly more pronounced for the valence bands).
These effects lead to a narrowing of the band gap, a change
of the sign of the electron affinity of {\it trans}-polyacetylene
and an overall reduction of the band widths. But there are also some
more subtle changes like the narrowing of the avoided crossing
of the 2nd and 3rd lowest valence bands close to the $X$ point
or the lifting of the accidental degeneracy of the highest two
valence bands at the $\Gamma$ point.

The value of the band gap of the infinite tPA single chain is
reduced by 2.31 eV from 6.42 eV (on the SCF level) down to 4.11 eV
when electron correlation is included, a 36.0\% reduction.
Similar values of the reduction of the SCF band gap, 2.35 and
2.38, are reported in two recent independent MP2 studies,
\cite{Sun96} and \cite{Ayala01}, respectively, on the $\pi$ bands
of isolated tPA chains.

The corrected values of the lowest ionization potential and the highest electron
affinity of {\it trans}-polyacetylene are 4.58 eV and 0.47 eV,
respectively. The positive sign of the electron affinity
(after inclusion of electron correlation) tells
that {\it trans}-polyacetylene is able to retain an extra electron
as is known from experiments \cite{Kaner89}.

\begin{table}
\refstepcounter{table}
\addtocounter{table}{-1}
\label{bandwidth}
\caption{Change of the band widths due to the correlation effects
(in eV).}
\vspace{5mm}
\centerline{
\renewcommand{\baselinestretch}{1.2}\normalsize
\begin{tabular}{p{3cm}|rrcr}
\hline
 &  &  & \multicolumn{2}{c}{reduction} \\
bands               & SCF width  \quad     & correlated width \; & absolute & relative \\
\hline
$\sigma$ valence    & 17.30  \; \; & 13.01 \qquad \; & 4.29 & 24.8 \% \\
$\pi$ valence       &  7.40  \; \; &  6.14 \qquad \; & 1.26 & 17.0 \% \\
$\pi$ conduction    &  4.86  \; \; &  4.67 \qquad \; & 0.19 &  3.9 \% \\
$\sigma$ conduction &  5.25  \; \; &  5.05 \qquad \; & 0.20 &  3.9 \% \\
\hline
\end{tabular}
}
\end{table}
\renewcommand{\baselinestretch}{1}\normalsize

The changes of the band widths are summarized in Table~\ref{bandwidth}.
There, the $\sigma$ valence band width refer to the width of all
four $\sigma$ valence bands while the $\sigma$ conduction band
width is that of the two $\sigma$ conduction bands only. One sees that
the relative reduction of the conduction bands widths is much
smaller than that of the valence bands. Since the dispersion
of the bands is mainly determined by the hopping matrix element between
nearest-neighbor sites (especially in 1D systems) one can conclude that
the hopping of holes is much stronger
affected by electron correlation than hopping of extra
electrons.

\section{Accuracy of the method}

Now that we have obtained numerical results for some physical
quantities which are characteristic for the investigated system the question
arises how accurate these results are and whether we are able to estimate
the accuracy of our method. In this section we show
that we can indeed provide some estimated error bars for the calculated
quantities.

As the correlation corrections to the local matrix elements
(the key quantities of our method) are composed of dozens
of increments we first estimate the error for the
individual increments. As can be seen from equations (\ref{IPcorrcorr}) and
(\ref{IPincr1})--(\ref{d_IPincr3}) each
individual increment is defined as the difference of several
correlation energies (correlation corrections to the total energies
of various states under various conditions which are calculated to evaluate a particular
increment). The uncertainty in the evaluation of these individual
increments essentially arises from the size-extensivity error of the
CI(SD) method. As shown in section~3.5 we are able to
reduce this error to about $10^{-4}$ of the value of the
correlation energy
(see last columns in Tables~\ref{correctionE0} and 
\ref{correctionEa}). In all our calculations these 
correlation energies are of the order of a few electron-volts.
Therefore, we estimate the error for a single increment
being a few meV. That is why in our calculations we simply omit all
increments whose absolute values are smaller
than 1 meV (see Tables~\ref{conver_pi} and \ref{conver_virt}).

The possible accumulated size-extensivity errors in the
correlation correction to some particular local matrix element
are expected to be not higher than the error for a single
increment times the number of increments entering the series.
Usually this amounts to few dozens of meV. 

Another kind of
error in the correlation corrections to LMEs
emerges from the truncation of the incremental series after
some order (after the third order in our case).
In section~4.1 we have explored the accuracy of the truncated
incremental scheme in detail. In the case of the most compact localized
orbitals ($\sigma$ bonds) the error produced by the incremental
scheme is of the same order as the accumulated size-extensivity
error (4--7\% of the total value of correlation correction to
a LME, see the discussion of Tables~\ref{CHincr} and \ref{CHCCincr}).
However, in the case of the more diffuse $\pi$ orbitals the error
due to the truncation of the series ($\sim$15\%, see the discussion
of Table~\ref{pi_incr}) is higher than the
estimated size-extensivity error for diagonal matrix element
($\sim$2\%). This indicates that for more diffuse orbitals the
fourth term of the incremental series (\ref{IPincr}) may give
noticeable contribution. Unfortunately, we can not calculate this term
explicitly as individual increments of the fourth order are usually smaller than
their size-extensivity error. Thus, we agree that the error
for correlation corrections to the diagonal matrix elements
IP$_{\rm pi,pi}$, EA$_{\rm pi^\ast,pi^\ast}$ and
EA$_{\rm CH1^\ast,CH1^\ast}$ are not more than 15\% and those for all
the other matrix elements not more than 10\% of absolute values of
the corrections.

Another question which we would like to discuss here concerns the
effect of polarization of infinite insulator due to the presence
of an uncompensated charge. The polarization of the infinite crystal
outside the explored cluster may also contribute to correlation
corrections. We can estimate such a contribution by a
continuum approximation as already done earlier by other authors,
e.g. in \cite{Horsch84}, \cite{Graef97} and \cite{Albrecht00}.

In this approximation far-range contributions to the reduction of 
the SCF energy of the ($N-1$)-
or ($N+1$)-electron system are attributed to the polarization
energy of an insulator in an electric field. This energy is equal
to

\begin{equation}
\label{Epolariz}
\delta E_i^{\rm pol} = \frac{1}{2} \int {\bf P\:E} \; {\rm d} V
\end{equation}
where ${\bf E}$ is the electric field,
${\bf P}=\frac{(\epsilon -1)}{4\pi} {\bf E}$ is the polarization
response and $\epsilon$ is the dielectric constant of the crystal. The subscript
$i$ in Eq.~(\ref{Epolariz}) denotes that this energy corresponds to
the charged system with a hole (an extra electron) being in the 
bond (antibond) $i$. In other words, we are estimating the polarization
effects on diagonal matrix elements here.

Let us say that starting with a distance $r_1$ from the center
of a localized orbital we can treat the electron in this orbital as
a point charge. Then the integration in Eq.~(\ref{Epolariz}) can be
divided in two parts:

\begin{equation}
\label{Epolariz2}
\delta E_i^{\rm pol} \; = \; \delta E_i (r_1) +
 \int_{r_1}^{\infty} {\rm d} r \; \frac{(\epsilon -1)}{4\pi} \;
\Bigl(\frac{e}{\epsilon r^2} \Bigr)^2 \; = \; \delta E_i (r_1) + C \: r_1^{-3}.
\end{equation}
Here we performed the integration for the case of one-dimensional system,
$\delta E_i (r_1)$ is the change of the energy due to electron
correlation for distances less than $r_1$ from the localized charge
(this value we calculate explicitly by the incremental scheme) and $C$
is a constant characteristic for the given material. Knowing
two $\delta E_i (r)$ values $\delta E_i (r_1)$ and $\delta E_i (r_2)$ for two different cut-off
radii $r_1$ and $r_2$ ($r_2 > r_1$) we can extract the 
tail correction $Cr_1^{-3}$ and by this estimate change of the
energy of the {\it whole} charged system

\begin{equation}
\label{Epolariz_inf}
\delta E_i(\infty) = \frac{\delta E_i(r_2)(r_2/r_1)^3 - 
\delta E_i(r_1)}{(r_2/r_1)^3 - 1}.
\end{equation}
Note the different power ($r^3$) which enter this relation
compared to the corresponding relation for a bulk crystal 
\cite{Albrecht00} $r^1$.

Thus, the correlation correction to diagonal matrix elements with
the polarization of the infinite crystal taken into account is

\begin{equation}
\label{IP_aa_polariz}
\delta {\rm IP}_{aa}^{\rm corr}(\infty) =
\frac{\delta {\rm IP}_{aa}^{\rm corr}(r_2)(r_2/r_1)^3 -
\delta {\rm IP}_{aa}^{\rm corr}(r_1)}{(r_2/r_1)^3 - 1}
\end{equation}
where $\delta {\rm IP}_{aa}^{\rm corr}(r_i)$, $i$=1, 2 are the correlation
correction to the matrix element IP$_{aa}$ when all bonds up to
distances $r_i$ from the bond $a$ are correlated. An analogous
formula holds for $\delta {\rm EA}_{rr}^{\rm corr}(\infty)$.

The hopping matrix element IP$_{ab}$ is one-half of the difference
between energies of the states $(\Phi_a + \Phi_b)/\sqrt{2}$ and
$(\Phi_a - \Phi_b)/\sqrt{2}$ which are two states with holes of two
different shapes delocalized within two bonds $a$ and $b$. Therefore,
the change of the hopping matrix element in continuum
approximation emerges due to the polarization of the insulator in the
electric field of a dipole. Substitution of the field of a point charge
in Eq.~(\ref{Epolariz2}) by that of a dipole leads to
the approximate formula for the correlation correction to the off-diagonal
matrix elements with the polarization of the entire crystal taken into account:

\begin{equation}
\label{IP_ab_polariz}
\delta {\rm IP}_{ab}^{\rm corr}(\infty) =
\frac{\delta {\rm IP}_{ab}^{\rm corr}(r_2)(r_2/r_1)^5 -
\delta {\rm IP}_{ab}^{\rm corr}(r_1)}{(r_2/r_1)^5 - 1}.
\end{equation}

Let us calculate explicitly the missing polarization effect in the
correlation correction $\delta {\rm IP}_{\rm pi,pi}^{\rm corr}$
caused by adding farther $\pi$ bonds
using the data on the decay of the second-order increments in
Table~\ref{conver_pi}. The total value of the correlation
correction in a cluster C$_{18}$H$_{20}$ (four
unit cells on both sides from the central one) as obtained by the incremental
scheme effectively is 2.007 eV and the cut-off
radius $r_2 = 4 a$. As the correlation effect associated with
the $\sigma$ bonds decay much faster than those arising from the
$\pi$ bonds, the change of the correlation correction when
the effective cluster increases from C$_{10}$H$_{12}$ to
C$_{18}$H$_{20}$ is mainly due to correlation of the electrons in
$\pi$ bonds. Thus, the value
$\delta {\rm IP}_{\rm pi,pi}^{\rm corr}(2a)$ is equal to
-1.871 eV. Using Eq.~(\ref{IP_aa_polariz}) we obtain the
value of the correlation correction in the infinite system
$\delta {\rm IP}_{\rm pi,pi}^{\rm corr}(\infty) = 2.026$ eV
compared to $\delta {\rm IP}_{\rm pi,pi}^{\rm corr}(4a)
= 2.007$ eV.
Therefore, the polarization of the outer part of the crystal
leads to less than 1\% change of the correlation correction which
can be neglected within the error bar of the incremental 
scheme. 

The continuum correction in the case of the off-diagonal matrix
elements is expected to be even less as the decay of
correlation effect with the distance is more rapid than for
diagonal ones (see Eqs.~\ref{Epolariz_inf} and \ref{IP_aa_polariz}). 
In fact for an effective C$_{16}$H$_{18}$ cluster (with $r=3.5 \: a$)
$\delta {\rm IP}_{\rm pi,pi,1}^{\rm corr}(3.5\: a) = -0.316$ eV while
for the smaller effective C$_{8}$H$_{10}$ cluster (with $r=1.5 \: a$)
$\delta {\rm IP}_{\rm pi,pi,1}^{\rm corr}(1.5\: a) = -0.148$ eV.
The resulting continuum correction is -0.002 eV leading to
$\delta {\rm IP}_{\rm pi,pi,1}^{\rm corr}(\infty) = -0.318$ eV.
Thus, we conclude
that the effect of polarization of the infinite
one-dimensional crystal outside the treated cluster is
negligibly small. 

Now we can estimate the accuracy of our method for the ultimate physical
quantities characterizing the investigated material. The shift
of the ionization potential due to electron correlation is
governed by two corrections: $\delta {\rm IP^{corr}} =
\delta {\rm IP_{0,pi,pi}^{corr}} - 
2\delta {\rm IP_{1,pi,pi}^{corr}}$.
This is easy to realize in terms of a H\"uckel (or tight-binding) model
where only diagonal matrix element and the hopping between
nearest-neighbor sites are considered.
Then, energy of the valence $\pi$ band as a function of
crystal-momentum $k$ is $\varepsilon_{\pi}(k)= -{\rm IP_{0,pi,pi}^{corr}}
-2{\rm IP_{1,pi,pi}^{corr}} \cos(ak)$. The top of this
band is at the $X$ point ($k=\pi/a$) and therefore the
ionization potential is given by ${\rm IP =
IP_{0,pi,pi}^{corr} - 2 IP_{1,pi,pi}^{corr}}$.
We have estimated the error for the correction
$\delta {\rm IP_{0,pi,pi}^{corr}}$ being equal to 15\% of its 
value (-2.0 eV) which is 0.3 eV. The error for the
correction $\delta {\rm IP_{0,pi,pi}^{corr}}$ is smaller by
one order of magnitude and is omitted. Thus, we consider the
ionization potential of tPA single chain to be
$\rm IP = 4.6 \pm 0.3$ eV. Approximately the same error bar
is obtained for the electron affinity: $\rm EA = 0.5 \pm 0.3$ eV.
Consequently, the error for the band gap is at most 0.6 eV
that is 15\% of its value. However, we have to emphasize that the
estimated error of 0.3 eV is an upper bound for the energy
of any band at any $k$-point which is in the order of 1\% of the total range
of band energies only.

As one can conclude, the two main reasons, which lead to the
error bar for correlation corrections up to 15\% of their
values, are i) the error due to truncation of the incremental
scheme after the third order and ii) the size-extensivity error
of the CI(SD) method for evaluation of correlation energies
of $(N\pm 1)$- and $N$-electron states. The truncation error of the
incremental scheme can not be reduced by taking into account
the next order because the values of the increments of the
fourth order are below the estimated size-extensivity error
for the entering correlation energies. Therefore, namely the latter error controls
the accuracy of our whole approach. To improve the
accuracy further one first needs a true size-extensive multireference
correlation method that could be multireference
coupled-cluster which is at the stage of development
at the present moment.

\section{Comparison to experimental data}

In this section we would like to compare our result
for the band gap to experimental data. One has to
emphasize that there are, of course, no experimental data on
{\it trans}-polyacetylene {\it single} chains and all data
available correspond to bulk
{\it trans}-polyacetylene and we have to
compare properties of a bulk system with
calculated results for the linear one.

One attempt to estimate the effect of the interchain coupling
on the band gap of tPA single chains was made by Ayala
{\it et al.} \cite{Ayala01}. They have compared the band
gap of an isolated single chain with the band gap of a system
consisting of two chains one precisely on the top of another
at a distance of 3.335 $\overset{\circ}{\rm A}$. In this
particular arrangement of the two chains the intermolecular
overlap of the $\pi$ orbitals (which
define the gap) is maximal. The chosen distance
corresponds to the minimum of the interaction energy of
the two stacked chains on the MP2 level of approximation.
The reduction of the band gap due
to interchain coupling was found to be 1.22 eV on the HF
level and further a 0.22 eV reduction (1/6 of the HF value)
was obtained when electron correlation was taken into
account (the M\o{}ller--Plesset perturbation method to
second order was used for this purpose). Thus, one can
conclude that the main reduction of
the band gap due to the interchain interaction already emerges
on the HF level.

We investigate this reduction for the true fishbone-like packing
of tPA chains in the bulk material as shown in Fig.~\ref{fishbone}
which was found experimentally \cite{Fincher82}. The unit cell of
bulk tPA consists of two C$_2$H$_2$ units of linear tPA (one on
each non-equivalent chain). The interchain interaction leads to
the splitting of the otherwise two-fold degenerate bands of single
tPA chain. This splitting at the $X$ point amounts to a 0.6 eV
reduction of the band gap of the bulk system as compared to that
of a single chain. Further reduction of the band gap due to
electron correlation in each of the four nearest-neighbor chains
is not expected to be noticeable because in the
experimentally-found packing of the chains the overlap of the
$\pi$ orbitals between neighbor chains is actually minimal. Thus,
based on the MP2 data presented in \cite{Ayala01} we estimate it
to be less than 1/6 of the HF reduction for each of the four
chains leading at most to a further 0.2 eV narrowing of the band
gap. Therefore, the account for the interchain splitting
reduces the calculated band gap of the bulk tPA down to 3.3 eV.

Two other effects should also be mentioned which can lead to a
reduction of the calculated band gap of tPA. Firstly, though our
basis set (VTZ) is already rather flexible we have not yet reached
the basis set limit. A more rich basis set will affect the
correlation corrections of the LMEs as a larger variational space
will be provided to the MRCI calculations. Secondly, the
long-ranged polarization of the infinite {\it bulk} system has to
be considered. There, one expects a more pronounced effect than in
the case of a one-dimensional system. A discussion of the
polarization effect on local matrix elements of different bulk
systems can be found in \cite{Graef97} and \cite{Albrecht00}.
However, both effects mentioned here are small compared to the
calculated electron correlation effects and each of them can lead
to a few tenth of an electron-volt reduction of the band gap only.

Next, we would like to address the experimental value of 2 eV for
the tPA band gap obtained from absorption spectra of bulk {\it
trans}-polyacetylene \cite{Fincher79}, \cite{Tani80} and
\cite{Leising88}. There is no clear indication in these
publications that the observed maximum of the absorbtion
coefficient at approximately 2 eV can really be assigned to the
interband transition rather than to an excitonic state. A
theoretical investigation of such gap states showed
\cite{Rohlfing99} that the energy of these excitons in tPA show up
approximately 0.4 eV below the band gap. Therefore, the
fundamental band gap of bulk {\it trans}-polyacetylene is expected
to be 2.4 eV.

Thus, taking into account the bulk effects in tPA which all lead
to a reduction of the band gap with respect to the one in a 1D
single chain and also the estimated error of our method the
calculated gap reaches the experimental value.

\chapter{Excitons}

In this Chapter we would like to show how our approach
to correlated band structures
can be extended to the study of the electron correlation
effects of {\it neutral} excitations in periodic
systems, namely excitons, such that standard 
quantum-chemical methods can be used. We refer here
to the process of the absorption of a light quantum by the
electrons of an extended non-conducting system
in its ground state i.e. the process of generating an excited 
state of the same {\it neutral} system.
In this case no electron leaves the system (in contrast
to the photoionization process which we refer to when
describing the valence bands).

In frames of the most simple model one can say that an electron
from the valence band $\nu$ and with band energy
$\varepsilon_{k\nu}$ is excited to the conduction band $\mu$ where
it occupies the virtual Bloch orbital $\psi_{k^\prime \mu}$ with
energy $\varepsilon_{k^\prime \mu}$. The wavefunction of such
a state can be obtained by applying the two operators $c_{k\nu}$
and $c^\dagger_{k^\prime \mu}$ entering the equations (\ref{PhiN-1k})
and (\ref{PhiN+1k}) to the ground-state wavefunction (\ref{PhiNk})
of the closed-shell system (we omit here the spin indices as we
suppose that the electron does not change its spin during such a
process). The operator $c_{k\nu}$ produces a hole in the valence
band $\nu$ and the operator $c^\dagger_{k^\prime \mu}$ creates an
electron in the conduction band $\mu$. If we neglect the
electron-hole interaction, the energy to produce shuch an excited state
is equal to $\varepsilon_{k^\prime \mu} - \varepsilon_{k \nu}$.
This way we describe a pure interband transition and the
threshold energy for such processes is equal to the fundamental
gap of the non-conducting system.

However, if we take into account the interaction between the
electron in the conduction band and the hole in the valence band,
the system may gain energy from the electron-hole interaction
when a bound electron-hole pair is formed. In other words, in the
extended system there exist {\it gap} states which correspond to
excitons (bound electron-hole pairs) travelling through the
crystal. Such gap states were found experimentally in many
semiconductors and insulators, in particular in {\it
trans}-polyacetylene \cite{Lauchlan81}, \cite{Orenstein82} and
\cite{Shank82}, the system which we refer to here. These states form
bands and we would like to use our approach to calculate these
excitonic bands. This means, we want to construct a set of localized
model electron-hole pair wavefunctions (forming our excitonic model 
space) which can be represented in finite clusters, and
calculate the corresponding local matrix elements (diagonal and
off-diagonal) in sufficiently large clusters of the crystal. These
{\it excitonic} LMEs are essentially the matrix elements of an
effective Hamiltonian obtained from an equation similar to
Eq.~(\ref{Heff_ij}), and we will get the excitonic bands by the
use of a (back) transformation analogous to Eqs.~(\ref{IPn}) and
(\ref{EAm}).

We exclusively refer here to bound excitons which carry a single
crystal momentum (associated with the center of mass motion) and
are otherwise regarded as a fixed entity. The appealing aspect of
this approach is that in quantum chemistry which deals with finite
systems an exciton is just an eigenstate of the given system like
the ground state, but different from it, and the same CI ansatz
and methodology as used for the correlated ground-state wavefunction
can be used. In fact, the CI equations are the same and one is
just looking at another root of these equations.

To describe a bound electron-hole pair of a crystal
in real space it is convenient to use local Wannier orbitals
(both occupied and virtual ones) to
construct the {\it local} excited
configurations. Let as consider one such configuration
$\Phi_{\bf{R} n}^{\bf{R}^\prime m}$ where an electron is removed
from the occupied Wannier orbital $\varphi_{\bf{R} n}$
centered at the unit cell with
lattice vector ${\bf R}$ and put into the virtual orbital
$\varphi_{\bf{R}^\prime m}$ centered at the unit cell with lattice vector
${\bf R}^\prime$ while the rest of the system remains unchanged
(i.e. we start again from the frozen orbital approximation).
Taking into account Eqs.~(\ref{E_0})--(\ref{EA_r}) one can readily
show that the energy to produce an excited state with this
configuration is equal to

\begin{eqnarray}
\label{exc_conf_energy}
E^{\rm exc} & = & \langle \Phi_{\bf{R} n}^{\bf{R}^\prime m} \: | \: H \: |
\: \Phi_{\bf{R} n}^{\bf{R}^\prime m} \rangle - E_0  \nonumber \\
& = & \varepsilon_m - \varepsilon_n -
\bigl(\langle {\bf R}^\prime m \: {\bf R} n \: | \: {\bf R}^\prime m \:
{\bf R} n \rangle - \langle {\bf R}^\prime m \: {\bf R} n \: | \:
{\bf R} n \: {\bf R}^\prime m \rangle \bigr)   \nonumber \\
& = & - {\rm EA}_m + {\rm IP}_n -
\bigl(\langle {\bf R}^\prime m \: {\bf R} n \: | \: {\bf R}^\prime m \:
{\bf R} n \rangle - \langle {\bf R}^\prime m \: {\bf R} n \: | \:
{\bf R} n \: {\bf R}^\prime m \rangle \bigr)
\end{eqnarray}
where we have used $| \:{\bf R} n \rangle$ instead of $| \:
\varphi_{{\bf R} n} \rangle$, and $\varepsilon_m$ and $\varepsilon_n$
are the energies of the Wannier orbitals (associated with the bands $\mu$ and
$\nu$, respectively) which are independent of the vectors of lattice
translation ${\bf R^\prime}$ and ${\bf R}$. The energy $E^{\rm
exc}$ is not equal to the difference of the ionization potential
IP$_n$ and the electron affinity EA$_m$ which defines the
interband transition but differs from it by the term
$\langle {\bf R}^\prime m \: {\bf R} n \: | \: {\bf
R}^\prime m \: {\bf R} n \rangle - \langle {\bf R}^\prime m \:
{\bf R} n \: | \: {\bf R} n \: {\bf R}^\prime m \rangle$. This
term decays with increasing distance between the centers of the
Wannier orbitals $\varphi_{\bf{R} n}$ and $\varphi_{\bf{R}^\prime
m}$ (as follows from Eq.~(\ref{v})) and is the interaction energy
between the electron and the hole. Such terms will be responsible
for the formation of bound electron-hole pairs when we
describe excited states of crystals (or big clusters) as linear
combinations of local configurations like $\Phi_{\bf{R}
n}^{\bf{R}^\prime m}$. The energies of such states can be
substantially reduced compared to the energy of the interband
transitions and thus the gap states of semiconductors and
insulators may emerge.

Below, for the sake of simplicity, we consider a single valence and
conduction band only that yields one occupied and one unoccupied
Wannie orbital per unit cell. Such a simplification is justified for
{\it trans}-polyacetylene, when excitations of the
valence $\pi$ electrons to the lowest conduction band of $\pi$
symmetry are considered. We then can omit the band indices $n$ and $m$
baring in mind that subscripts (superscripts) in 
$\Phi_{\bf{R} n}^{\bf{R}^\prime m}$ correspond to the hole
(electron) orbitals. Also, again we will refer to a one-dimensional
system as the excitons in the conjugated $\pi$ system of polymers
are essentially confined to
a single polymer chain \cite{Friend99}.

Let us start to describe a bound electron-hole pair by writing an
excitonic wavefunction in the form of a linear combination of local
configurations $\Phi_{R}^{R^\prime}$:

\begin{equation}
\label{general_ansatz}
\Psi^{\rm exc} = \sum_{RR^\prime} \alpha_R(R^\prime)
\Phi_{R}^{R^\prime} = \sum_{R} \sum_{d} \alpha_R(d)
\Phi_{R}^{R+d}
\end{equation}
where we introduced the relative distance $d$ of the centers of the
Wannier orbitals $\varphi_R$ and $\varphi_{R^\prime}$: 
$d=R^\prime - R$. Here we have to make a few remarks.
Let us focus for a moment on a wavefunction where the hole
is pinned to a certain unit cell, $R$, say. Then the CI ansatz
(\ref{general_ansatz}) for the exciton is reduced to

\begin{equation}
\label{fixed_hole}
\Psi_R^{\rm exc} = \sum_{d} \alpha_R(d)
\Phi_{R}^{R+d}
\end{equation}
Firstly, if we set all coefficients $\alpha_R(d\neq 0)$ to zero and
$\alpha_R(0)=1$ we get a so-called Frenkel exciton \cite{Frenkel31}
where the excitation is restricted to one unit cell. Yet, we
use the ansatz (\ref{fixed_hole}) which allows us to describe
excitons in the general case when the atomic structure of the given crystal is
explicitly taken into account. Secondly, it is possible to produce
$N^{\rm virt}$ excitonic states using the ansatz (\ref{fixed_hole}) where
$N^{\rm virt}$ is the number of virtual WOs in the system. However,
we focus our attention on the energetically lowest excited state
which corresponds to the gap state and is usually well separated
from the other excitonic states by some energy. Thirdly, doubly
excited configurations like $\Phi_{R,R^\prime}^{R+d, R+d^\prime}$ may, of course,
also contribute to the wavefunction of the excited state but here
we aim to construct our model space, which is built up by singly excited configurations
which describe the electron-hole pair as being a two-particle object.
The effect of doubly (and higher) excited configurations on the energy of
excited states can be taken into account later by the MRCI procedure.

The representation of the exciton-state wavefunction given by
the right-hand-side expression in
Eq.~(\ref{fixed_hole}) suggests to plot the distribution of
the electrons relative to the fixed hole. As the electron-hole 
interaction energy (the two last terms in
Eq.~(\ref{exc_conf_energy})) contained in each configuration
decays with the distance between the electron and the hole orbital,
one expects configurations with small distances $d$
to dominate the wavefunction (\ref{fixed_hole}). This
expectation is based on the fact that the electron tries to stay as
close as possible to the hole.

An investigation of such electron-hole pairs in a
{\it trans}-polyacetylene single chain was presented in
Ref.~\cite{Rohlfing99} which was done in the framework of a different
(recently developed) approach based on solving the
Bethe--Salpeter equation of the two-particle Green's function
and using the $GW$ approximation for the self-energy operator. That
approach allowed the authors to visualize the distribution of the
electron relative to the hole fixed somewhere in real space (see
Fig.~\ref{e_h_pair}). From this Figure one concludes that
such localized objects can indeed be found in periodic
non-conducting systems and the probability to find the electron
of this pair at some place decays with the distance from the fixed hole.
Hence, a finite cluster can be used to represent this
localized excitonic state.

\begin{figure}
\centerline{ \psfig{figure=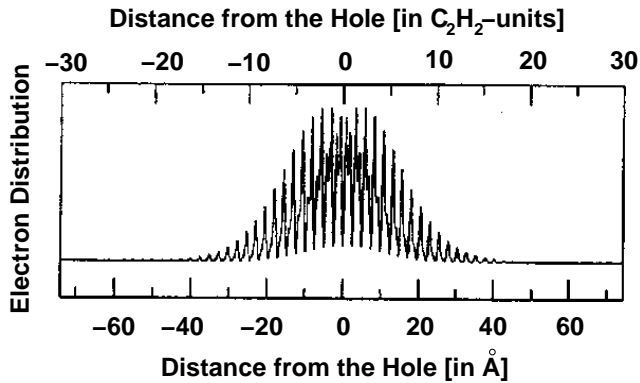,width=3.3in} }
\caption{Electron-hole wavefunction of the exciton state of a tPA
single chain showing the distribution of the electron relative
to the hole which is fixed at some arbitrary position on the chain
(at 0~{\AA}). The figure is taken from
Ref.~\cite{Rohlfing99} where the calculations were done within the
$GW$ approximation and using the Bethe--Salpeter equation.}
\label{e_h_pair}
\end{figure}

As follows from Fig.~\ref{e_h_pair}, the dimension of a single electron-hole
pair in a tPA chain is approximately 30 unit cells that implies that
a C$_{60}$H$_{60}$ cluster is required to properly describe such an
exciton within our cluster approach. Presently
we can not stand calculations for such big clusters of
{\it trans}-polyacetylene with reasonable basis sets. Therefore, all
our calculations presented in this Chapter are done for a much simpler system
consisting of 9 H$_2$ molecules which are arranged analogously to the system of
4 H$_2$ molecules considered in section~3.5 (see Fig.~\ref{4H2}) but
with the distance of 3~{\AA} between neighbor
H$_2$ molecules to allow the exciton to travel along the chain. 
This system represents a 9-unit-cell cluster of an
infinite $({\rm H}_2)_x$ chain which can be considered as a simple model
for the $\pi$ system of a tPA chain.
In our $9 \times {\rm H}_2$ cluster we first generate localized occupied and
virtual molecular orbitals (in fact, bonding and anti-bonding orbitals of the
H$_2$ molecules) and then compute the energetically lowest excited state
when only the configurations $\Phi_5^{5+d}$ with the hole fixed at the central
molecule H$_2$ (the 5-th one) and the electron in any of anti-bonding orbitals in
the 9 H$_2$ chain are used (by a reference space only CI-calculation using
the MOLPRO program package \cite{MOLPRO}, \cite{Werner88}, \cite{Knowles88},
\cite{Knowles92}).
The squares of the coefficients $\alpha_5(d)$ of these configurations,
being the probabilities to find an electron in the localized virtual
orbital centered in the unit cell $R^\prime = 5+d$, 
are presented in Table~\ref{e_h_pair_9H2}.
We find the same type of electron distribution around the fixed hole
as reported for a tPA single chain in Ref.~\cite{Rohlfing99} (see
Fig.~\ref{e_h_pair}). Therefore, the electron-hole interaction,
which enters the state energy via terms analogous to the two last
terms on the right-hand side of Eq.~\ref{exc_conf_energy}, leads to
the formation of well-bound electron-hole pair and such a (pinned) pair
in a crystal can be described in the form of the linear combination
(\ref{fixed_hole}) with only a finite amount of local excited 
configurations $\Phi_R^{R+d}$.

\begin{table}
\refstepcounter{table}
\addtocounter{table}{-1}
\label{e_h_pair_9H2}
\caption{Sqares of coefficients $\alpha_5(d)$ of the configurations
$\Phi_5^{5+d}$ entering the local excitonic wavefunction (\ref{fixed_hole})
in a $9 \times {\rm H}_2$ cluster of an
$({\rm H}_2)_x$ chain with the hole being fixed at the central (5-th)
H$_2$ molecule.}
\vspace{5mm}
\centerline{
\renewcommand{\baselinestretch}{1.2}\normalsize
\begin{tabular}{c|ccccc}
\hline
 $d$ &
 0 &
 $\pm  a$ &
 $\pm 2a$ &
 $\pm 3a$ &
 $\pm 4a$ \\
\hline
 $(\alpha_5(d))^2$ & 0.951 & 0.024 & $1.54 \times 10^{-4}$ &
$5.84 \times 10^{-7}$ & $1.79 \times 10^{-9}$\\
\hline
\end{tabular}
}
\end{table}
\renewcommand{\baselinestretch}{1}\normalsize

Such localized states given by Eq.~(\ref{fixed_hole}) are not
eigenstates of the periodic system as they are degenerated. To
remove the degeneracy we proceed as in chapter~II of
Ref.~\cite{Knox} that is the pinned-hole excitonic wavefunctions
$\Psi_R^{\rm exc}$ are treated as rigit objects and a set of LCAO
ansatz is made for the wavefunction of the excited state in
the one-dimensional crystal (like in a tight-binding model):

\begin{equation}
\label{fixed_hole_cryst}
\Psi_K^{\rm exc} = \sum_R {\rm e}^{iKR} \Psi_R^{\rm exc} =
\sum_R {\rm e}^{iKR} \sum_{d} \alpha_R(d)
\Phi_{R}^{R+d}.
\end{equation}
Here $K$ is the crystal momentum of pair made up by the hole and the electron
being distributed among virtual WOs in the close vicinity of this
hole. The ansatz (\ref{fixed_hole_cryst}) takes into account the
periodic structure of the crystal and the excited states of the
crystal can be referred to
"particles" of neutral electronic excitations (which are called
{\it excitons}) travelling through the crystal with wave
vector $K$. The energies to produce such excited states
$\varepsilon_K^{\rm exc}$ regarded as a function of $K$ form the {\it
excitonic} bands which can be obtained from the finite-cluster
calculations.

In the $9 \times {\rm H}_2$ cluster we open all 9 bonding and 9
anti-bonding orbitals to construct the singly excited
configurations $\Phi_{a}^s$ which form the model space and 81 
excited states can be obtained
as linear combinations of these configurations:

\begin{equation}
\label{exc_cluster}
\Psi_i^{\rm exc} = \sum_{a=1}^9 \sum_{s=1}^9 \alpha_{i \:a}^{\;\; s}
\Phi_{a}^s
\end{equation}
where index $a$ ($s$) denotes an occupied (virtual) localized
molecular orbital. Actually this equation is just the finite-cluster
version of the crystalline ansatz given in Eq.~(\ref{general_ansatz}).
Among all these excited states we find 9 states which are
energetically well-separated from the rest and lower than all other 
excited states. These 9 states $\Psi_i^{\rm exc}$, $i=1,\ldots,9$ 
are strongly-bound excitons we are looking for here and can
be regarded as some linear combinations of 9 suitable localized 
excited states. These 9 "delocalized" states $E_i^{\rm exc}$ are
easily identifiable since the energies of these states only spread over an energy
interval of 1.4~eV (this value is approximately equal to the
excitonic band width) while they are separated by a gap of 2.1~eV from
all the other excited states. In fact, we focus our attention here on 
the lowest excitonic band which
corresponds to the "ground state" of an exciton.

To calculate the excitonic band in the framework of our approach we have
to define local excitonic matrix elements (which are stransferable from
clusters to the infinite periodic system):

\begin{equation}
\label{LME_exc}
{\rm EX}_{ij} = H^{\rm exc}_{ij} -  \delta_{ij} E_0
= \langle \Phi_i^{\rm exc} \: | \: H \: | \:
\Phi_j^{\rm exc} \rangle - \delta_{ij} E_0
\end{equation}
in terms of some {\it localized} excitonic wavefunctions $\Phi_i^{\rm exc}$.
The latter are obtained by a unitary transformation applied to
the $N^{\rm exc}$ delocalized excitonic wavefunctions $\Psi_i^{\rm
exc}$ in the cluster

\begin{equation}
\label{loc_exc_cluster}
\Phi_i^{\rm exc} = \sum_{j} \Psi_j^{\rm exc} u_{ji}
\end{equation}
where the unitary matrix $u_{ji}$ is determined by minimizing 
the total spatial spread of the electron-hole pairs

\begin{equation}
\label{exc_functional} I^{\rm exc} = \sum^{N^{\rm exc}}_{i=1}
(\sigma^{\rm exc}_i)^2
\end{equation}
(cf. Eq.~(\ref{FBfunctional})). Below we provide one possible definition of
the spatial spread of the excitonic wavefunction in terms of
singly excited configurations.

Note that the introduction of an electron-hole pair with a pinned
hole was only used to develop the idea of the
quantum-chemical description of a well-bound exciton in terms of
singly excited configurations. However, in both, crystals and
clusters there exist no such fixed hole, and electrons and holes
together are delocalized. Hence, it seems natural to
describe the bound electron-hole pairs in terms of {\it excitonic}
coordinates, i.e. a coordinate $\bar{R}$ of the "center of mass" of the
exciton and a relative coordinate $\bar{r}$. They can be defined as 
the population-weghted
average of the "centers of mass" and the relative coordinate
$R_a^s$ and $r_a^s$ of each of the configurations $\Phi_a^s$ 
entering the excitonic wavefunction:

\begin{equation}
\label{exc_coords} \bar{R}_i=\sum_{a,s}R_a^s \: (\alpha_{i
\:a}^{\;\; s})^2, \qquad \bar{r}_i=\sum_{a,s}r_a^s \: (\alpha_{i
\:a}^{\;\; s})^2
\end{equation}
(cf. Eq.~(\ref{el_position})) where

\begin{equation}
\label{conf_coords} R_a^s = \frac{1}{2}(R_a + R_s) \qquad {\rm
and} \qquad r_a^s = \frac{1}{2}(R_a - R_s),
\end{equation}
$R_a$ denotes the center of the bonding orbital
$\varphi_a$ and $R_s$ the center of the
anti-bonding orbital $\varphi_s$.

The spread of each electron-hole pair is defined by the average
distance between the electron and the hole (given by $\bar{r}_i$) 
and by the average deviation of the "center of
mass" of each contributing configuration $R_a^s$ from the
"center of mass" $\bar{R}_i$ of the exciton itself:

\begin{equation}
\label{exc_spread} (\sigma^{\rm exc}_i)^2 = (\bar{r}_i)^2 + \sum_{a,s} (R_a^s
- \bar{R}_i)^2 \: (\alpha_{i \:a}^{\;\; s})^2
\end{equation}
(cf. Eq.~(\ref{orb_spread})).

Having obtained the unitary matrix $u_{ji}$ which corresponds to
the minimum of the functional $I^{\rm exc}$ in
Eq.~(\ref{exc_functional}) or some other measure for the extent of 
the exciton, we get $N^{\rm exc}$
localized excitonic wavefunctions $\Phi_i^{\rm exc}$. Then, the
matrix elements of the Hamiltonian $H^{\rm exc}_{ij}$ can be
calculated analogously to the elements of the effective
Hamiltonian for hole or electron attachment states as
given by Eq.~(\ref{Heff_ij}):

\begin{equation}
\label{H_exc} H^{\rm exc}_{ij} = \sum_{i^\prime} u_{i^\prime i} \:
E_{i^\prime}^{\rm exc} \: u_{i^\prime j}.
\end{equation}

In the periodic system one expects to find one such localized excitonic
wavefunction $\Phi_{R}^{\rm exc}$ in each unit cell (with
$R$ being the translation vector of this unit cell) which all together
correspond to the lowest excitonic band of the crystal. Note, that here the center of
the exciton is its "center of mass" $\bar{R}_R$ in contrast to the
position of the fixed hole in the ansatz (\ref{fixed_hole}).
The local excitonic matrix elements in the periodic system are then
defined as

\begin{equation}
\label{LME_exc_cryst}
{\rm EX}_{R} = H^{\rm exc}_{R} -  \delta_{0R} E_0
= \langle \Phi_0^{\rm exc} \: | \: H \: | \:
\Phi_R^{\rm exc} \rangle - \delta_{0R} E_0
\end{equation}

The diagonal matrix element EX$_{0}$ is the (Brillouin zone averaged) energy of
the exciton and the off-diagonal elements EX$_{R}$ are the hopping
matrix elements between the reference (zero) unit cell and the unit cell at $R$.
As the cluster represents a finite part of the crystal one can
associate the excitonic LMEs EX$_{ij}$ obtained in the cluster
(for the two localized excited configurations $\Phi_i^{\rm exc}$
and $\Phi_j^{\rm exc}$ with centers $\bar{R}_i$ and $\bar{R}_j$,
respectively) with the crystalline LMEs EX$_{\bar{R}_j - \bar{R}_i}$. The
closer the centers of the chosen localized excitonic wavefunctions
$\Phi_i^{\rm exc}$ and $\Phi_j^{\rm exc}$ are to the center of the
cluster the better the LMEs EX$_{ij}$ from the cluster approximate
the corresponding crystalline LMEs EX$_{R}$.

The excitonic band finally can be obtained by the local excitonic
matrix elements (\ref{LME_exc_cryst}) being subject to a transformation
similar to Eqs.~(\ref{IPn}) and (\ref{EAm}):

\begin{equation}
\label{EXn} {\rm EX} (k) = \sum_R {\rm e}^{ikR} {\rm
EX}_{R}.
\end{equation}
As the local off-diagonal matrix elements EX$_R$ are expected to
decay with increasing distance $R$ between the centers of the localized
electron-hole pairs $\Phi_0^{\rm exc}$ and $\Phi_R^{\rm exc}$, the
series (\ref{EXn}) can be truncated at some cut-off radius 
$R_{\rm cut}$ such that the omitted part of the series is negligible
with respect to the terms already summed up. Therefore, one only needs a
finite number of the off-diagonal matrix elements to calculate the
excitonic band with a given accuracy, and both, diagonal and
off-diagonal excitonic matrix elements can be obtained from finite
clusters.

\chapter{Conclusions and perspectives}

\vspace{-0.3cm}

The main result of the present thesis is the implementation of a
multireference configuration interaction method with singly and
doubly excited configurations to account for the electron correlation
effects in both neutral and charged infinite periodic systems, in
particular in excited electron hole and attached electron states.
Our approach is one of the first approaches for solids which is
purely based on the correlated wavefunction of the system.
Sufficiently large basis sets (valence triple zeta) to describe
the correlations especially in the anionic ($N+1$)-electron
systems are used. As an output we obtain the quasiparticle band
structure (all valence and the lowest conduction bands) of a
non-conducting material where electron correlation is
systematically taken into account which allows to extract the
ionization potential, the electron affinity, the band gap and the
band widths of the material as those properties which can be
measured in experiments. All these quantities are obtained with
{\it controlled} accuracy. Thus, the approach provides highly
desired information on electronic properties of extended systems
on a very high level of sophistication and, as we believe, will
find many interesting applications. We have applied it to {\it
trans}-polyacetylene single chains and obtained complete and {\it
quantitatively} correct information on electronic properties of
the investigated system. Further, an extension of the above
approach to excitons (i.e. optical excitations) in crystals is
developed which allows to use standard quantum-chemical methods to
describe the electron-hole pairs and to finally obtain {\it
excitonic} bands.

Our method starts from the many-body wavefunction of the infinite
closed-shell system obtained in the Hartree--Fock approximation.
Local quantities in {\it real} space (namely, localized
electron orbitals and local matrix elements) are then defined
which are well-transferable from finite clusters to the infinite
system. By the use of these local matrix elements derived from
clusters the SCF band structure of the extended system is
accurately reproduced. Exploiting the local character of the
electron correlation we perform correlation calculations by the
MRCI(SD) method in the finite clusters for the $N$-, ($N-1$)- and
($N+1$)-electron states and as a result obtain the change of the
local SCF matrix elements due to the correlation effects which
again turned out to be well-transferrable from the clusters to the
periodic system.

To reach the current stage of implementation of our approach the
following challenging problems were solved. A proper treatment of
the ($N+1$)-particle states requires {\it localized} virtual SCF
orbitals which are essentially the same for both infinite periodic
systems and clusters. Such orbitals can not be generated in a
cluster by any standard localization scheme applied to some ad hoc
set of the canonical virtual orbitals of the cluster (in contrast
to the occupied orbitals). Therefore, we first generate a set of
localized virtual Wannier orbitals in the crystal which
correspond to bands of our interest and then project them onto the
space of the virtual canonical orbitals of a proper cluster. This
way we are able to generate localized virtual molecular orbitals
and the relevant local SCF matrix elements in the clusters which
are used further in the correlation calculations.

To systematically account for the correlation effects the
incremental scheme was exploited. However, this scheme (originally
designed for the ground and hole states calculations) has the
problem of the emergence of a divergently large number of
increments when bigger clusters are considered. To handle this
problem the scheme for the hole states was reformulated in terms
of orbital groups. Also, the scheme was extended to the case of
attached-electron states.

We showed that within the multireference configuration interaction
method one can also handle the problem of satellite states. This
problem appears in systems with a relatively small band gap and
leads to severe instabilities in the calculations when many-body
perturbation theory is used to account for the correlation
effects. In the MRCI method the satellite states can be treated
together with the true hole or attached-electron states by a
proper modification of the reference space however by the cost of
increasing computational effort.

The use of configuration interaction methods with singly and
doubly excited configurations in combination with the incremental
scheme requires a very accurate handling of the size-extensivity
error intrinsic to all truncated CI methods. We could show that
the known Pople correction applied to the correlation energy of
the neutral closed-shell system reduces this error to the required
level. However, there did not exist any correction for the
size-extensivity error of the MRCI(SD) method, which in case of
the charged open-shell systems could lead to a comparable
precision. In the present work we have developed such a
correction. The analytic formula for it is derived and its
performance is checked. The new correction allowed us to extract
the final results with a reasonable accuracy.

The whole approach is focused on obtaining local matrix elements
which correspond to correlated wavefunctions of the system and
contain information on the $N$-, ($N-1$)- and ($N+1$)-electron
states of a system when electron correlation is systematically
included. Using these matrix elements we were able to compile a
band structure (both valence and conduction bands) of the chosen
crystal which fully takes into account correlation effects. From
this band structure we extract correlation-corrected values of some
characteristic electronic properties of the system which can be
measured experimentally. They are the band gap, the band widths, and
the ionization potential and electron affinity. We applied our
method to an infinite {\it trans}-polyacetylene single chain. We
found that upon inclusion of electron correlation, the calculated
results approved substantially as compared to those on the HF
level and that they are in good accordance with the experimental
data once the effect of interchain interaction and long-range
polarization in {\it bulk} tPA are also taken into account.

As a remarkable feature of our method we emphasize the possibility
to estimate the error for the obtained quantities and to
systematically improve the results to any desired accuracy by just
increasing the computational effort. For the calculated correlated
band structure of a {\it trans}-polyacetylene single chain which
spans an energy range of approximately 30 eV the estimated error
for any band energy at any $k$-point does not exceed 0.3 eV.

Comparison of our results on a tPA single chain to the
experimental data on the bulk crystal, composed of weakly
interacting chains, indicates that the study of correlation
effects in polymers can not be reduced completely to the treatment
of isolated chains. There is clear evidence that accounting for
the mutual interaction of the tPA chains in the bulk material
improves the theoretical results with respect to the experimental
data. The correlation of electrons in neighbor chains has to be
considered for that purpose. This can be done with our approach.
However, to calculate these effects explicitly is beyond the scope
of the present study.

Finally, a new wavefunction-based approach which allows to
describe well-bound excitons in crystals by means of standard
quantum-chemical methods is established. Exploiting the finite
dimension of a bound electron-hole pair (extending over several
unit cells) we can restrict ourselves to finite clusters of a
crystal to calculate the relevant properties of such a compact
object. A localization scheme to generate localized excitons in
clusters is proposed. Then, the local {\it excitonic} matrix
elements as obtained in the finite clusters can be transferred to
the periodic system, and, as the ultimate result of this approach,
the {\it excitonic} bands can be calculated.

At the end we would like to discuss some perspectives of our
approach. The present study opens the possibility to apply highly
accurate standard quantum chemical method, designed to account for
the correlation effects in finite molecules, to a wide range of
crystalline insulators and semiconductors being periodic systems.
The approach provides information on the correlated wavefunction
of the system and gives {\it quantitatively} correct results for
band energies with a controlled precision which can systematically
be improved by increasing the computational cost. With such a
combination of highly desired features our approach has no
analogue at the present moment.

In the present work the method was applied to a semiconducting
polymer, i.e. to a linear system. However, this study was going
hand-in-hand with the application of the same methodology to bulk
diamond carried out in our laboratory. By now, that investigation
of bulk diamond has also been finished successively demonstrating
that our approach can equally well be applied to true
three-dimensional periodic infinite non-conducting systems.

As discussed in section 3.4, the main computational efforts of our
approach goes into the evaluation of a large number of individual
increments (roughly two hundred) which corresponds to the
calculation of several hundreds of the $N$-, ($N-1$)- and
($N+1$)-electron states (using a limited number of configurations
only) providing us with the information on the energies and the
wavefunctions. Such an amount of information requires an
automatization of the calculations and data processing. An
appropriate general-purpose computer tool, which automatically
generates the input files for the correlation calculations with
the MOLPRO program package, picks up all the relevant data from
the output files and also computes the matrix elements of the
effective Hamiltonians and the size-extensivity corrections to the
correlation energies, evaluates the individual increments, checks
for missing increments and properly sums up the increments of the
correlation correction of some particular quantity, have been
already developed in our laboratory. Also, some code has been
designed to assist the user in generating clusters from a given
crystal. Nonetheless, to become a routine tool used by many
researchers studying solids a complete automatization of our
method is required. By this we mean a further automatization of
the projection of the Wannier orbitals generated by the CRYSTAL
code onto the clusters, a dynamic and automatic control of the
convergence of the increments with the distance between the active
bonds and (by this) a proper choice of suitably-sized clusters for
the correlation calculations, a grouping of molecular bonds, the
summing up of the increments to both the SCF and the
correlation-corrected local matrix elements and finally the
compiling of the bands by the use of these LMEs. Many programm
tools to tackle these tasks for some particular classes of systems
have already been established but most of them are not yet fully
general-purpose codes.

On the other hand, there is an increasing activity in the
implementation of quantum-chemical correlation methods for
extended systems (MP2, say) into existing packages for the SCF
calculations which account for the electron correlation directly
in the periodic systems using local representations of the
single-particle wavefunctions and exploiting the predominantly
local character of the correlation effects. The main difference of
these approaches to our one is the requirement of {\it all} the
virtual orbitals in the periodic system being localized. Then, by
some cutoff criteria on the mutual distance of the localized
orbitals the problem is effectively reduced to the case of
correlation calculations in clusters (or some other suitable
finite subspace). Thus, our experience will be very useful for the
developers of such new codes.

Let us finally turn to our approach to excitons in crystals which
was established in the present work. Here, a completely new
direction in the use of standard quantum-chemical program packages
is opened. Further development and an explicit implementation of
this approach seems to be a very attractive task since this way
the optical properties of non-conducting crystals can be studied
by the already existing highly accurate quantum-chemical methods
providing quantitatively correct results. There are many open
question in this new field of research and, as we believe, they
can be answered soon.


\baselineskip=12pt
\addtocontents{toc}{\vspace{15pt}}
\addtocontents{toc}{\hspace*{-17pt}{\bf Bibliography} \hfill
{\bf \arabic{page}} \newline}
\bibliography{Thesis}
\bibliographystyle{alpha}



\addtocontents{toc}{\addvspace{20pt}}
\addtocontents{toc}{\hspace*{-5pt}
{\bf Acknowledgements} \hfill
{\bf \arabic{page}}}
\pagestyle{plain}
\pagenumbering{arabic}
\setcounter{page}{111}

\newpage
\renewcommand{\baselinestretch}{1}\normalsize

\centerline{\Large \bf Acknowledgements}
\vspace{1cm}

I am deeply grateful to Prof.~Dr.~Peter Fulde, director of the
Max Planck Institute for the Physics of Complex Systems in
Dresden, for his kind invitation to work in this
modern international research centre with its impressive computing
facilities and very friendly working atmosphere. He
drew my attention to this fascinating field of
research which brings together physics, chemistry
and computer science and for that I would like to give him
my special thanks. I am also grateful to him for the
encouragement and continuous interest in my work during
the whole time of my study at the Institute and for the many
illuminating and helpful discussions.

I am very grateful to my advisor Dr.~habil.~Uwe
Birkenheuer, head of the Quantum Chemistry group, for
introducing me to my current research field
and for his constant and patient guidance of my work.
Many of the ideas presented in this thesis emerged
in discussions with him. I came to appreciate his
very kind personality. He could always find the time
for the listening to my little problems and never
refused to provide any kind of help or advice.
Also I am indebted to him for
the developed computer tools which
enormously speeded up the data processing.
Owing to all mentioned above it was possible to complete
the study in a rather limited amount of time.

I would like to thank all my co-laborators
Dr.~Dmitry Izotov, Dr.~Malte von Arnim, Dr.~Walter
Alsheimer, Dr.~Liudmila Siurakshina, J.~Prof.~Dr.~Martin
Albrecht, Dr.~Christa Willnauer and Dr.~Beate Paulus for
providing codes and many helpful discussions.

I am also thankful to Prof. Dr. Hermann Stoll and
Prof. Cesare Pisani for their very helpful discussions.

Separately I would like to thank Dr.~Uwe Birkenheuer,
Dr.~Aleksey Bezuglyj and Prof.~Peter Fulde for the
critical reading of the manuscript of this thesis.

I also gratefully acknowledge the generous financial 
support by the Max Planck Society.

Last, but not least, I am deeply grateful to my parents
Aleksey and Galina, my wife Evgeniya and my son Alyosha for
their understanding and support during the whole time
of my work in Dresden.

\baselineskip=15pt

\end{document}